%% file: psource.tex
\documentclass[12pt,onecolumn]{mn2e}

\newcommand{\usepdfeps}{pdf}
\usepackage{graphicx}
\usepackage{rotating}
\usepackage{hyperref}
\usepackage{multirow}

\newcommand{\changed}[1]{#1}

\title[Statistical properties of polarized radio sources at high frequency etc]
  {Statistical properties of polarized radio sources at high frequency and their impact on CMB polarization measurements}
\author[Battye {\it et al.}]
  {R.A.~Battye, I.W.A.~Browne, M.W.~Peel,  N.J.~Jackson, C. Dickinson\\
   Jodrell Bank Centre for Astrophysics, School of Physics and Astronomy, \\
    The University of Manchester, Oxford Road, Manchester, M13 9PL}
\date{Released September 2009}

\def\LaTeX{L\kern-.36em\raise.3ex\hbox{a}\kern-.15em
    T\kern-.1667em\lower.7ex\hbox{E}\kern-.125emX}

\begin{document}

\label{firstpage}

\maketitle

\begin{abstract}
We have studied the implications of high sensitivity polarization
measurements of objects from the WMAP point source catalogue made
using the VLA at 8.4, 22 and 43~GHz. The fractional polarization of
sources is almost independent of frequency with a median of $\approx 2$
per cent and an average, for detected sources, of $\approx 3.5$ per
cent. These values are also independent of the total intensity over the narrow
range of intensity we sample.  Using a contemporaneous sample of 105
sources detected at all 3 VLA frequencies, we have investigated the
spectral behaviour as a function of frequency by means of a 2-colour
diagram. Most sources have power-law spectra in total intensity, as
expected. On the other hand they appear to be almost randomly
distributed in the polarized intensity 2-colour diagram. This is
compatible with the polarized spectra being much less smooth than
those in intensity and we speculate on the physical origins of this.
We have performed an analysis of the correlations between the
fractional polarization and spectral indices including computation of
the principal components. We find that there is little correlation
between the fractional polarization and the intensity spectral
indices. This is also the case when we include polarization
measurements at 1.4~GHz from the NVSS. In addition we compute 45
rotation measures from polarization position angles which are
compatible with a $\lambda^2$ law. We use our results to predict the
level of point source confusion noise that contaminates CMB
polarization measurements aimed at detecting primordial gravitational
waves from inflation.  We conclude that some level of source
subtraction will be necessary to detect $r\sim 0.1$ below 100~GHz and
at all frequencies to detect $r\sim 0.01$. We present estimates of the
level of contamination expected and the number of sources which need
to be subtracted as a function of the imposed cut flux density and
frequency.
\end{abstract}


\section{Introduction}

Synchrotron radiation is the primary emission mechanism for sources
detected in the radio band. Hence, radio sources are likely to be
linearly polarized with a fractional polarization of typically a few
percent, depending on the level of order of the magnetic field in the
emission producing regions. Large-scale polarization catalogues exist
for sources selected at low frequencies such as 1.4~GHz (NVSS - Condon
et al. 1998) with $\approx 2\times 10^5$ detections. At higher
frequencies Jackson et al. (2007) have looked at flat spectrum sources
in the CLASS survey (Myers et al., 2003; Browne et al.,
2003) and have $\approx 5000$ detections of 8.4~GHz
polarizations. In recent times more information has become available
at even higher frequencies. Ricci et al. (2004) have detected significant
polarization in 170 sources at 18.5~GHz, while Massardi et al. (2007)
present the results from follow-up of the AT20G survey (Sadler et al.,
2006). They have detected polarization at 22~GHz in 218 sources with
total intensity greater than 0.5Jy at 22~GHz with additional
information at 4.8 and 8.4~GHz. \changed{Murphy et al. (2010) present the full AT20G survey containing 5890 sources stronger than 40mJy at 20GHz. Most sources have near-simultaneous 5 and 8~GHz measurements and 1559 have detections of polarised emission at one or more frequencies.} In addition, Agudo et al (2009) have
presented measurements of 146 sources at 86~GHz made using the IRAM
telescope and Lopez-Caniego et al. (2009) have created a catalogue of
polarized sources detected from the WMAP maps between 22 and 90~GHz.

Our main motivation for studying the properties of polarized radio
sources is to assess their potential contamination of Cosmic Microwave
Background (CMB) polarization data. Measurements have been made of the
so-called E-mode polarization which is produced by scalar density
fluctuations (Kovac et al. 2002, Readhead et al. 2004, Montroy et
al. 2005, Page et al.  2007, Brown et al. 2009, Chiang et
al. 2009). Future instruments are aimed at detecting the B-mode
signature of primordial gravitational waves produced by inflation
(see, for example, Baumann et al. 2009) which is often quantified in
terms of the scalar-to-tensor ratio, $r$.  These will require not only
high signal-to-noise, but also exquisite control of systematics and
separation of astrophysical foregrounds. To date the only work on this
issue (de Zotti et al 1999, Mesa et al 2002, Tucci et al., 2004) has
relied on extrapolation of polarized source counts from frequencies
which are factors of $\approx 20$ lower than where the CMB observations
will take place.

We will focus on the implications of the measurements we have obtained
for a sample of 203 radio sources extracted from the WMAP 22~GHz
catalogue which was complete to $\approx 1{\rm Jy}$. Most of these sources were also detected 
in the NVSS (Condon et al., 1998) 
at 1.4~GHz and 71 were detected at 86~GHz using the IRAM 
telescope by Agudo et al. (2009); we will make use of this information
later. We studied sources in the region
with declinations greater than $-34^{\circ}$ and all were observed at
22 and 43~GHz using the Very Large Array (VLA). A subset of 134 were
also observed at 8.4~GHz. Observations were missing for 3 sources due
to misidentifications and a further 7 sources were deemed
inappropriate to include in the present statistical analysis for a
variety of reasons such as the source being very extended and being heavily 
resolved in our VLA observations. Polarized emission was detected for
123, 169 and 167 sources at 8.4, 22, 43~GHz respectively and 105 were
detected at all 3 frequencies. No polarization bias correcttion was applied 
since all detections are greater than $5\sigma$ and many are substantially 
stronger. This ``contemporaneous sample''
provides the first direct information at frequencies relevant to CMB
measurements allowing more accurate extrapolations to be made. The
details of the observations and the catalogue are presented in Jackson
et al. (2010).

\section{Statistical Properties of the contemporaneous sample}

In this section we will focus on the sample of 105 sources in which
polarized emission was detected at all 3 frequencies. Since all these
measurements were made on a single day they provide a snapshot of the
polarized source population and are unlikely to be significantly
affected by the variability that can be important for such sources.
Although it will be biased in some complicated way toward objects
which had strong polarized emission at the time of observation, this
sample is particularly useful for investigating the properties of the
sources as a function of frequency.  In the following section we will
describe our analysis of all our measurements plus those made at 8.4~GHz
as part of the CLASS program, at 1.4~GHz from the NVSS and at 86~GHz using the IRAM 
telescope which were
taken at different times and could be affected by variability.

\subsection{Summary of basic properties}

We define the intensity spectral index, with $S\propto\nu^{\alpha}$
between frequencies $\nu_1$ and $\nu_2$ by $\alpha^{\nu_2}_{\nu_1}$,
and the corresponding spectral index for polarized intensity, defined
in terms of the Stokes parameters $Q$ and $U$ by $P=\sqrt{Q^2+U^2}$,
as $\beta^{\nu_2}_{\nu_1}$. The fractional polarization at frequency
$\nu$ is defined to be $\Pi_\nu=P_\nu/S_\nu$. In an analogous way to
the total  and polarized intensity, one can also define a
spectral index for the fractional polarization,
$\gamma^{\nu_2}_{\nu_1}$.

\begin{figure}
\centering
\includegraphics[width=6cm]{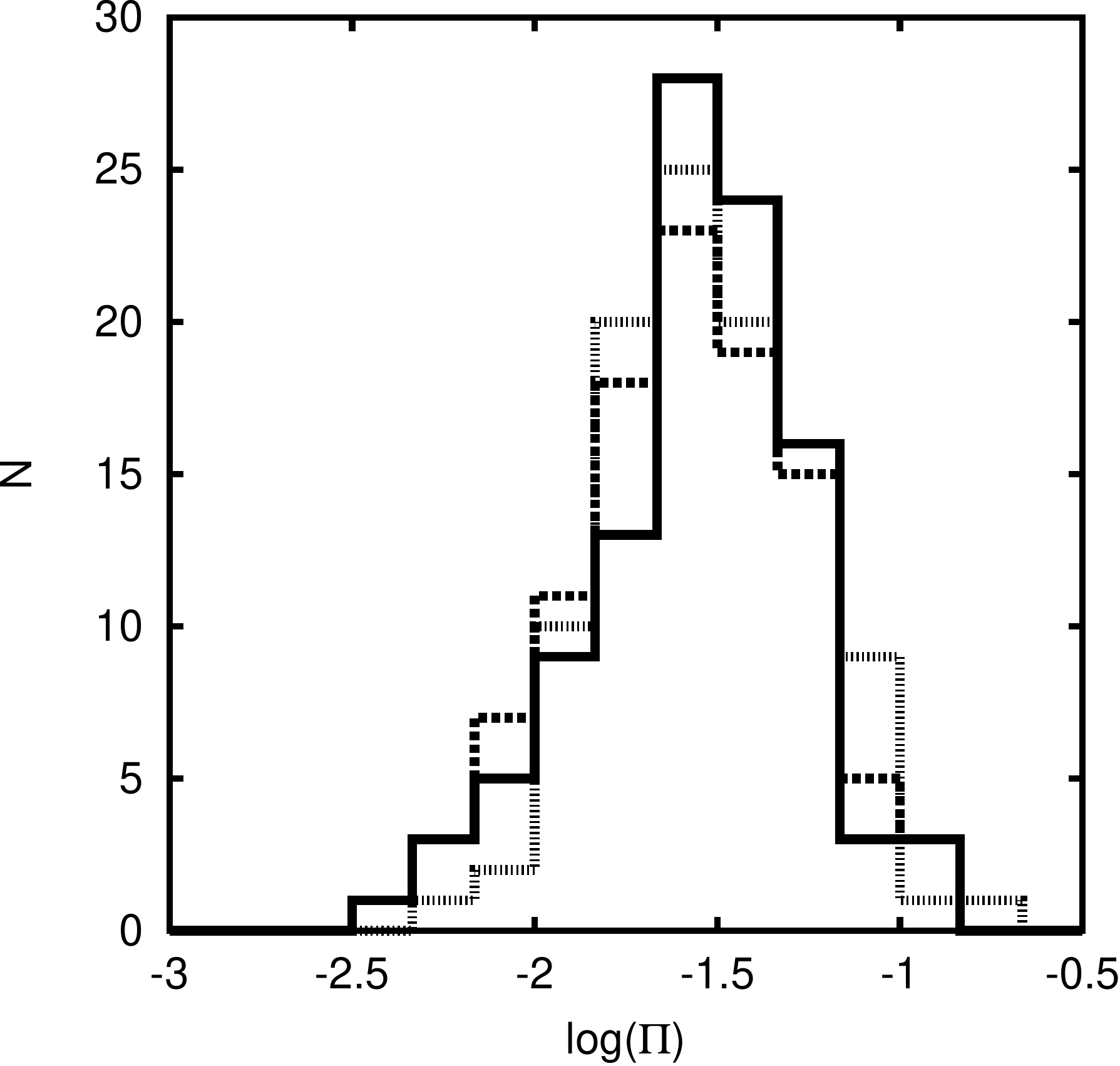}
\caption{Histogram of the number sources in the contemporaneous sample
as a function of the logarithm of the fractional polarization,
$\log(\Pi)$. The solid line is for 8.4~GHz, the dashed line is 22~GHz
and the dotted line 43~GHz. The distributions of fractional
polarizations appear to be relatively independent of frequency. It
should be possible to model each of them as a Gaussian in
$\log(\Pi)$ with a reasonable degree of accuracy as discussed in section 5.}
\label{fig:frachist}
\end{figure}

Histograms of the fractional polarization are presented in
Fig.~\ref{fig:frachist}. It is clear that the distributions only vary
very slightly, if at all, as a function of frequency. The mean
fractional polarization is computed to be $100\langle\Pi\rangle=3.3$,
3.3 and 3.7 at 8.4, 22, 43~GHz and the r.m.s. is
$100\langle\Pi^2\rangle^{1/2}=3.9$, 4.1 and 4.7.
Approximately the same level of fractional polarization was
seen in the sample studied by Massardi et al. (2007). The results
suggest that fractional polarization is approximately constant as a
function of frequency, although there is an increase in the dispersion
as frequency increases, something that is also seen in the Massardi et
al (2007) data.

\begin{figure}
\centering
\includegraphics[width=5.5cm]{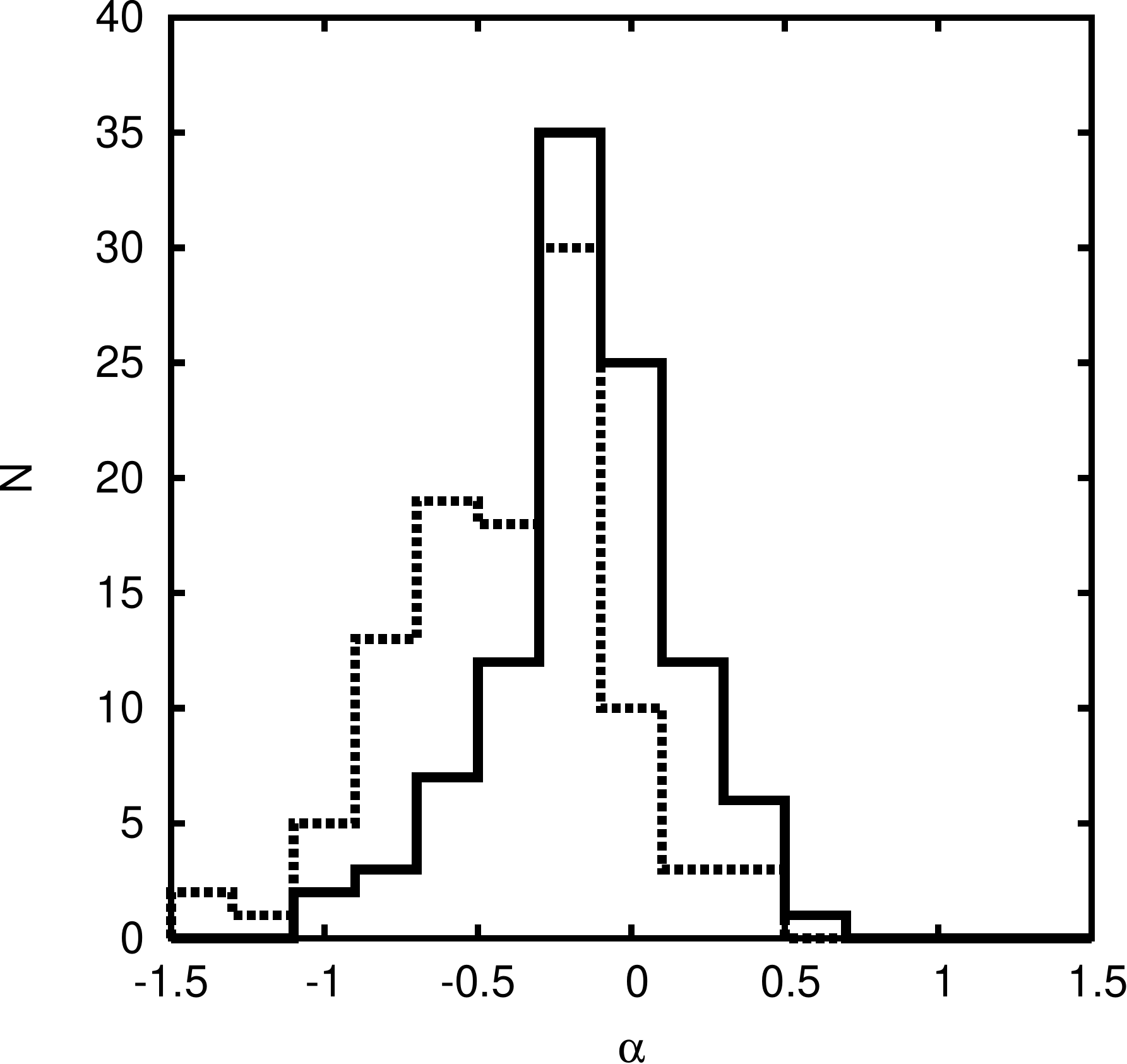}
\includegraphics[width=5.5cm]{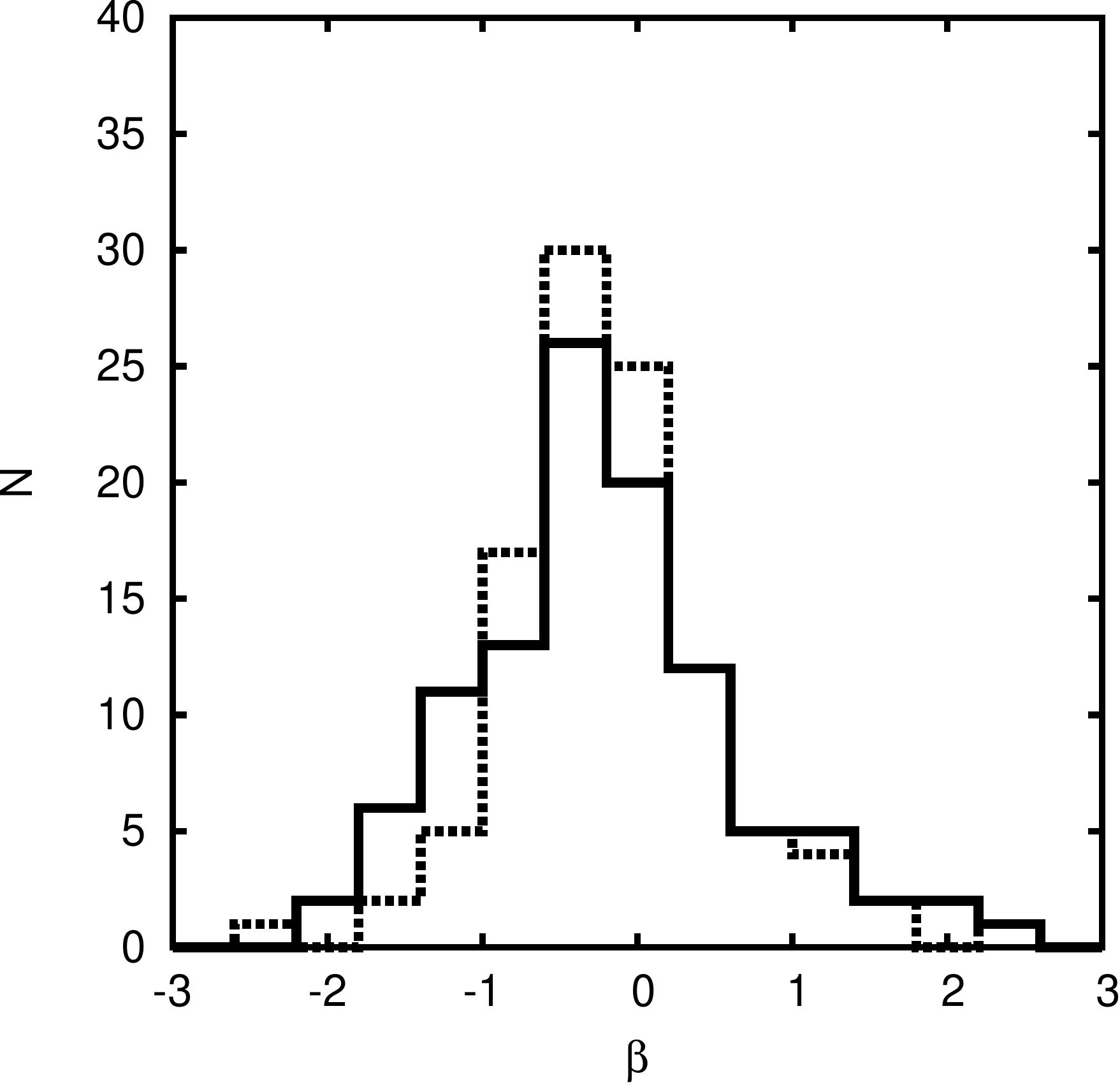}
\includegraphics[width=5.5cm]{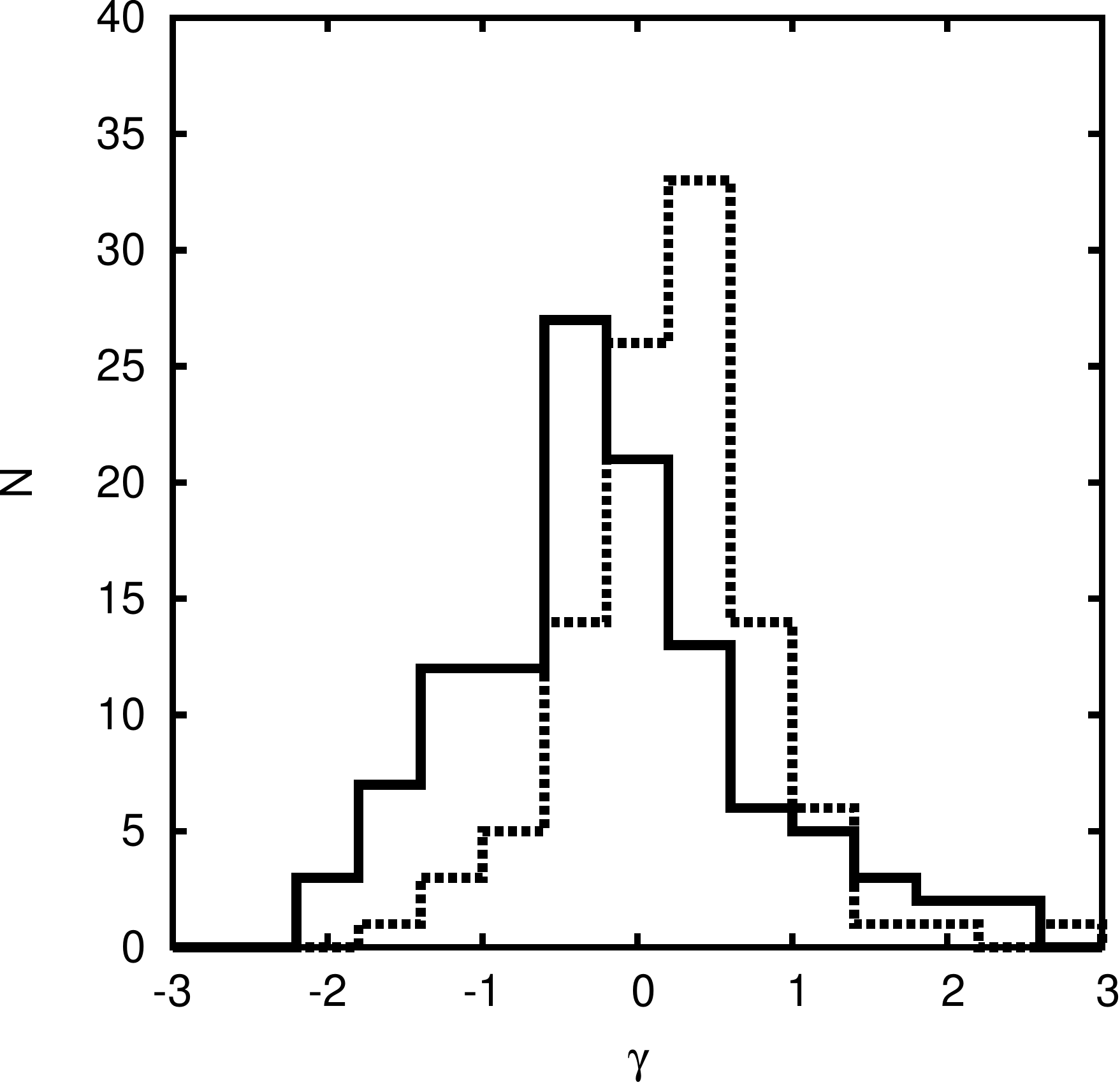}
\caption{Histograms of the number of sources in the contemporaneous
sample as a function of the spectral indices for the intensity (left), polarized intensity (middle) and fractional polarization (right). Solid lines are
the spectral indices measured between 8.4 and 22~GHz and the dashed
lines between 22 and 43~GHz. The intensity histograms are offset for
each other, whereas those for the polarized intensity are very similar
over the two different frequency ranges.}
\label{fig:spechist}
\end{figure}

Histograms of the intensity, polarized intensity and fractional
polarization spectral indices are presented in
Fig.~\ref{fig:spechist}. There is a shift in the intensity spectral
index distribution as one goes from 8.4-22~GHz to 22-43~GHz indicating,
as expected, that on average the intensity spectrum is steepening as
one goes to higher frequencies. The polarized intensity spectral index
distributions appear to be consistent with no significant variation with
frequency. The mean and r.m.s spectral indices within the sample were
computed to be $\alpha_{8.4}^{22}=-0.17\pm 0.36$,
$\alpha_{22}^{43}=-0.44\pm 0.47$, $\beta_{8.4}^{22}=-0.23\pm 0.85$,
$\beta_{22}^{43}=-0.20\pm 0.81$, $\gamma_{8.4}^{22}=-0.06\pm 0.84$ and
$\gamma_{22}^{43}=0.24\pm 0.67$ where the dispersion is quantified by the 
standard deviation within the sample and is not the error on the mean 
(which would be smaller by a factor of $1/\sqrt{N}$). Of course the $\alpha$s, $\beta$s and
$\gamma$s are not independent, thus since $\beta$ is approximately the
same for the two frequency ranges and $\alpha$ becomes more negative
for the higher frequency range, $\gamma$ will be more positive in the
higher range.

\subsection{Two colour diagrams}

\begin{figure}
\centering
\includegraphics[width=5.5cm]{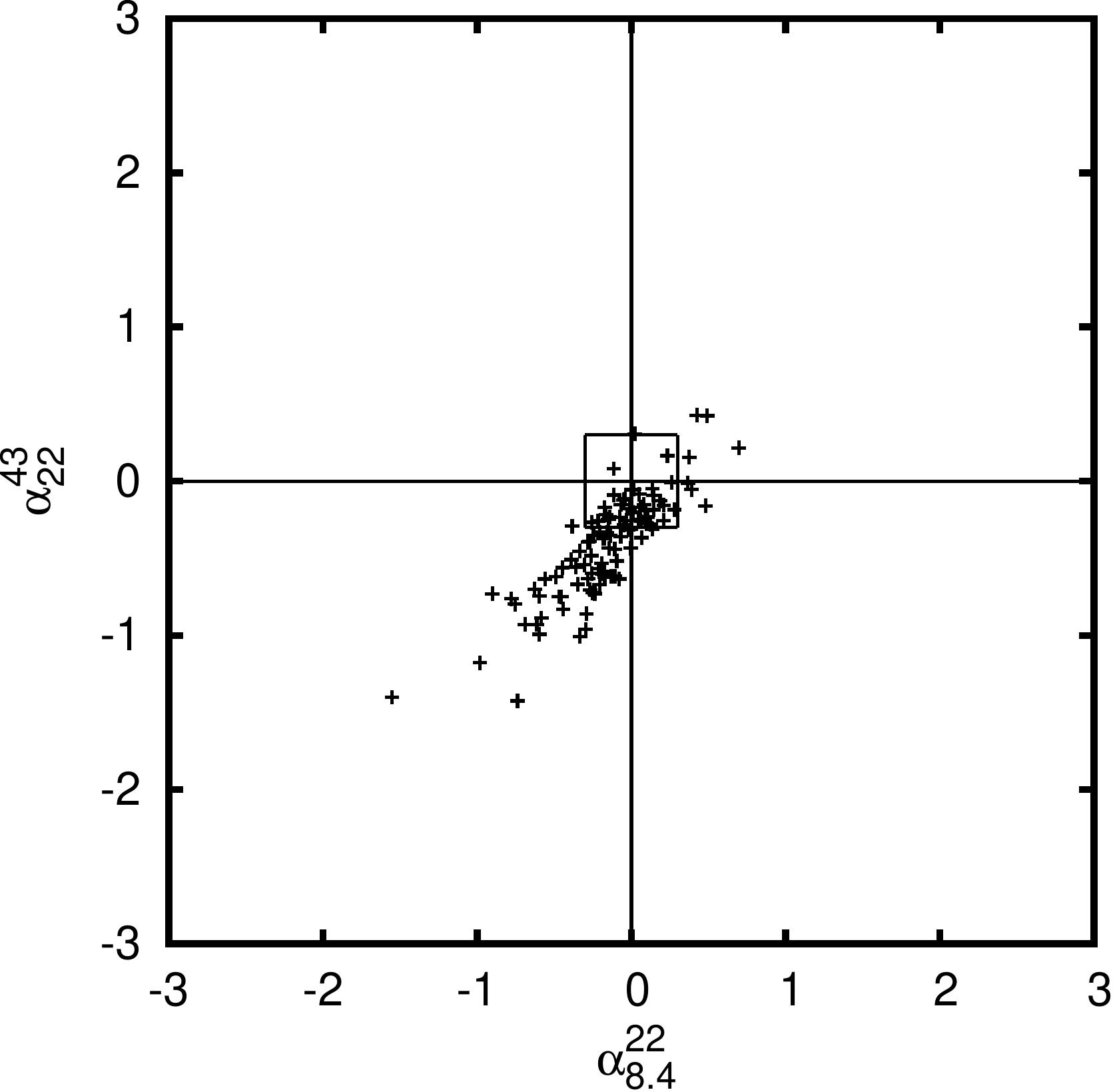}
\includegraphics[width=5.5cm]{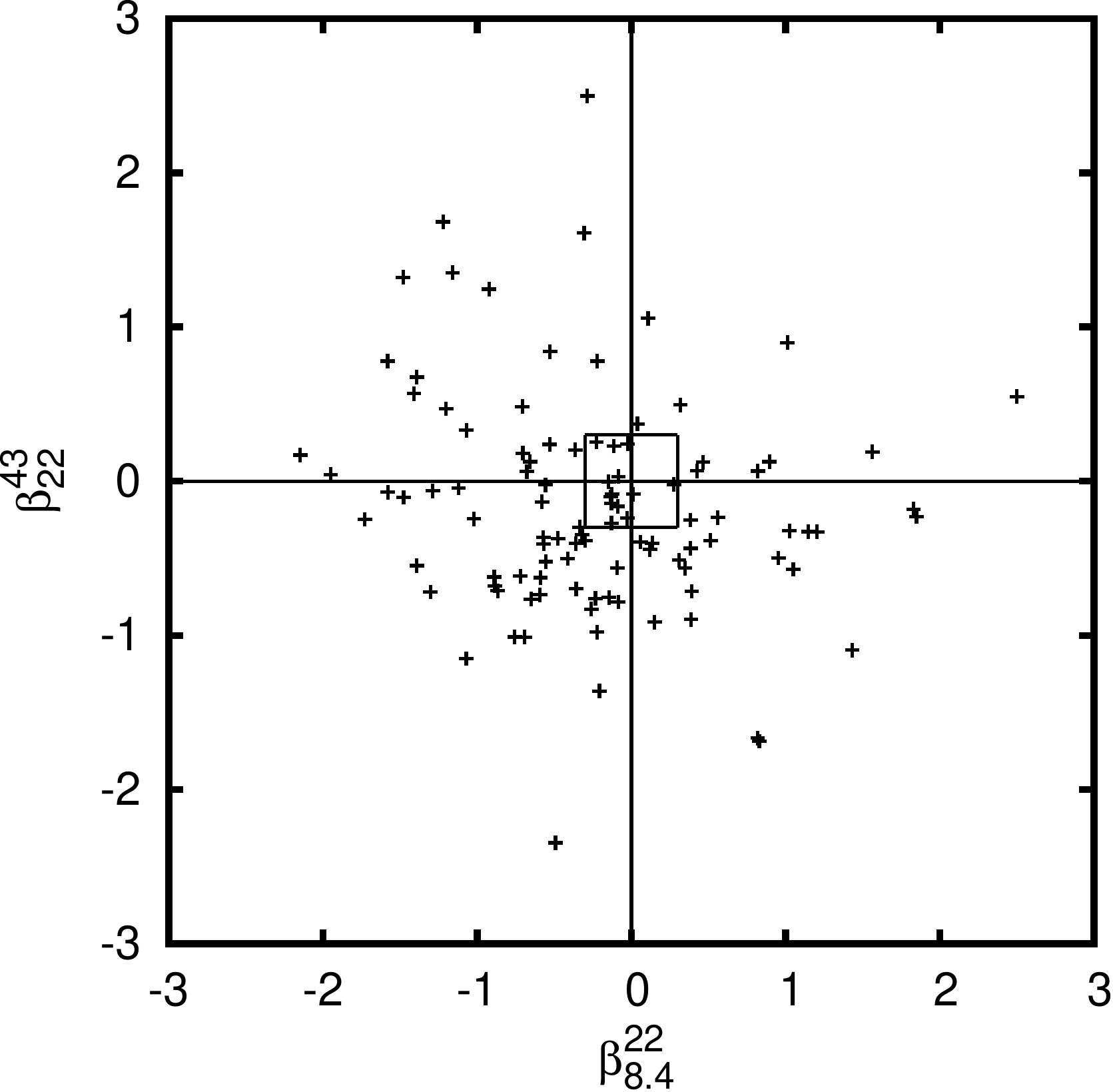}
\includegraphics[width=5.5cm]{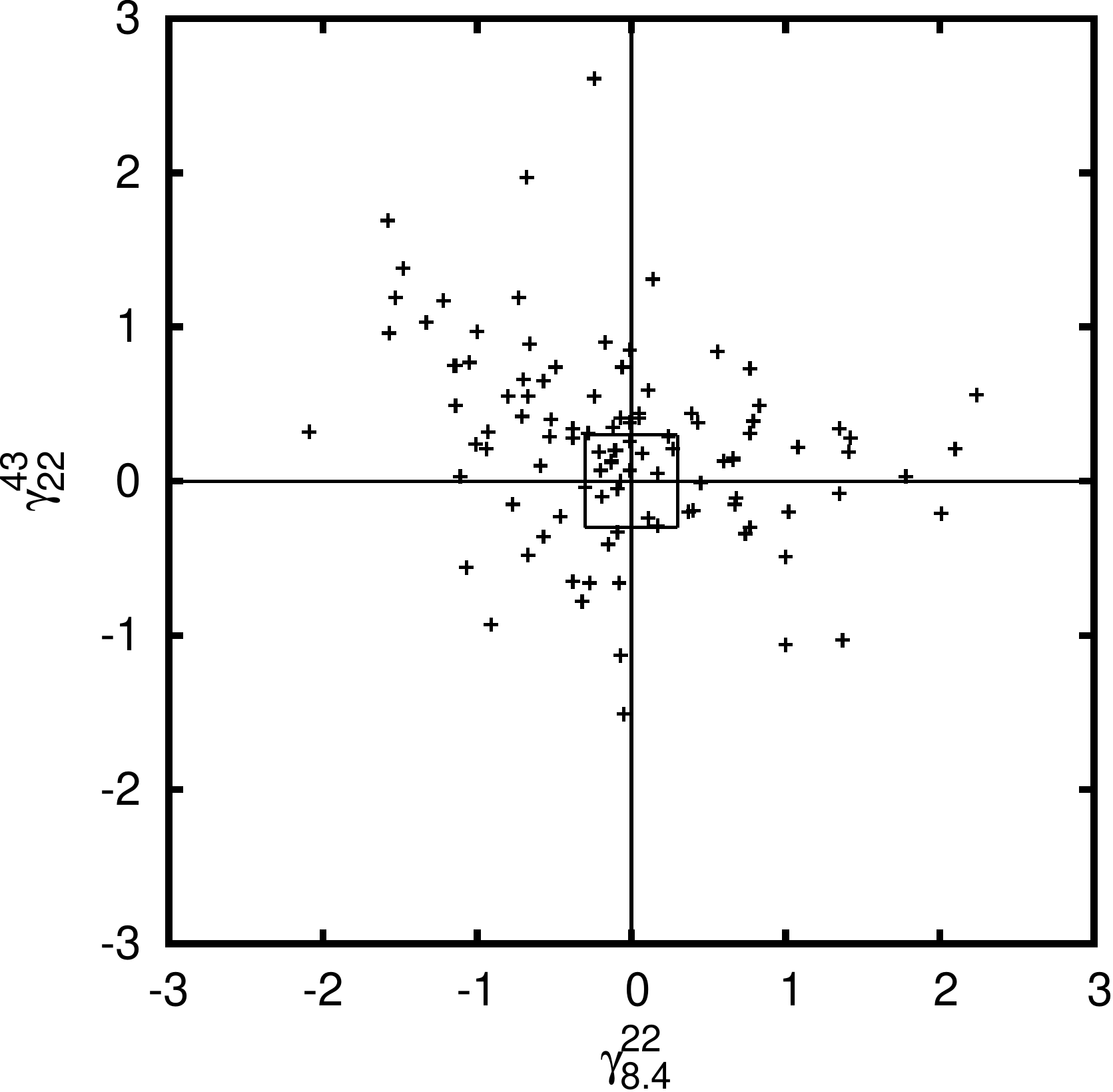}
\caption{The two colour diagram (8.4-22-43) for intensity (left), polarized intensity (middle) and fractional polarization (right) for the contemporaneous sample. The upper-left hand column quadrant corresponds to upturn (U) sources, the upper-right hand quadrant to inverted (I) sources, the lower-left hand quadrant to steep spectrum (S) sources and the lower-right hand quadrant to peaked spectrum (P) sources. The box in the middle corresponds to the flat spectrum region (F). There is a strong correlation between $\alpha_{8.4}^{22}$ and $\alpha_{22}^{43}$, very little correlation between $\beta_{8.4}^{22}$ and $\beta_{22}^{43}$ and an anti-correlation between $\gamma_{8.4}^{22}$ and $\gamma_{22}^{43}$. There are a number of upturn sources in both  polarized intensity and fractional polarization in marked contrast to intensity.}
\label{fig:twocol}
\end{figure}

In Fig.~\ref{fig:twocol} we present two colour diagrams for the
intensity and polarized emission for the spectral indices between 8.4
and 22~GHz, and between 22 and 43~GHz. These are larger versions of
plots presented in Fig.~\ref{fig:corrspec}. We define the
region $|\alpha_{8.4}^{22}|<0.3$ and $|\alpha_{22}^{43}|<0.3$ to be
flat spectrum in intensity. Excluding this region, the 4 quadrants of
the plot correspond to steep spectrum ($\alpha_{8.4}^{22}<0$ and
$\alpha_{22}^{43}<0$), peaked ($\alpha_{8.4}^{22}>0$ and
$\alpha_{22}^{43}<0$), inverted ($\alpha_{8.4}^{22}>0$ and
$\alpha_{22}^{43}>0$) and upturn ($\alpha_{8.4}^{22}<0$ and
$\alpha_{22}^{43}>0$). We will denote the upturn, inverted, flat, peaked and steep spectrum 
regions by the letters $U$, $I$, $F$, $P$ and $S$ respectively. Similar definitions can be made for the spectral
indices of the polarized intensity and fractional polarization.

Most of the sources have roughly power-law spectra in total intensity;
there are very few peaked sources, a few inverted sources and no
sources whose spectra upturn. The numbers of
sources in the different categories are
$(U,I,F,P,S)=(0,5,37,5,58)$. There is a strong correlation
(correlation coefficient 0.83) between the two intensity spectral
indices and most spectral indices are in the range
$-1.5<\alpha<0.5$.

The situation is very different for the polarized spectral
indices. There appears to be no correlation (correlation coefficient
-0.11) between the two polarization spectral indices, with each of the
4 quadrants in Fig.~\ref{fig:twocol} being populated by significant
numbers of sources. The numbers in the different categories are
$(U,I,F,P,S)=(21,10,13,22,39)$. Also worthy of note is the fact that
there are a number of sources which have upturn spectra in
polarized flux density in this frequency range. 

The range of fractional polarization spectral indices is also large,
but there is now an anti-correlation (correlation coefficient -0.59) between
the indices in the two frequency ranges. 
The numbers of sources in each of the categories are
$(U,I,F,P,S)=(39,21,18,13,14)$. More sources have increasing
fractional polarizations with frequency than decreasing and, if such a
trend is continued to frequencies $\approx$100~GHz, this could have
significant implications for CMB observations.

In order to explore the complex interaction between the spectral
behaviour in intensity, polarized intensity and fractional
polarization we have computed two matrices as done in Massardi et al.
(2007). These matrices are presented in
Tables~\ref{tab:pol} and \ref{tab:fracpol}. Each of the columns corresponds to a spectral type in intensity
and each of the rows the equivalent in either polarized intensity or
fractional polarization. 

\subsection{Correlations between fractional polarization and spectral indices}
\label{sect:indices}

\begin{figure}
\includegraphics[scale=0.3]{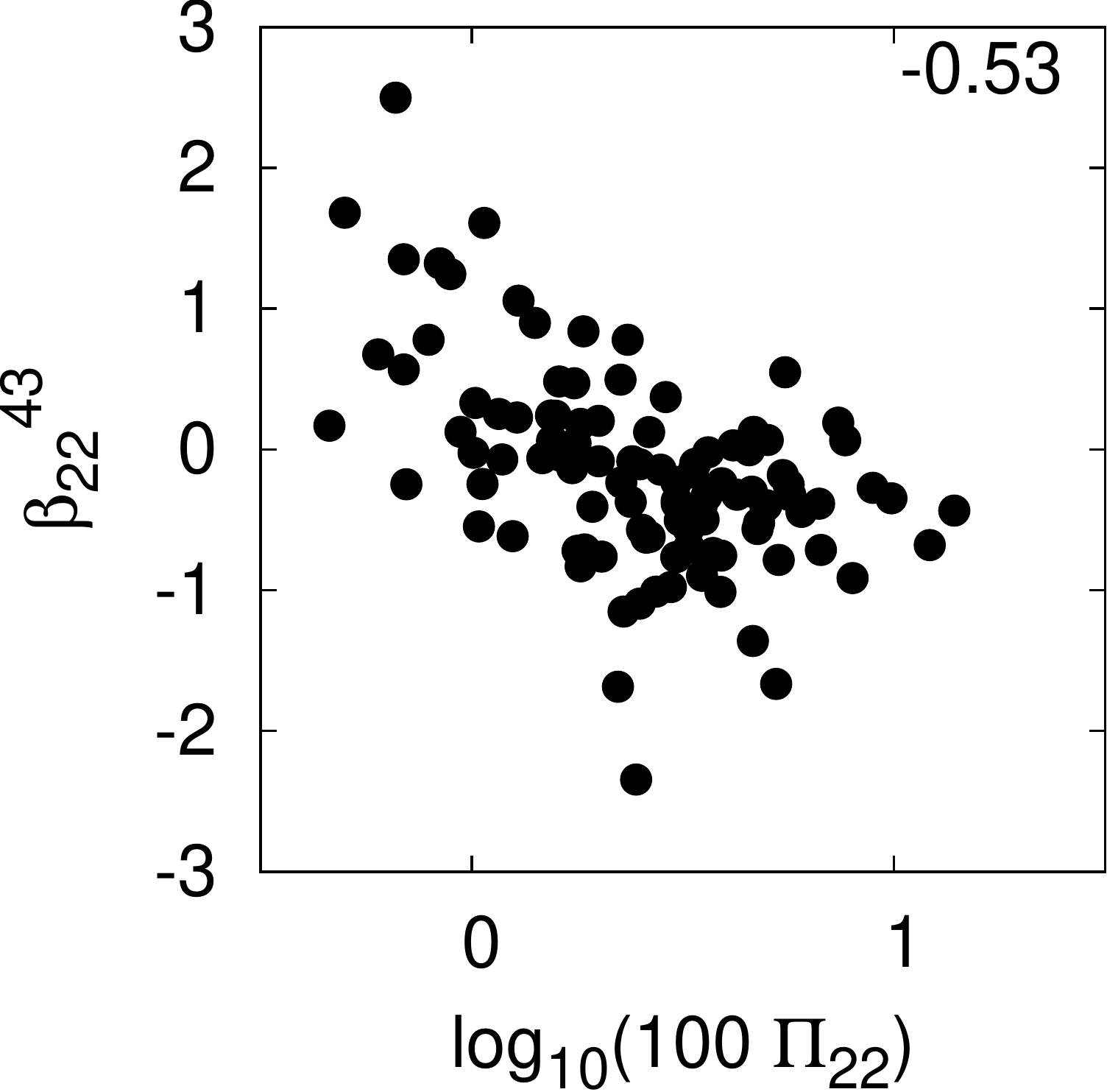}
\includegraphics[scale=0.3]{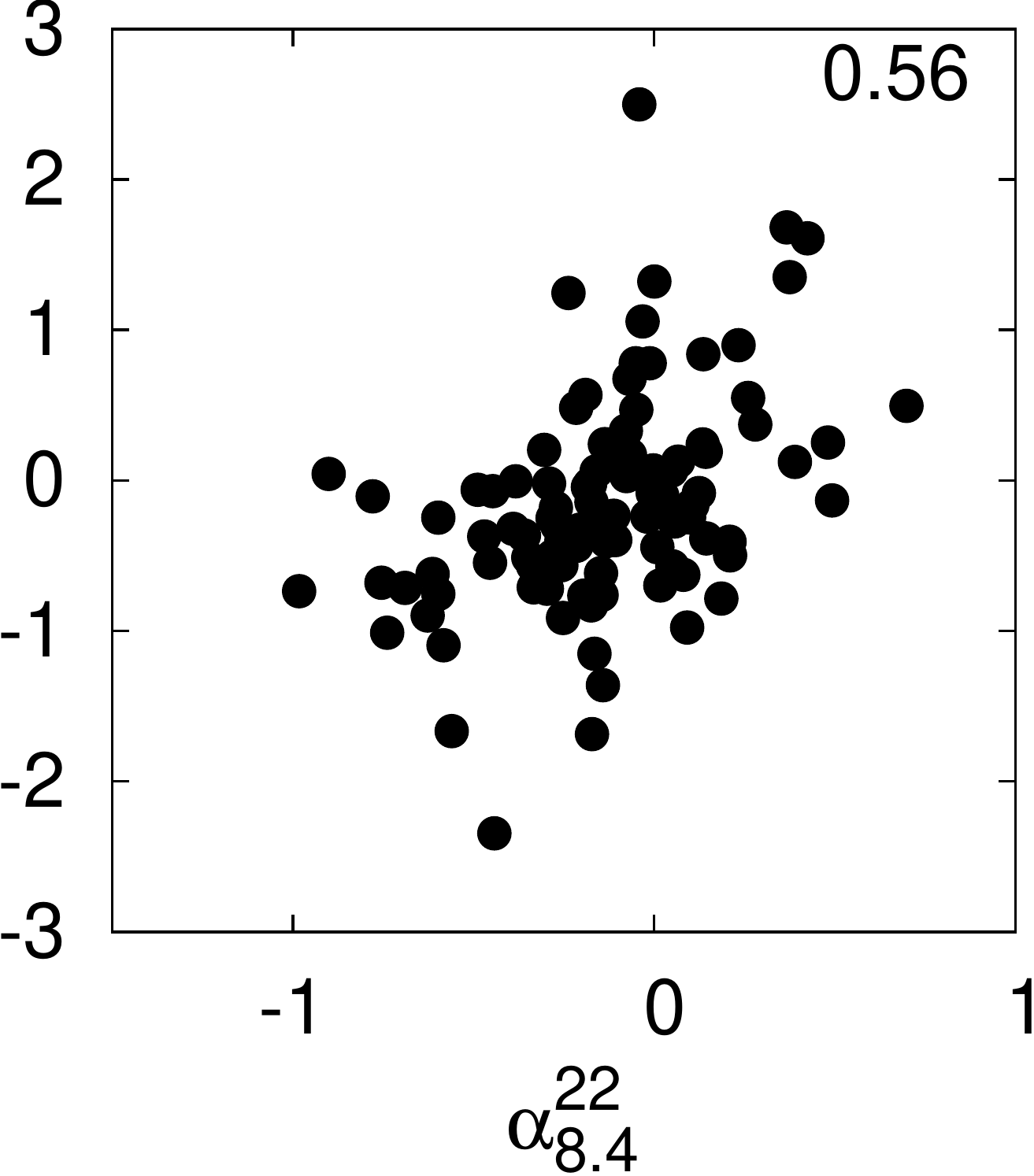}
\includegraphics[scale=0.3]{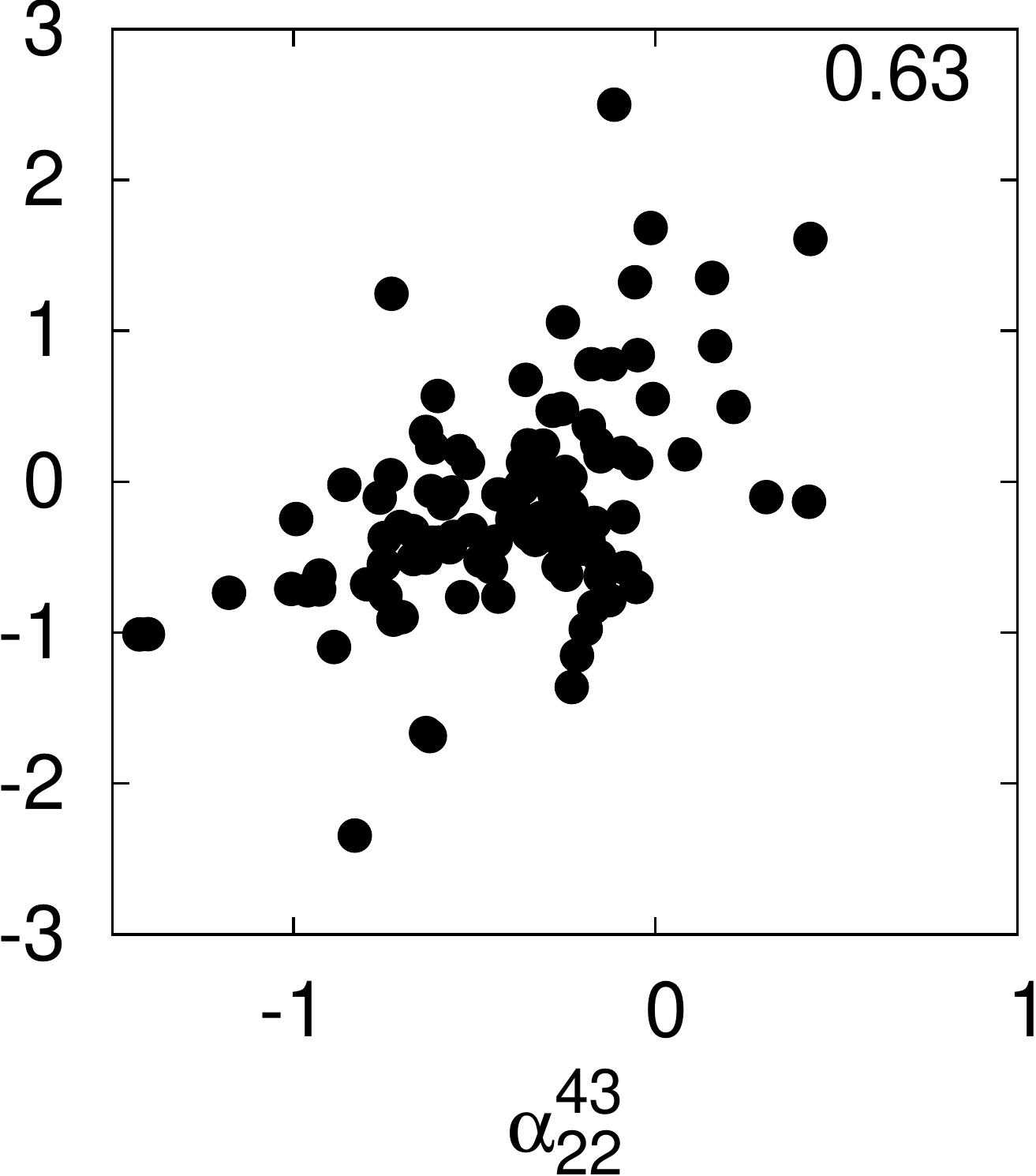}
\includegraphics[scale=0.3]{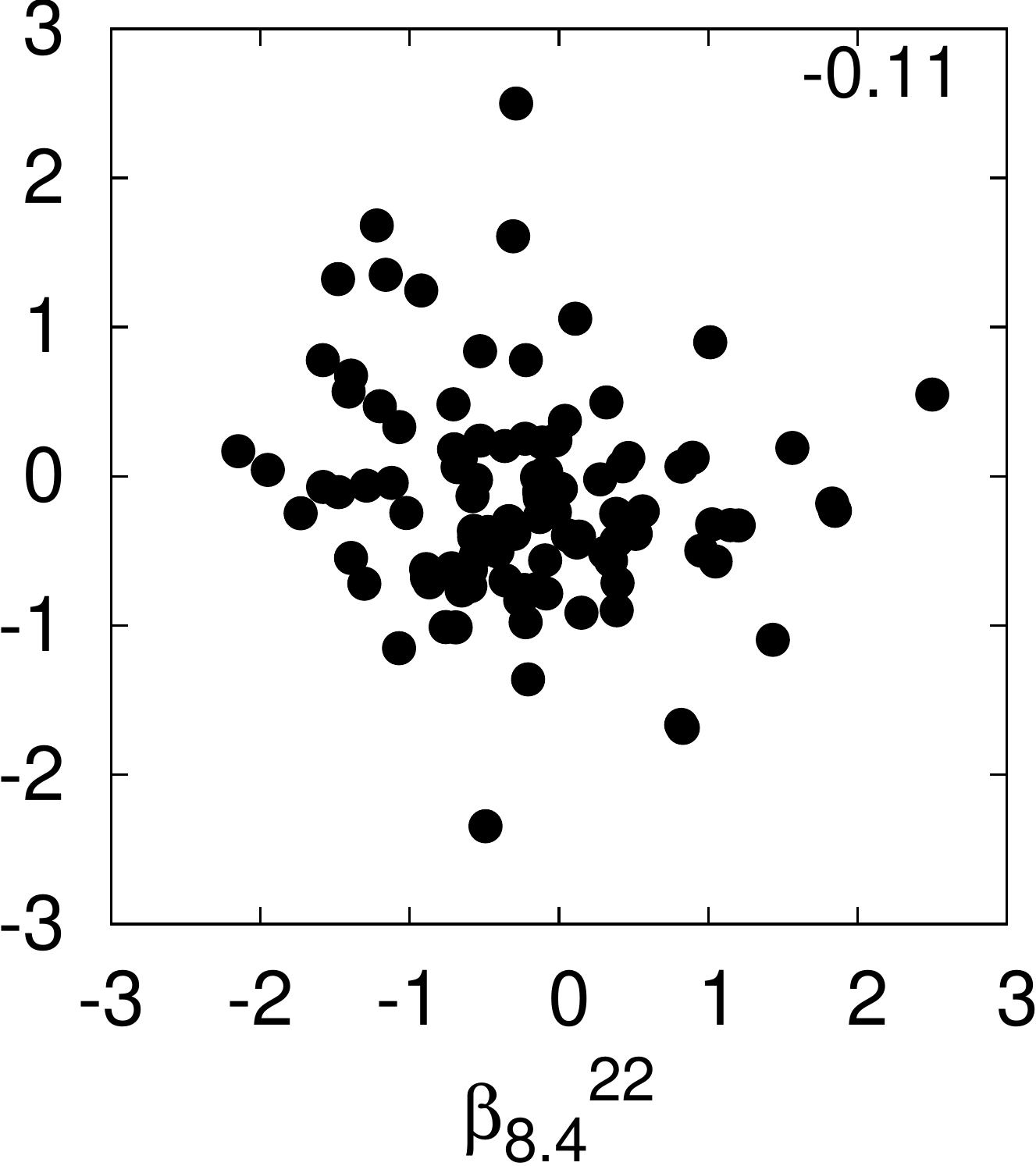}\\
\includegraphics[scale=0.3]{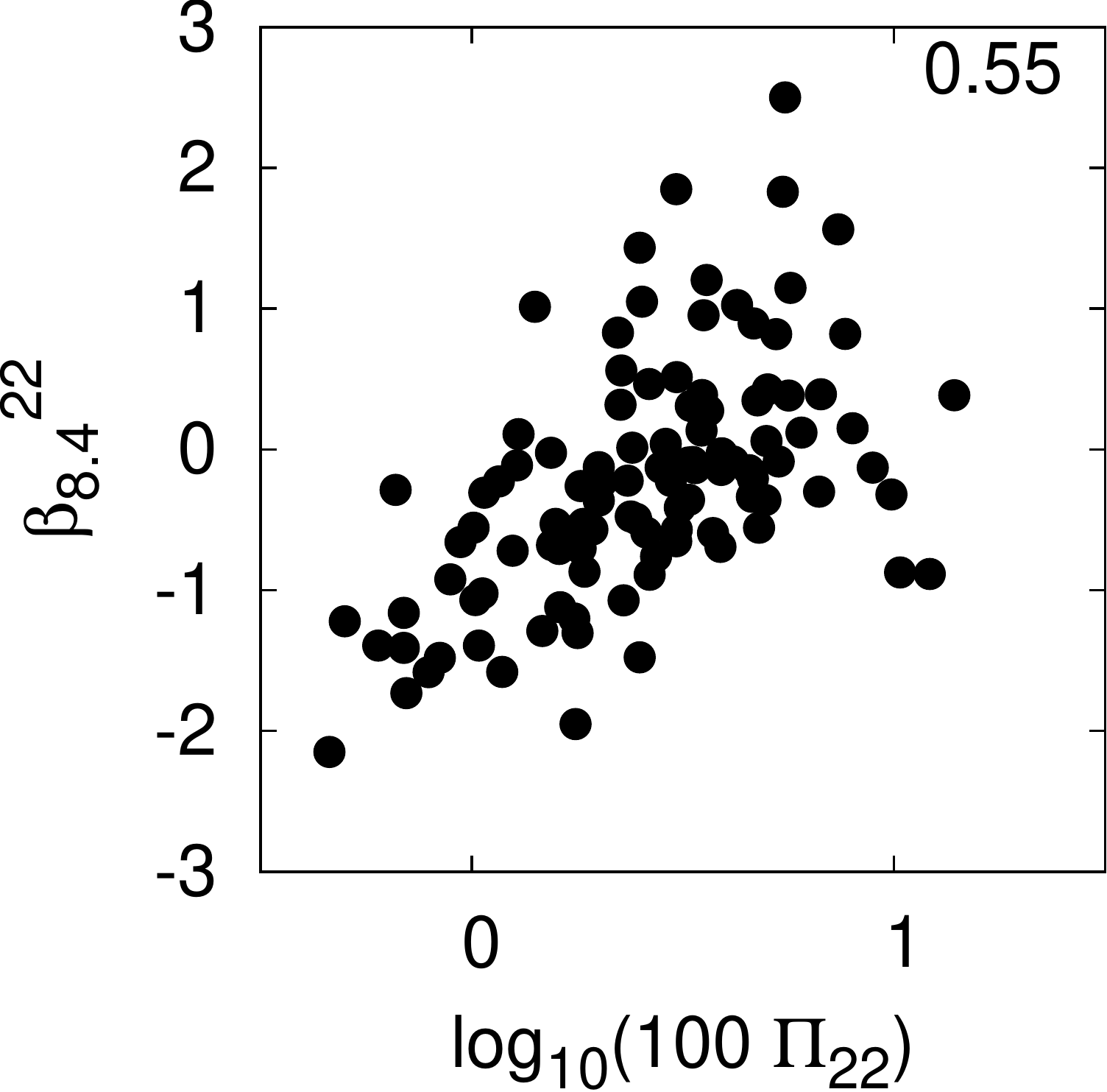}
\includegraphics[scale=0.3]{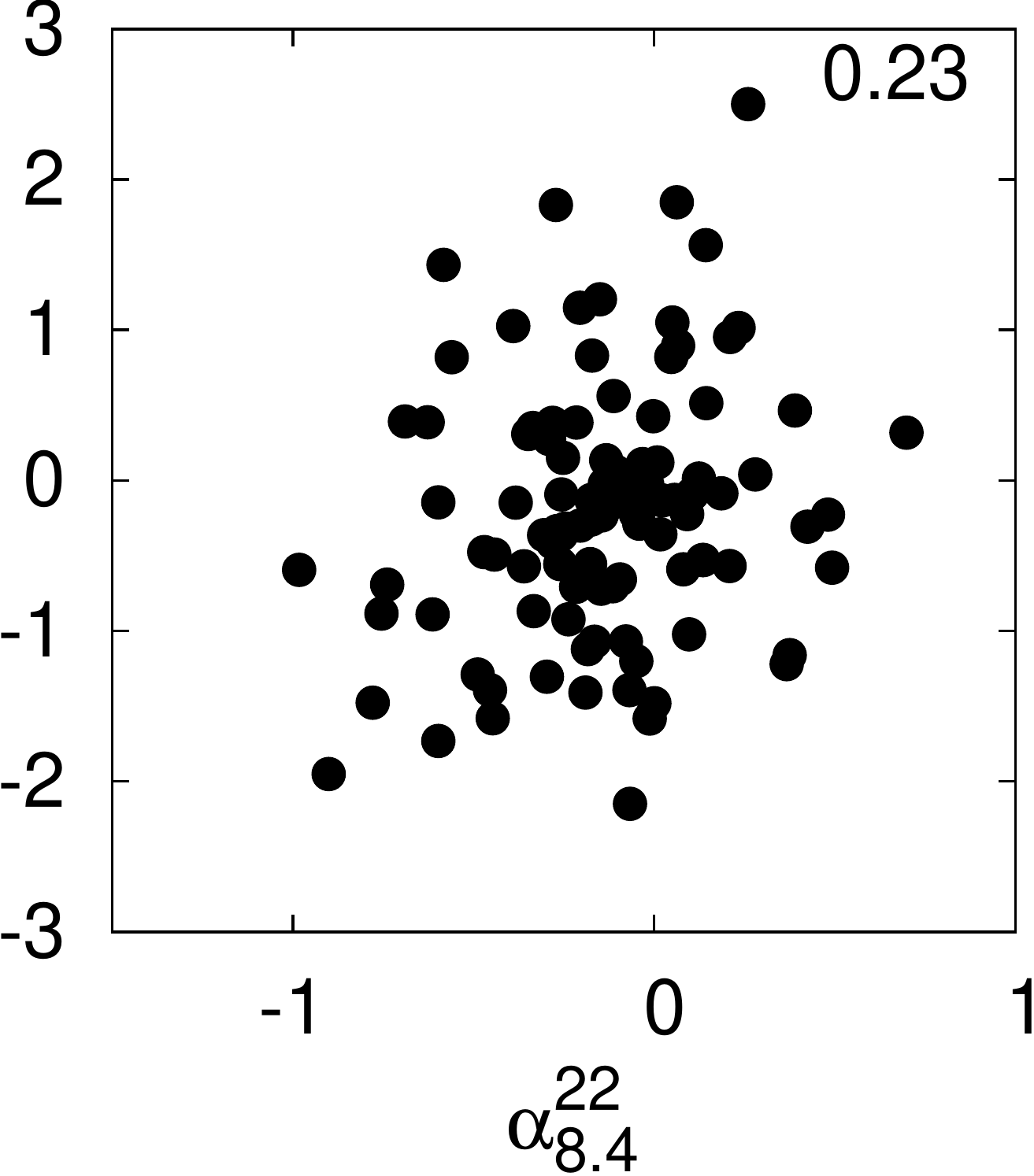}
\includegraphics[scale=0.3]{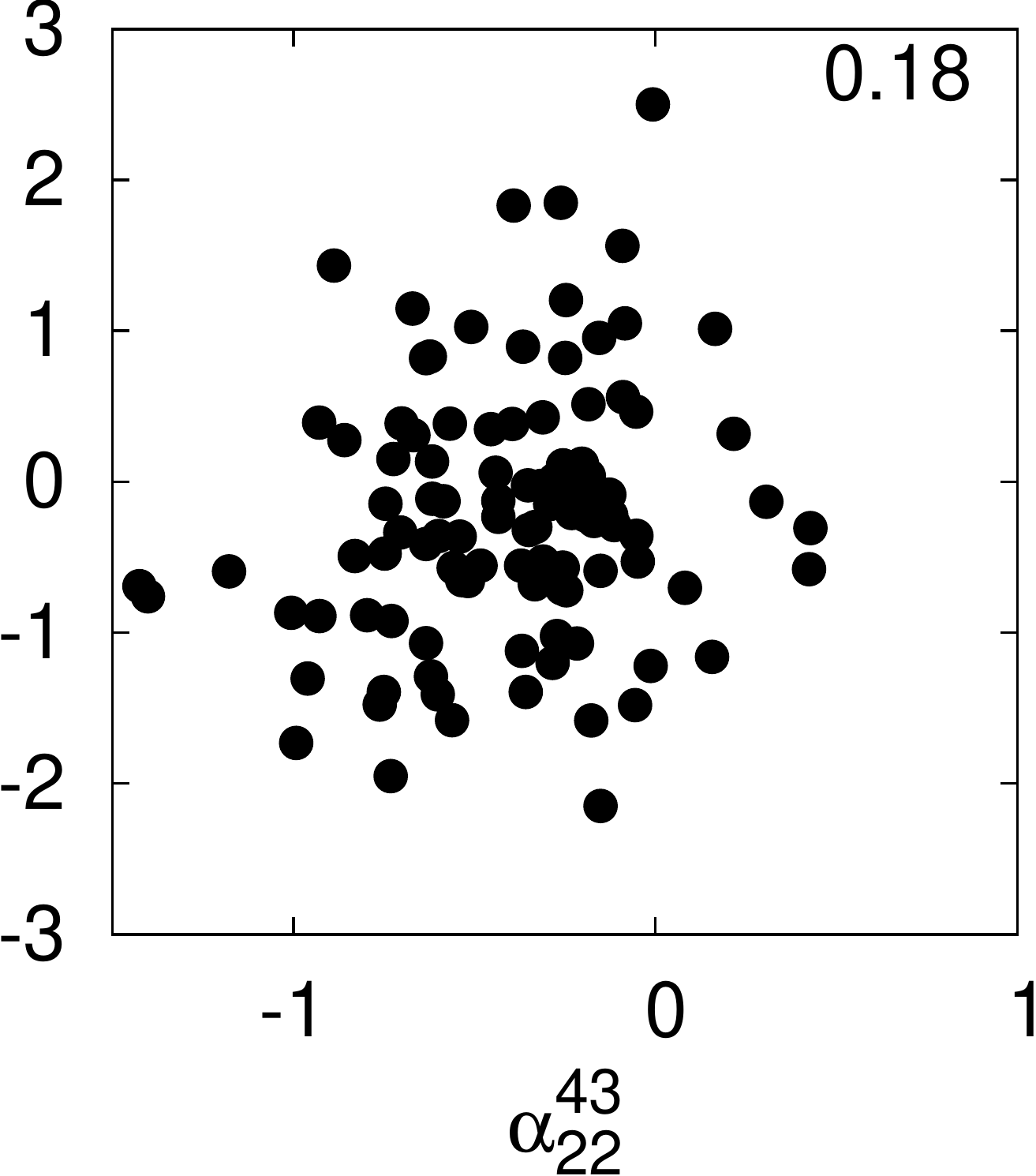}\\
\includegraphics[scale=0.3]{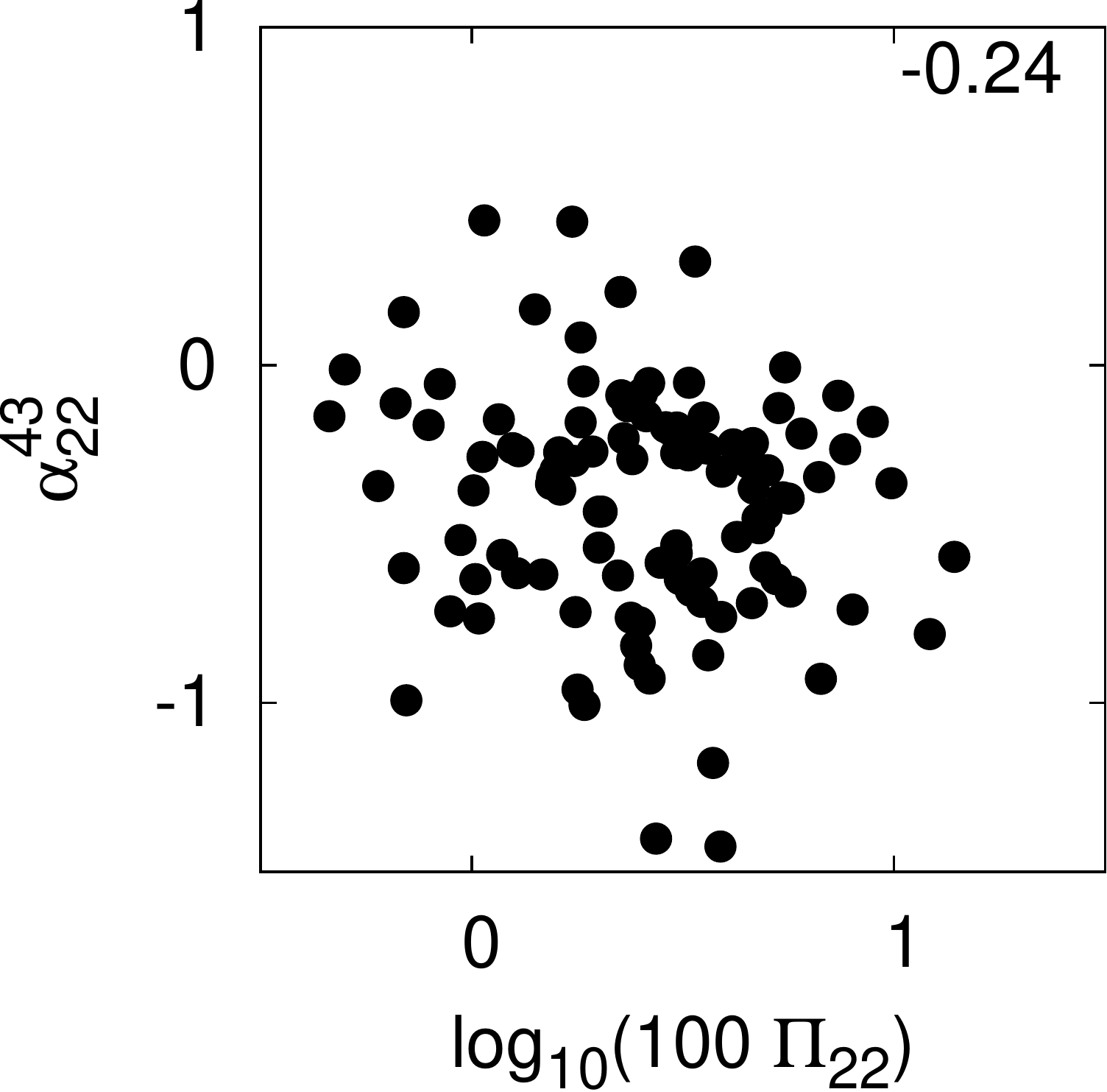}
\includegraphics[scale=0.3]{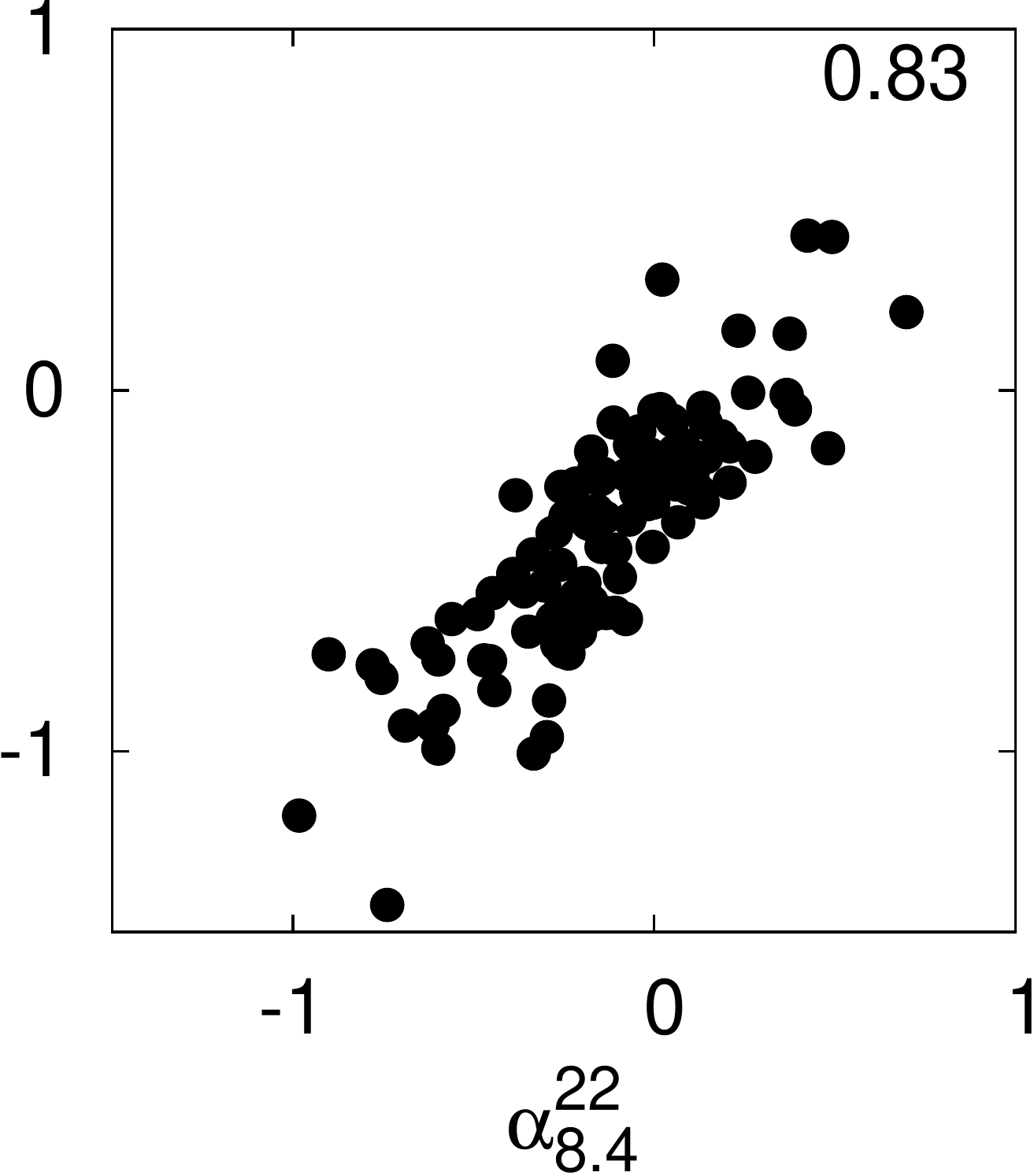}\\
\includegraphics[scale=0.3]{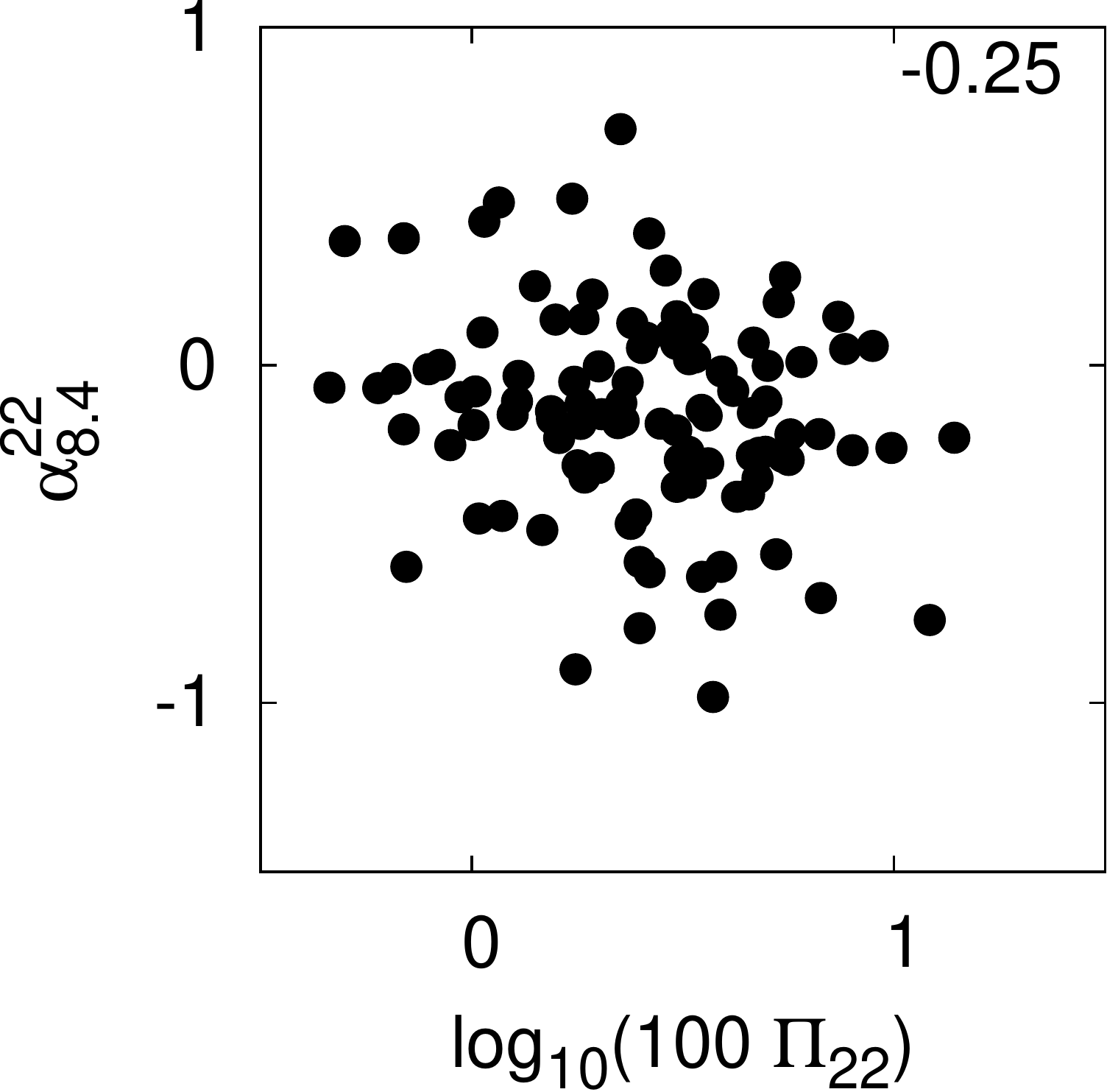}
\caption{Correlations between 5 derived quantities, $\log_{10}\Pi_{22}$, $\alpha_{8.4}^{22}$, $\alpha_{22}^{43}$, $\beta_{8.4}^{22}$ and $\beta_{22}^{43}$ for the contemporaneous sample, along with a table of correlation coefficients. The strongest correlation is between $\alpha_{8.4}^{22}$ and $\alpha_{22}^{43}$. }
\label{fig:corrspec}
\end{figure}

We can also perform a correlation analysis between the fractional
polarization, $\Pi$, and the spectral indices $\alpha$ and
$\beta$. There are many possible correlations which could be presented
between 3 fractional polarizations and 6 spectral indices. However,
not all are independent. Extracting out the overall intensity flux
scale $\approx 1{\rm Jy}$ which is natural when considering ratios (or
logs of ratios), we have chosen to present correlations between 5
quantities, the logarithm of the fractional polarization at 22~GHz,
$\Pi_{22}$, and the spectral indices, both in intensity and
polarization, to the other two frequencies, $\alpha_{8.4}^{22},
\alpha_{22}^{43}$ and $\beta_{8.4}^{22}, \beta_{22}^{43}$. The
results of this are presented in Fig.~\ref{fig:corrspec}.

\begin{table}
\begin{center}
\begin{tabular}{|c|c|c|c|c|c|}
\hline & U & I & F & P & S\\ \hline\hline U & 0 & 2 & 9 & 2 & 8\\ I &
 0 & 1 & 6 & 2 & 1 \\ F & 0 & 1 & 4 & 1 & 7 \\ P & 0 & 0 & 7 & 0 & 15
 \\ S & 0 & 1 & 11 & 0 & 27 \\ \hline\hline
\end{tabular}
\end{center}
\caption{Matrix containing the number of sources with a given spectral type in intensity (columns) and polarized intensity (rows) for the contemporaneous sample. U, I, F, P, S signify the upturn, inverted, flat, peaked and steep spectrum spectral types defined in the text.}
\label{tab:pol} 
\end{table}

\begin{table}
\begin{center}
\begin{tabular}{|c|c|c|c|c|c|}
\hline
 & U & I & F & P & S\\
\hline\hline
U & 0 & 3 & 11 & 3 & 22\\
I & 0 & 0 & 6 & 1 & 14 \\
F & 0 & 0 & 6 & 1 & 11 \\
P & 0 & 0 & 5 & 0 & 8 \\
S & 0 & 2 & 9 & 0 & 3 \\
\hline\hline
\end{tabular}
\end{center}
\caption{Matrix containing the number of sources with a given spectral type in intensity (columns) and fractional polarization (rows) for the contemporaneous sample. U, I, F, P, S signify the upturn, inverted, flat, peaked and steep spectrum spectral types defined in the text.}
\label{tab:fracpol} 
\end{table}

We have performed a principal component analysis (PCA) of this combination of data using 5 variables ($\log_{10}\Pi_{22}$, $\alpha_{8.4}^{22}$, $\alpha_{22}^{43}$, $\beta_{8.4}^{22}$, $\beta_{22}^{43}$). The principal components are 
\begin{eqnarray}
c_1&=&-0.30\log_{10}\Pi_{22}-0.77\alpha_{8.4}^{22}+0.55\alpha_{22}^{43}+0.08\beta_{8.4}^{22}-0.07\beta_{22}^{43}\,,\nonumber\\
c_2&=& 0.92\log_{10}\Pi_{22}-0.22\alpha_{8.4}^{22}+0.24\alpha_{22}^{43}-0.19\beta_{8.4}^{22}-0.15\beta_{22}^{43}\,,\nonumber\\
c_3&=&-0.01\log_{10}\Pi_{22}-0.50\alpha_{8.4}^{22}+0.67\alpha_{22}^{43}+0.20\beta_{8.4}^{22}-0.50\beta_{22}^{43}\,,\nonumber\\
c_4&=& 0.12\log_{10}\Pi_{22}+0.23\alpha_{8.4}^{22}+0.29\alpha_{22}^{43}+0.88\beta_{8.4}^{22}+0.26\beta_{22}^{43}\,,\nonumber\\
c_5&=&-0.24\log_{10}\Pi_{22}+0.21\alpha_{8.4}^{22}+0.33\alpha_{22}^{43}-0.37\beta_{8.4}^{22}+0.81\beta_{22}^{43}\,.\\
\end{eqnarray}
These are listed in order of ascending variance, that is, the first component has the lowest variance. We find that $c_1=0.36\pm 0.16$, $c_2=-1.51\pm 0.21$, $c_3=-0.22\pm 0.36$, $c_4=-0.61\pm 0.86$ and $c_5=0.13\pm 0.93$. The lowest variance component suggests that there is a complicated connection between the fractional polarized intensity and the total intensity and polarized intensity spectral indices.

This can be further investigated by performing a PCA using just two ($\alpha_{22}^{43}$, $\alpha_{8.4}^{22}$) and three ($\log_{10}\Pi_{22}$, $\alpha_{22}^{43}$, $\alpha_{8.4}^{22}$) variables. We find that the lowest variance components for these two cases are $-0.81\alpha_{8.4}^{22}+0.58\alpha_{22}^{43}=-0.12\pm 0.17$ and $-0.02\log_{10}\Pi_{22}-0.81\alpha_{8.4}^{22}+0.59\alpha_{22}^{43}=-0.08\pm 0.16$, respectively. It is clear that there is a strong correlation between the intensity spectral indices and they are almost independent of the the fractional polarization. Hence, the connection between the fractional polarization and the spectral indices can only be established once the polarized intensity spectral indices are included.

\subsection{Rotation measures}

It is clear from Figure 6 of Jackson et al. \changed{(2010)} and from their
Table 2 that the polarization position angles measured for many of the
sources are not the same at all three frequencies. If the source is
simple and there is a simple Faraday screen then one would expect the
measured position angle to satisfy $\phi=\phi_{\rm int}+{\rm
RM}\lambda^2$ where $\phi_{\rm int}$ is the intrinsic position angle
and RM is the rotation measure. However, many sources are not
compatible with this simple law and there are several possible reasons
for this:

\begin{itemize}
\item it could arise from the polarization position angle of a
single component changing as it moves from being optically thick to being
optically thin;

\item the source could have multiple components each
having different spectral indices and each with different
intrinsic polarization position angles and/or suffering different
Faraday rotations;

\item it could result from the different resolutions of the
observations at the different frequencies being more or less sensitive
to different parts of the source having different polarization position angles;

\end{itemize}
or a combination of these reasons.

We have looked for the $\lambda^2$ dependence in all those sources
with three contemporaneous detections of polarization. The method we
use is to compute the rotation measures necessary to be compatible
with each of the two angle differences $\phi_{22}-\phi_{43}=7.8\times
10^{-3}{\rm deg}({\rm RM}/{\rm rad}\,{\rm m}^{-2})$ and
$\phi_{8.4}-\phi_{22}=6.2\times 10^{-2}{\rm deg}({\rm RM}/{\rm
rad}\,{\rm m}^{-2})$. The errors on the combined rotation measure and
the intrinsic position angle are computed by combining errors in
quadrature. Those which are compatible within the $2\sigma$ errors are
deemed to be acceptable. The angle difference $\phi_{8.4}-\phi_{22}$
is more sensitive to rotation than $\phi_{22}-\phi_{43}$ and hence we
allow for $\pm 180^{\circ}$ or $\pm 360^{\circ}$ to be added to
$\phi_{8.4}-\phi_{22}$ in order to account for possible
ambiguities inherent in estimating rotation measures for 3 angle
measurements.

We find satisfactory $\lambda^2$ fits for 45 of the 105 sources and
the resulting rotation measures are presented in
Table~\ref{tab:rm}. We caution against believing every individual
rotation measure since the process of adding $\pm 180^{\circ}$ and
$\pm 360^{\circ}$ can lead to some spurious matches to the $\lambda^2$
law but the fact that the position angle measurements typically have
error bars $<\pm10^{\circ}$ will make this unlikely.

\begin{figure}
\centering
\includegraphics[width=8cm]{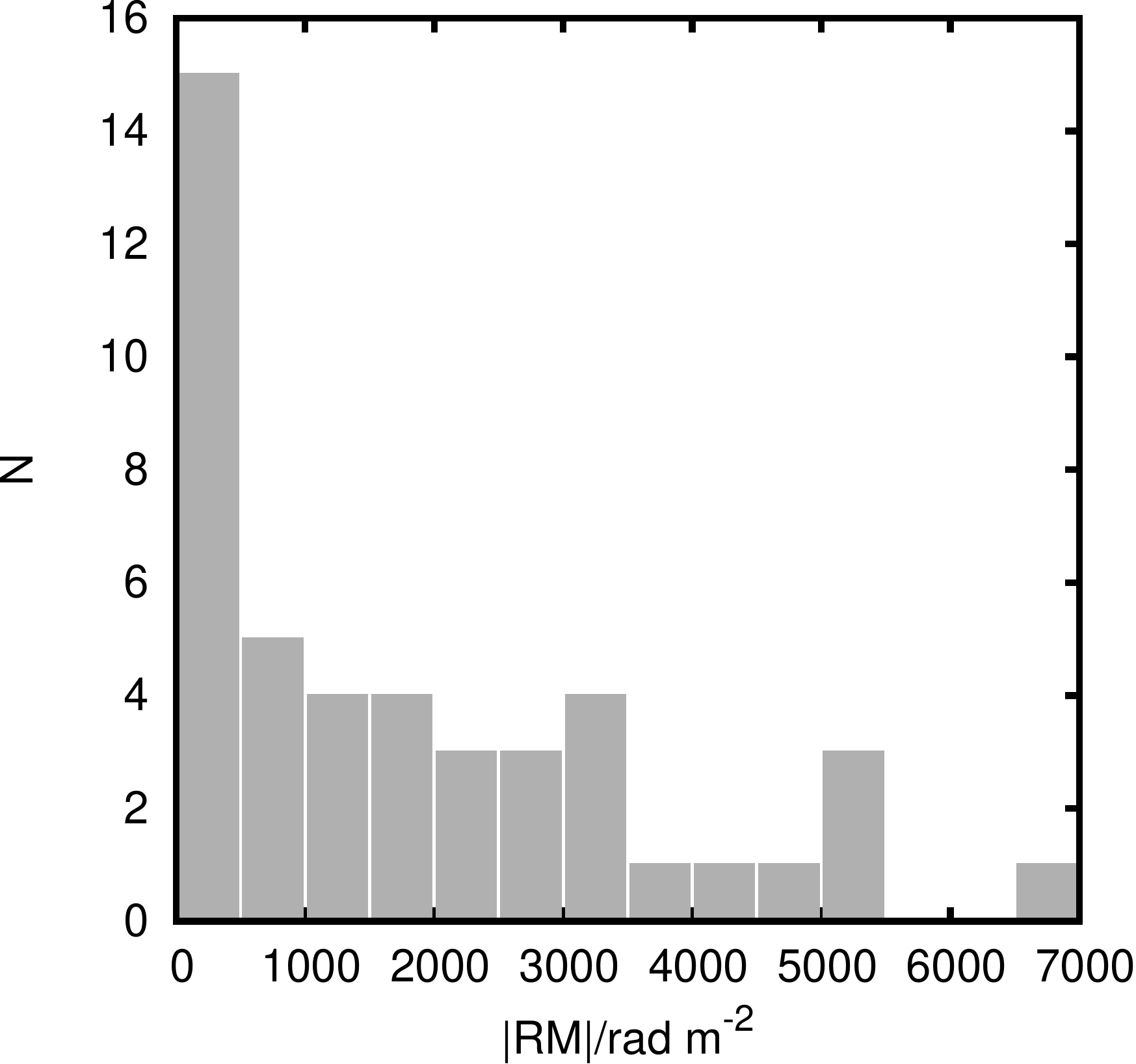}
\caption{Histogram of the estimated modulus of the rotation measures for the 45 sources listed in Table~\ref{tab:rm}. Note that there are a significant number of sources with $|{\rm RM}|>1000{\rm rad}\,{\rm m}^{-2}$.}
\label{fig:rm}
\end{figure}

We have plotted a histogram of $|{\rm RM}|$ in Fig.\ref{fig:rm}. This
appears to show strong statistical evidence for a significant number
of compact sources which have intrinsic rotation measures in excess of
$1000\,{\rm rad}\,{\rm m}^{-2}$ and at high galactic latitudes such
high rotation measures cannot be of Galactic origin. This result is
not unexpected since VLBI measurements, for example Zavala \& Taylor
(2003), find evidence of rotation measures of this order in the
milliarcsecond cores of compact radio sources. In most of our targets
the emission is dominated by such compact cores and therefore we might
expect comparable rotation measures for the whole source and the
compact core alone.

We have compared our results with those available in the
literature. There are only four sources common to our list and to the
Zavala \& Taylor (2003) list. For one they could not get a
satisfactory $\lambda^2$ fit, for two we both have low rotation
measures of order a few $100\,{\rm rad}\,{\rm m}^{-2}$ and for the
last we have a value of $(587\pm 14)\,{\rm rad}\,{\rm m}^{-2}$ and
they quote $1433\,{\rm rad}\,{\rm m}^{-2}$. There is no obvious way to
reconcile the two values for this last source with a single rotation
measure and we suggest that the source probably has different
components suffering different rotation measures. Many of the sources
in Table~\ref{tab:rm} are in the list presented in Taylor et
al. (2009) which were computed using the two-bands around 1.4~GHz in
the NVSS. There is very little correspondence between the two.  It is
clear that it is difficult to make correspondence between rotation
measures computed using observations at different frequencies and
resolutions.

Though we think it is probably not astrophysically significant we note
that there is statistical evidence for a bias in our sample toward
positive values of the rotation measures with there being 30 positive
values and only 15 negative.  It is also noticeable that there is a
long line of positive rotation measures having right ascensions (RAs)
between 13h and 18h.  A relatively small ($\approx +80{\rm m}^{-2}$)
systematic Galactic contribution could be an explanation. 
Alternatively we note that a $5^{\circ}$ systematic error in the position
angle difference $\phi_{8.4}-\phi_{22}$, for example arising from
mis-calibration, would correspond to an error in the estimated
rotation measure of $\pm 80\,{\rm rad}\,{\rm m}^{-2}$ and hence such a
systematic shift in the position angles could reduce, but not
completely remove, the anomaly.

\begin{table}
\begin{center}
\begin{tabular}{|c|c|c|c|c|c||}
\hline
 Name & $\phi_{8.4}$ &  $\phi_{22}$ & $\phi_{43}$ & RM/${\rm rad}\,{\rm m}^{-2}$  & $\phi_{\rm int}$ \\
\hline\hline
\input det.tex
\hline\hline
\end{tabular}
\end{center}
\caption{Estimated rotation measurements and intrinsic position angles for 45 sources which appear compatible with the $\lambda^2$ law. Notice that there is a substantial region of positive rotation measure in the RA range 13h to 18h.}
\label{tab:rm} 
\end{table}

\section{Statistical properties of all detected sources including information from other surveys}

In this section we will use all the measurements we have made
(including non-detections) plus information from the NVSS (Condon et
al, 1997) at 1.4~GHz, additional information at 8.4~GHz from CLASS
(Jackson et al., 2007) and measurements made at 86~GHz using the IRAM
telescope (Agudo et al., 2009).  In total we have 191, 154, 169, 167
and 71 significant detections of polarization at the five
frequencies. Before continuing, we note that the inclusion of this
extra information could possibly lead to some extra scatter in the
properties of the sources for two reasons. Firstly the observations
were not taken at the same time as our own and therefore they could be
affected by source variability. In addition the measurements were made
at different resolutions. For example, those made by CLASS used a
different array configuration (VLA A-configuration as opposed to
D-configuration) which has 35 times higher resolution. This could lead
to polarized emission being resolved out for sources with extended
emission. This was pointed out by Jackson et al. (2009); in addition
to the comments made there, we point out that the average polarization
of the 31 (out of 154) sources for which we are using a CLASS
measurement, the average polarization percentage
$100\langle\Pi\rangle\approx 2$ where it is $\approx 3$ for the whole
sample. \changed{In addition to extra scatter, we draw attention to the fact that the sample is now much more heterogeneous and this could affect the interpretation of the results. For example, we caution against making strong conclusions from the 86~GHz data which is just an incomplete sub-sample with lower signal-to-noise ratio than the other observations.}

\begin{figure}
\centering
\includegraphics[height=6cm]{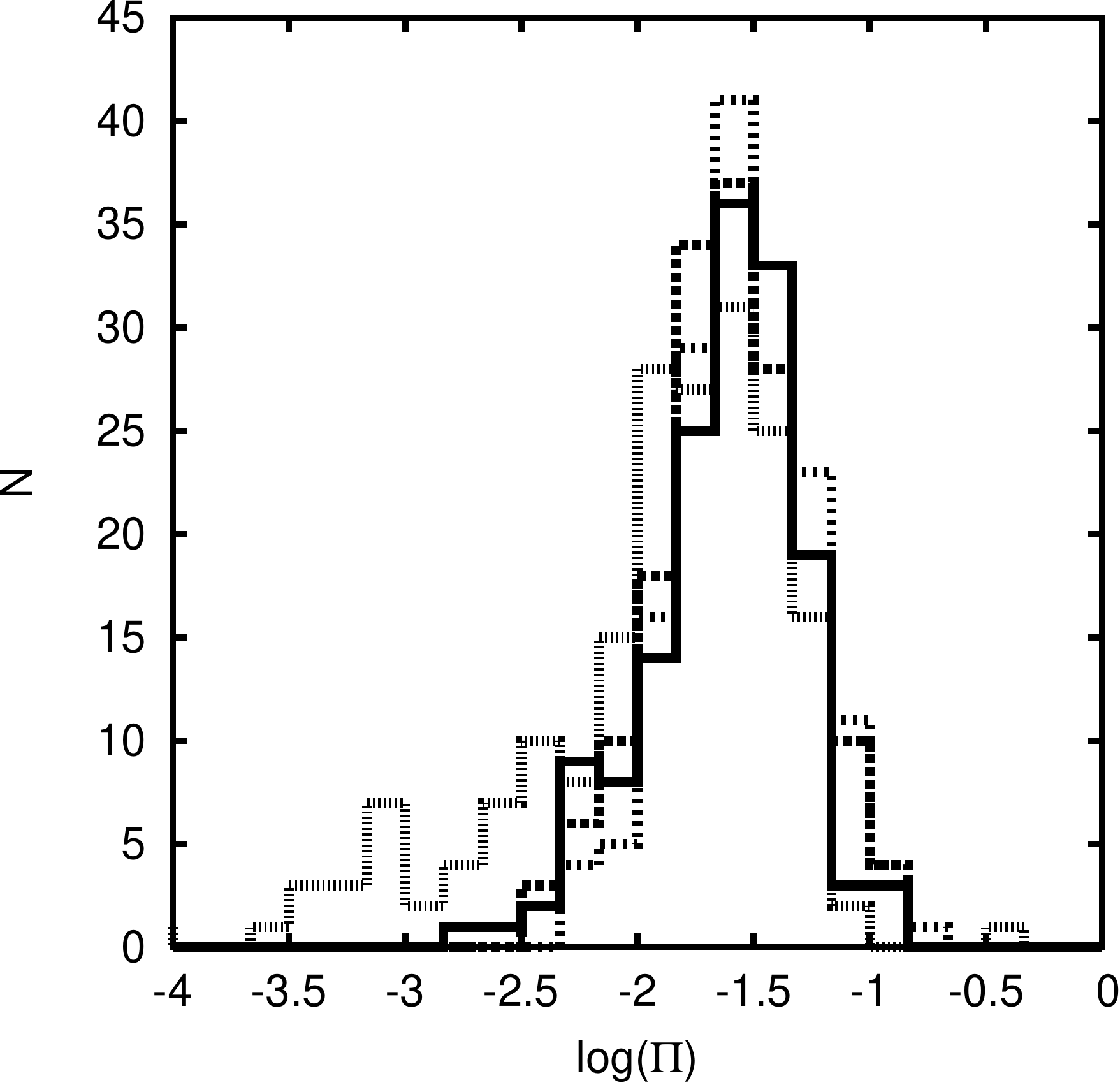}
\includegraphics[height=6cm]{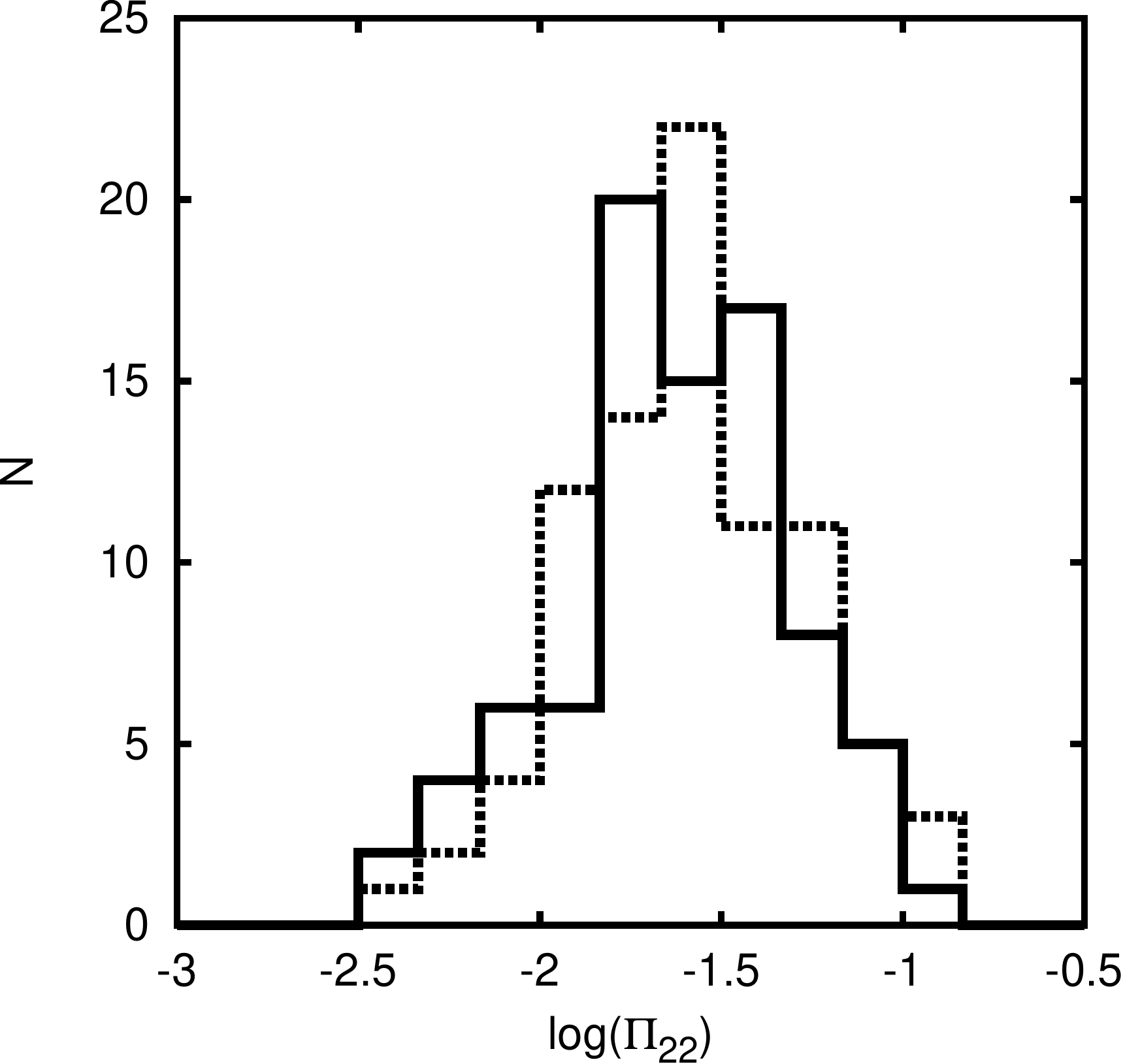}
\caption{On the left: histogram of the number of sources as a function of the logarithm of the fractional polarization, $\log_{10}(\Pi)$, for all detections including those detected at 8.4~GHz by CLASS only and at 1.4~GHz by NVSS. The fine dotted line is for 1.4~GHz, the solid line is 8.4~GHz, the dashed line is 22~GHz and the dotted line 43~GHz. On the right is a split of the detections at 22~GHz into those with $S>1.4{\rm Jy}$ (solid line) and $<1.4{\rm Jy}$ (dashed line).} 
\label{fig:frachist_all}
\end{figure}

The statistical properties of the fractional polarization at the 5
frequencies are tabulated in Table~\ref{tab:whole}. The values
$\langle\Pi\rangle$ and $\langle\Pi^2\rangle^{1/2}$ (which are
computed only from the detections) at 8.4, 22 and 43~GHz are similar
to those found for the contemporaneous sample - although the values
for 8.4~GHz are slightly lower due to the inclusion of the CLASS
sources. \changed{In addition, we have also performed a survival analysis using the ASURV package (Lavalley et al. 1992) which attempts to take into account the upper limits on the fractional polarization when computing the statistical properties. We find that the estimates for $\langle\Pi\rangle$ and $\langle\Pi^2\rangle$ are in agreement with those obtained from detections only.} We have also included the median fractional polarization,
$\Pi_{\rm med}$, taking into account the non-detections. This will be
a more accurate representation of the true distribution than
$\langle\Pi\rangle$ and we will use this in
section~\ref{sec:prediction} to normalize the probability distribution
for $\Pi$. The numbers are again compatible with there being no
evolution in the properties between 8.4~GHz and 43~GHz. The situation
is different at 1.4~GHz and 86~GHz. We see that $\langle\Pi\rangle$,
$\langle\Pi^2\rangle^{1/2}$ and $\Pi_{\rm med}$ are significantly
lower  at 1.4~GHz and are higher at 86~GHz. These results
are illustrated in the left hand panel of Fig.~\ref{fig:frachist_all}
where we plot the distribution of $\Pi$ as a function of frequency for
those frequencies where the number of sources are approximately equal,
that is 1.4-43~GHz. The histograms at 22~GHz and 43~GHz are similar to
those for the contemporaneous sample while that at 8.4~GHz is slightly
modified by the inclusion of the CLASS sources. The histogram at
1.4~GHz is very different to the others. The overall distribution is
shifted to lower values of $\Pi$ and there is a long tail to lower
values. This in not unexpected as it has been long known that Faraday
depolarization becomes an important factor at frequencies around
1.4~GHz (e.g. Garrington et al., 1988)

The range of values of total intensity for our sample is relatively
low making it difficult for us to investigate the variation of
statistical properties with total intensity - this would require a
comparison sample of data which was complete to a lower flux density
limit, say to $\sim 100{\rm mJy}$. Nonetheless we have attempted to
investigate this issue.  In the right hand panel of
Fig.~\ref{fig:frachist_all} we present histograms of sub-samples split
into two almost equal parts either side of 1.4Jy. The two
distributions appear to be almost identical, suggesting that, within
this sample for which the total intensity is in the range $\approx $0.5Jy
to $\approx $4.5Jy, there is no evidence for evolution. \changed{Gawro\'nski et al.
(2010)} point out that a variation of average polarization properties
with flux density might be expected because the proportion of steep
spectrum sources increases as the flux density limit is
decreased. They compared the spectral index distributions of sources
selected at 15 and 22~GHz over a wide range of total
intensity flux density. 
We would expect that some evolution in the
fractional polarization distribution will eventually emerge as sources with lower
flux densities are studied and samples become increasingly dominated
by steep spectrum extended sources.

\begin{table}
\begin{center}
\begin{tabular}{|c|c|c|c|c|c|c|}
\hline
 $\nu/{\rm GHz}$ & $N_{\rm det}$  & 100$\langle\Pi\rangle$ &  100$\langle\Pi^2\rangle^{1/2}$ & $100\Pi_{\rm med}$ & $100\langle\Pi\rangle_{\rm ASURV}$ & $100\langle\Pi^2\rangle^{1/2}_{\rm ASURV}$ \\
\hline\hline
1.4 & 191 & 2.2 & 3.9 & 1.6 & & \\
8.4 & 154 & 2.9 & 3.5 & 2.5 & $2.9\pm 0.2$ & $3.6\pm 0.9$\\
22  & 169 & 3.1 & 3.9 & 2.0 & $3.0\pm 0.2$ & $3.8\pm 1.0$\\
43  & 167 & 3.5 & 4.3 & 2.3 & $3.5\pm 0.2$ & $4.3\pm 1.3$\\
86 & 71 & 4.5 & 5.1 & 4.0 & & \\
\hline\hline
\end{tabular}
\end{center}
\caption{Statistical information for the fractional polarizations of
the whole sample at the different frequencies.  Those made using the survival analysis tool ASURV are also presented for 8.4-43GHz. $N_{\rm det}$ is the
number of detections at that frequency. At 8.4~GHz there are 123
detections from the present observations and 31 taken from the CLASS
sample which used the VLA at a different resolution (A-configuration
as opposed to D-configuration). All
detections at 1.4~GHz are taken from NVSS and those at 86~GHz from
Agudo et al. (2009).}
\label{tab:whole} 
\end{table}

\begin{figure}
\centering
\includegraphics[height=6cm]{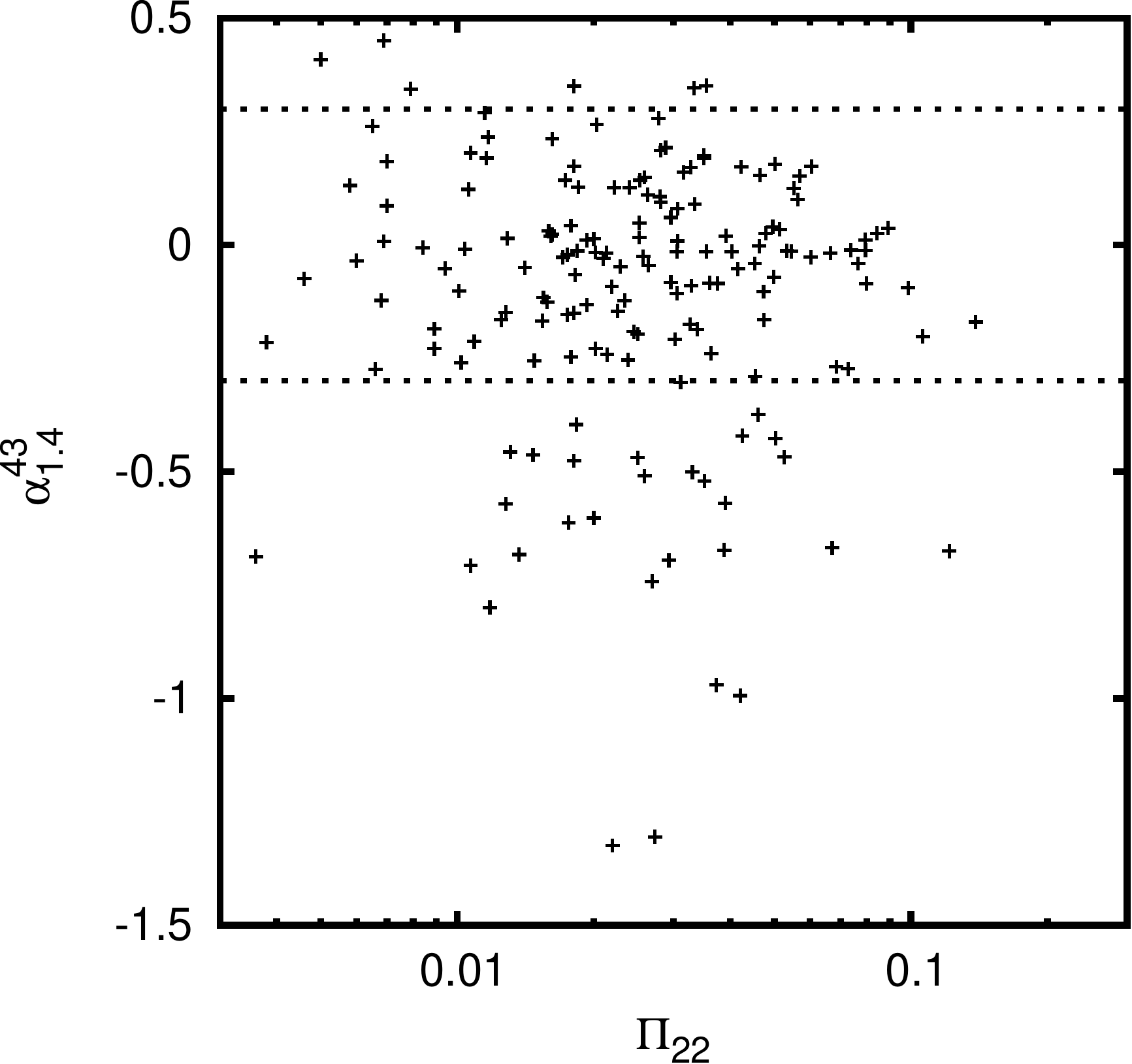}
\includegraphics[height=6cm]{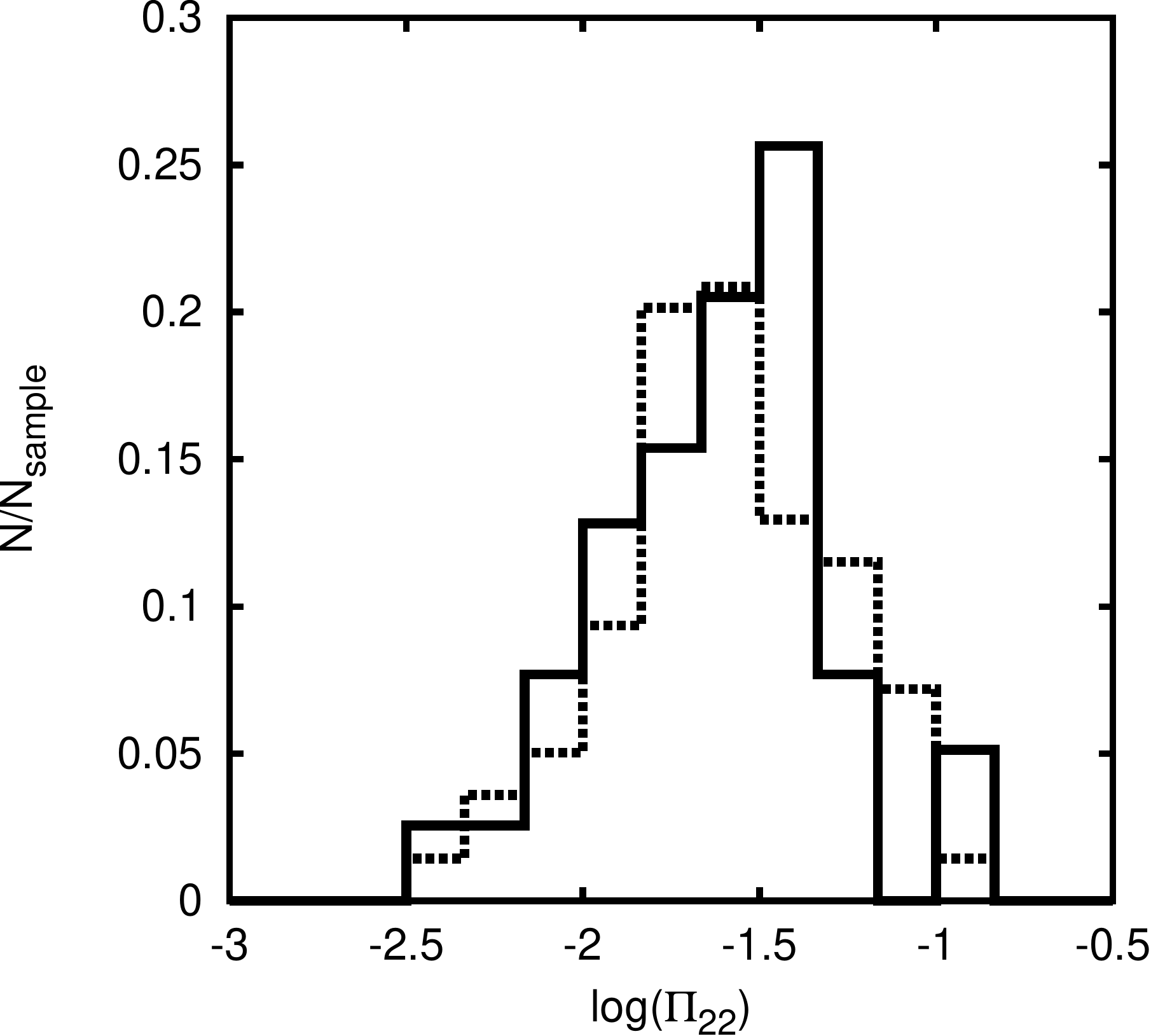}
\caption{On the left: scatter plot of the fractional polarization at 22~GHz ($\Pi_{22}$) against the intensity spectral index from 1.4~GHz to 43~GHz ($\alpha_{1.4}^{43}$). The two horizontal lines enclose the flat spectrum region $|\alpha_{1.4}^{43}|<0.3$. On the right is a histogram of the number of sources (divided by the number of sources in the sample) as a function of fractional polarization at 22~GHz ($\Pi_{22}$) for two samples. The dashed line is the flat spectrum sample with $|\alpha_{1.4}^{43}|<0.3$ and the solid line is the steep spectrum sample with $|\alpha_{1.4}^{43}|>0.3$. There appears to be no strong correlation between $\Pi_{22}$ and $\alpha_{1.4}^{43}$, and the histograms for the two samples are almost identical.}
\label{fig:frachist_split}
\end{figure}

Previous analyses of the polarization properties of sources in samples
selected at lower frequencies than ours have suggested that steep
spectrum sources typically have larger values of $\Pi$ than compact
flat spectrum sources (e.g. Gardner and Whiteoak, 1969). This is likely to be related to the high
level of magnetic order in the synchrotron producing regions in
extended regions of steep spectrum sources. We have attempted to
search for any evidence for this in our sample. In the left hand panel of
Fig.~\ref{fig:frachist_split} we have plotted the fractional
polarization at 22~GHz, $\Pi_{22}$, against the spectral index between
1.4 and 43~GHz, $\alpha_{1.4}^{43}$. We have chosen this specific
range in order for it to be optimal in separating steep and flat
spectrum sources. We see that the vast majority of sources in the
sample (130 out of 169) are flat spectrum $(|\alpha_{1.4}^{43}|<0.3)$
making it difficult to see the expected correlation between $\Pi_{22}$
and $\alpha_{1.4}^{43}$. We have also plotted histograms of $\Pi_{22}$
for the flat and steep spectrum sources, normalized by the numbers
sources in each of the samples. The number of sources in the steep
spectrum group is much lower and hence the random errors on their
histogram are larger. Given this it seems reasonable to conclude that
there is no evidence for any difference in the properties of the two
samples from our data.

\begin{figure}
\centering
\includegraphics[height=6cm]{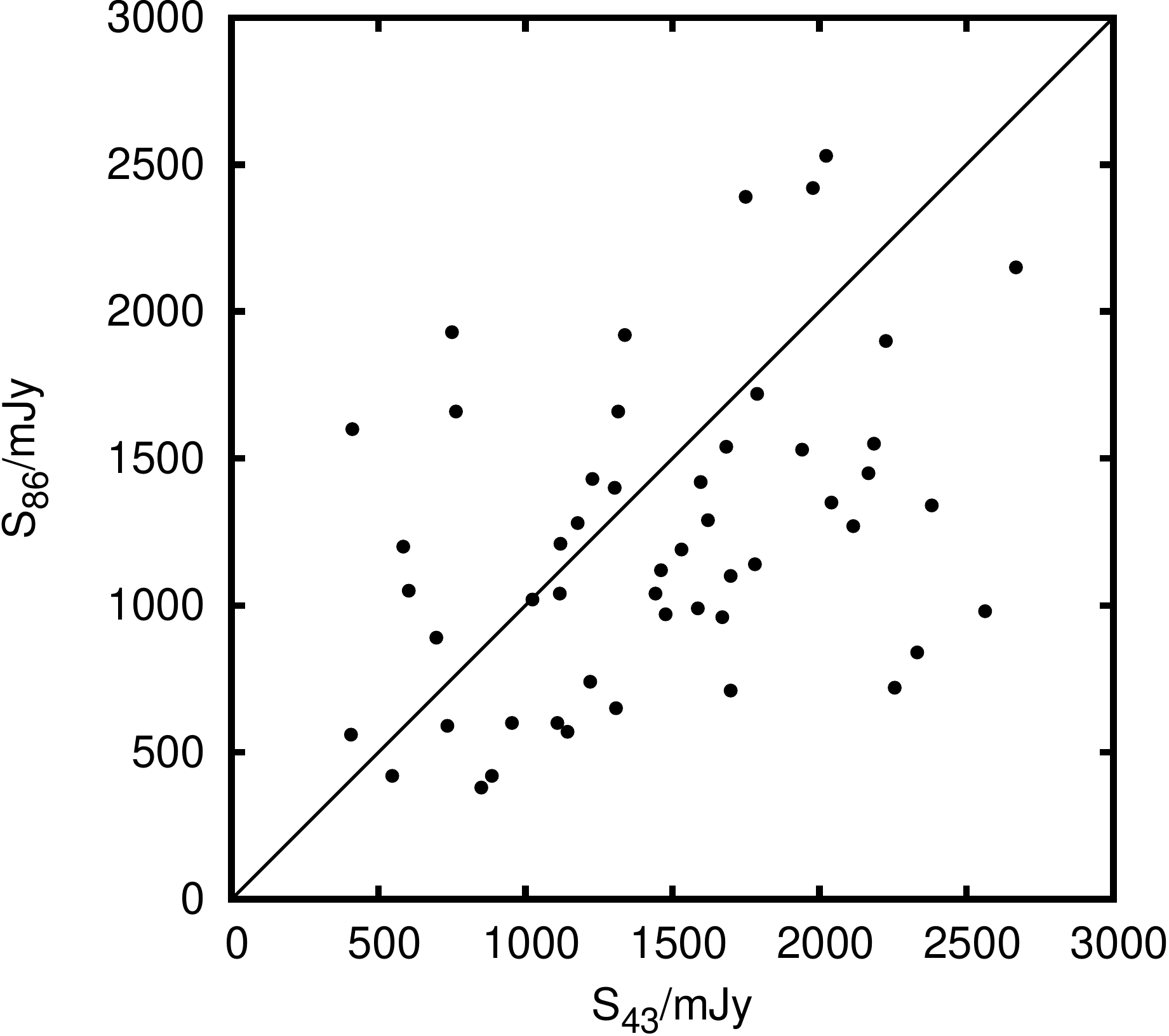}
\includegraphics[height=6cm]{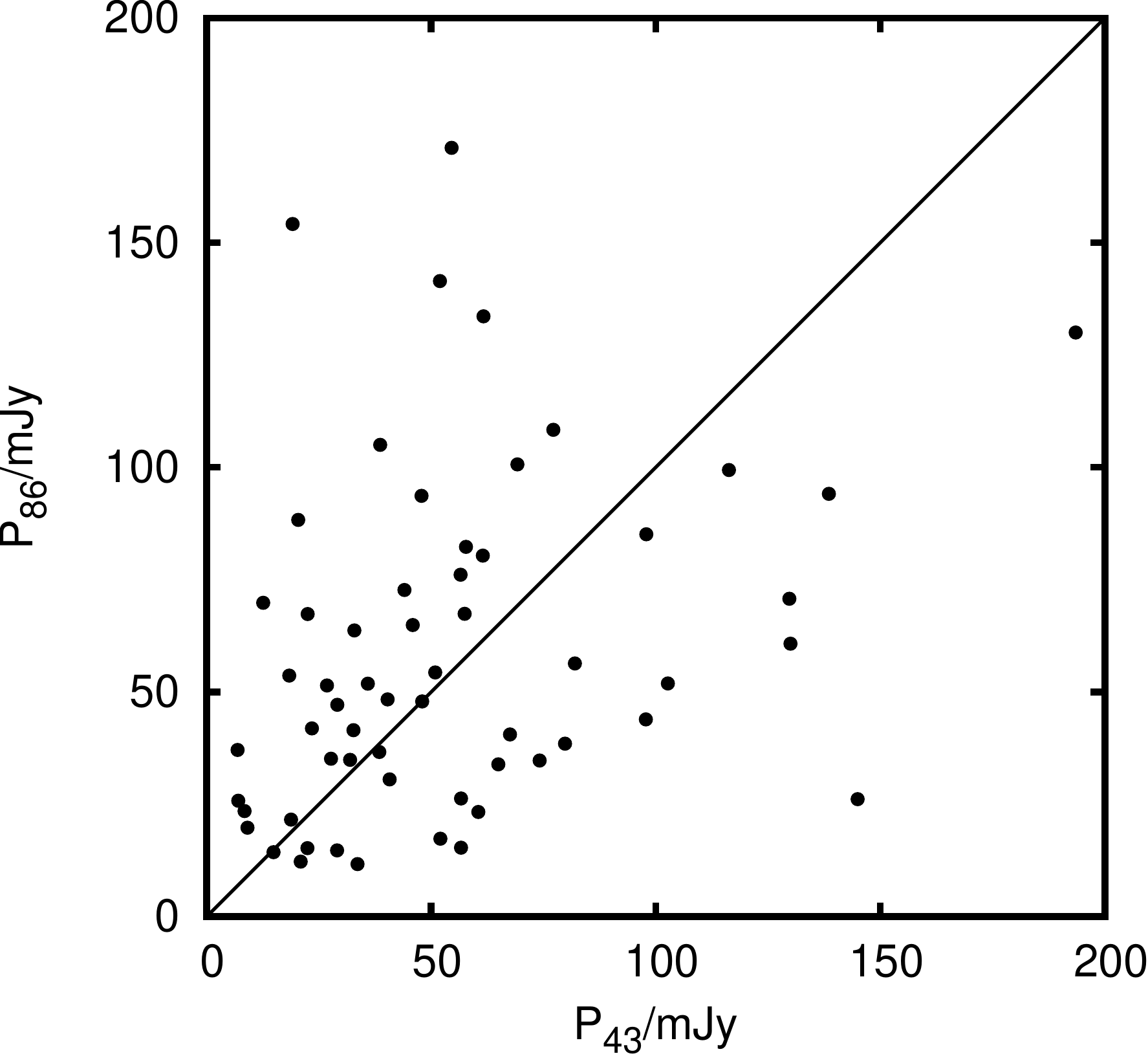}\\
\includegraphics[height=6cm]{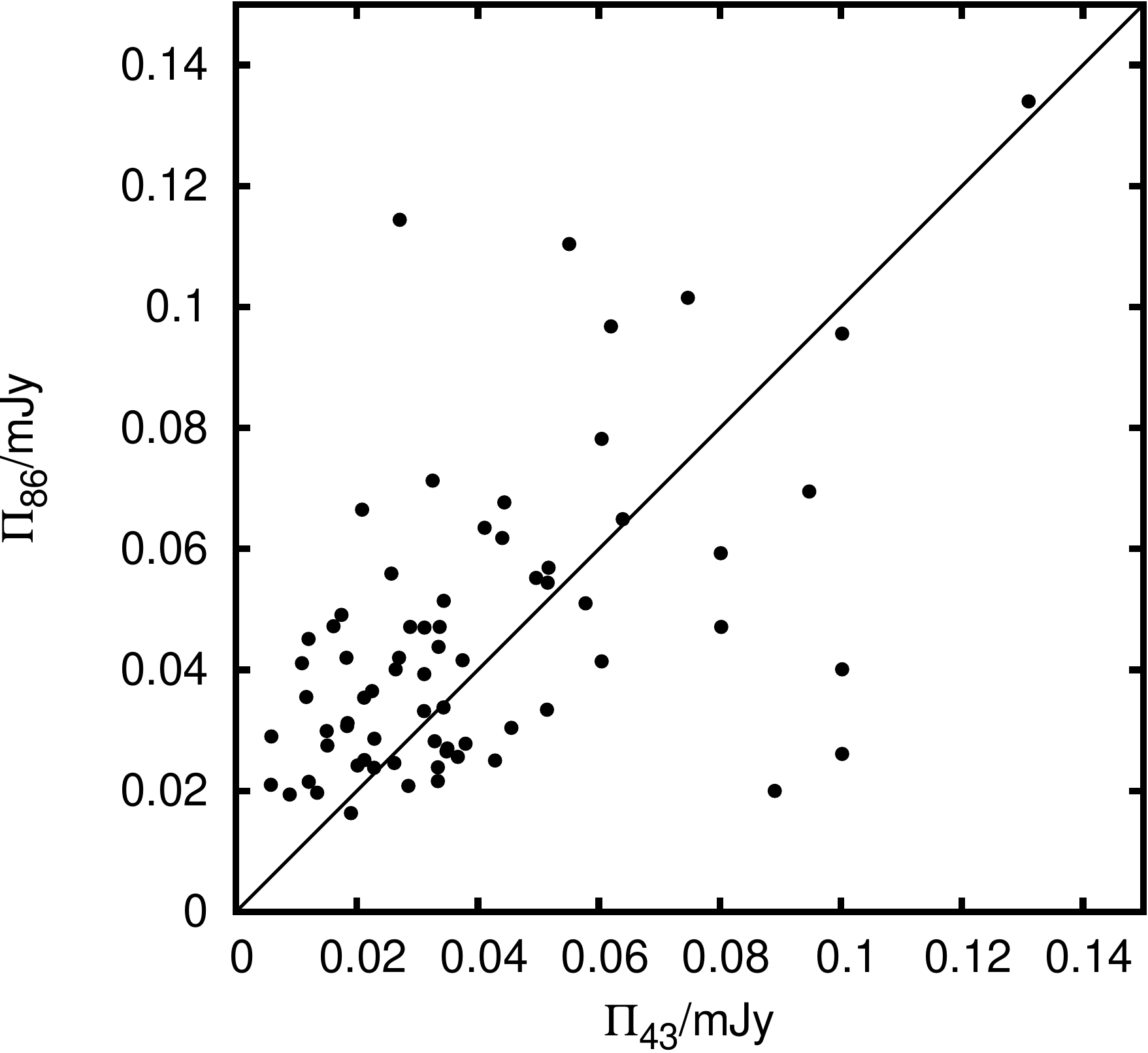}
\includegraphics[height=6cm]{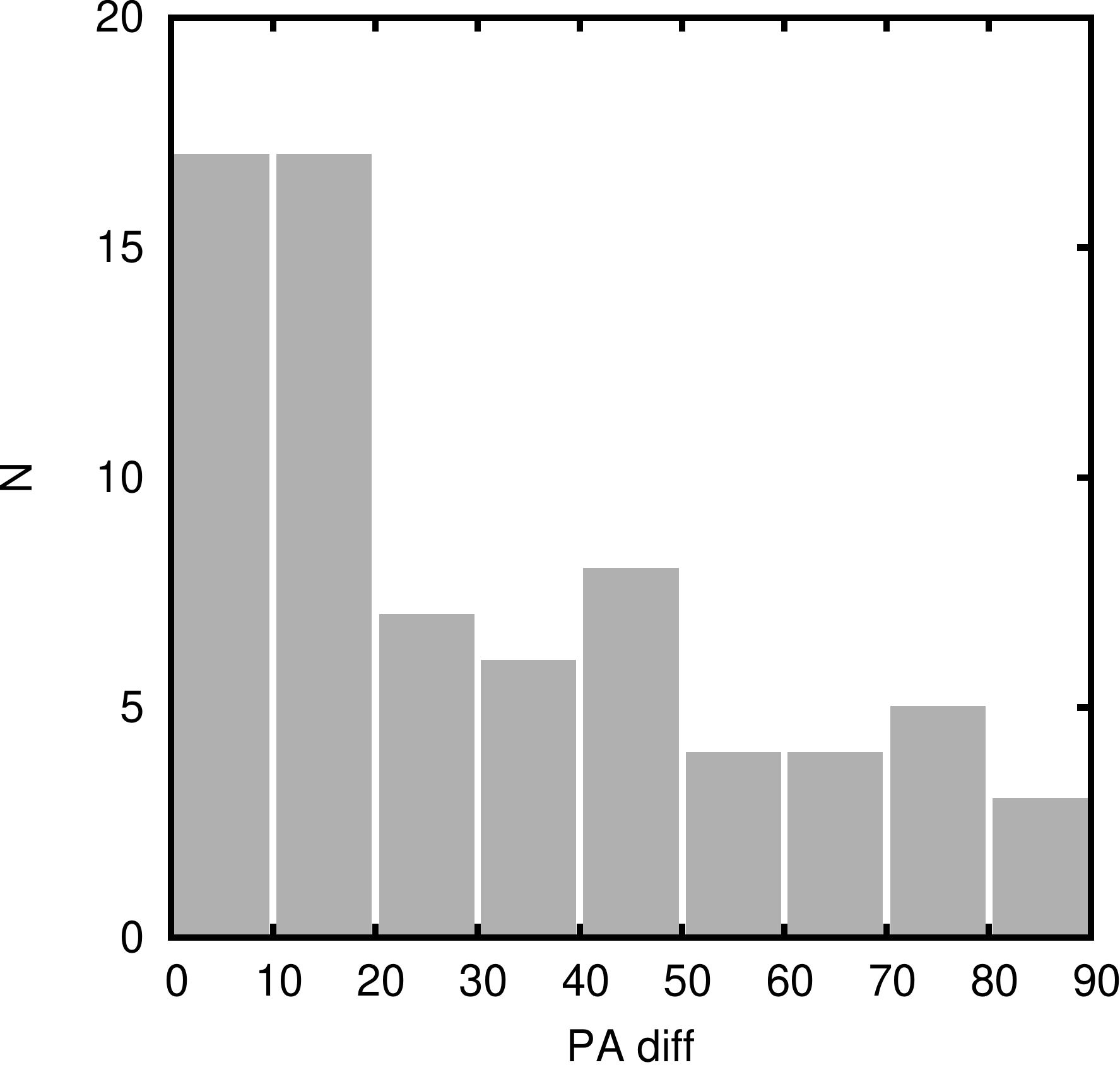}
\caption{A comparison between the properties of sources measured at 43 and 86~GHz: (top left) total intensity; (top right) polarized intensity; (bottom left) fractional polarization; (bottom right) histogram of the difference in the polarization position angle. There are clear correlations between the measured properties at the two frequencies.}
\label{fig:qvw}
\end{figure}

\begin{figure}
\centering
\includegraphics[width=6cm]{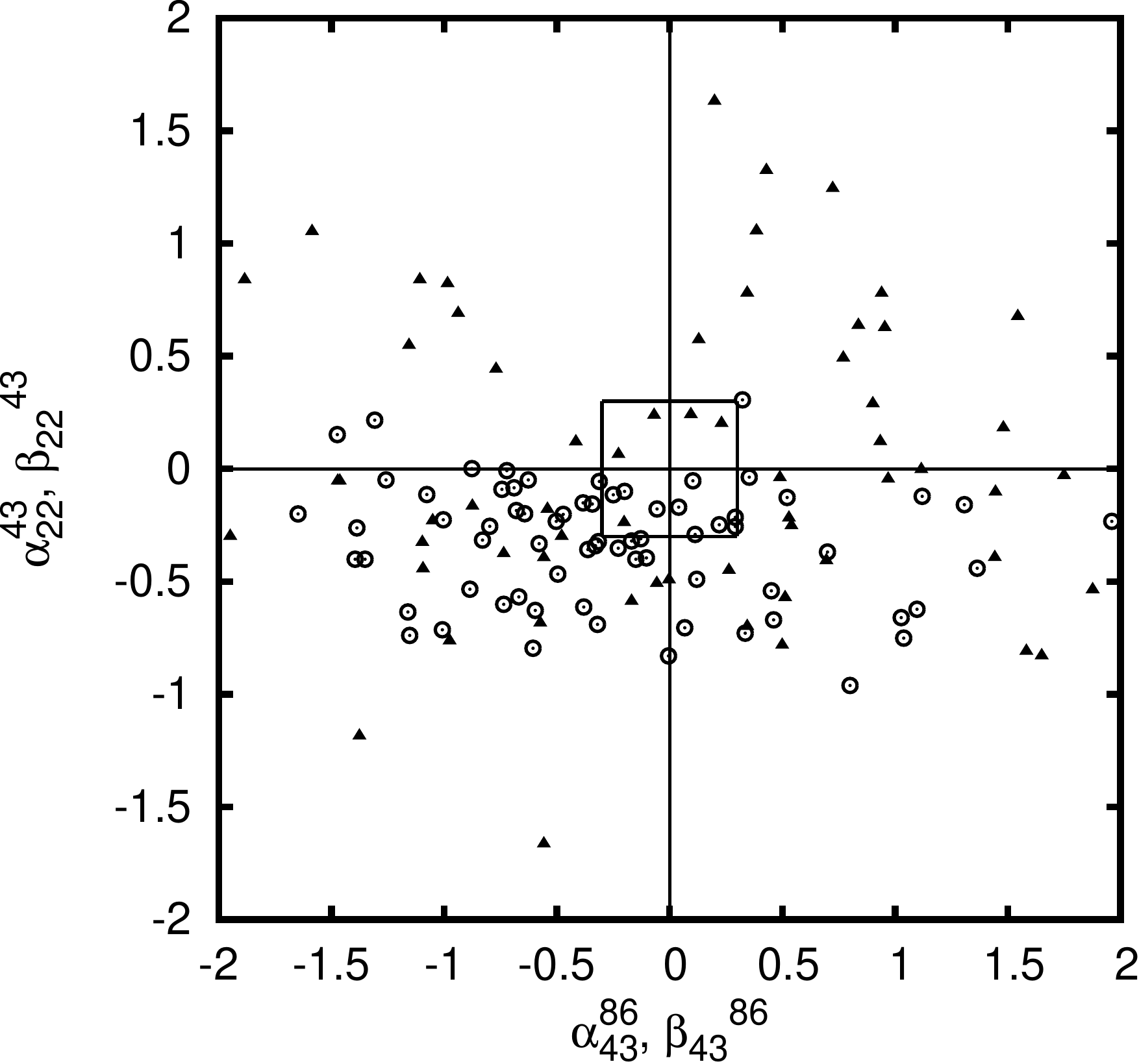}
\includegraphics[width=6cm]{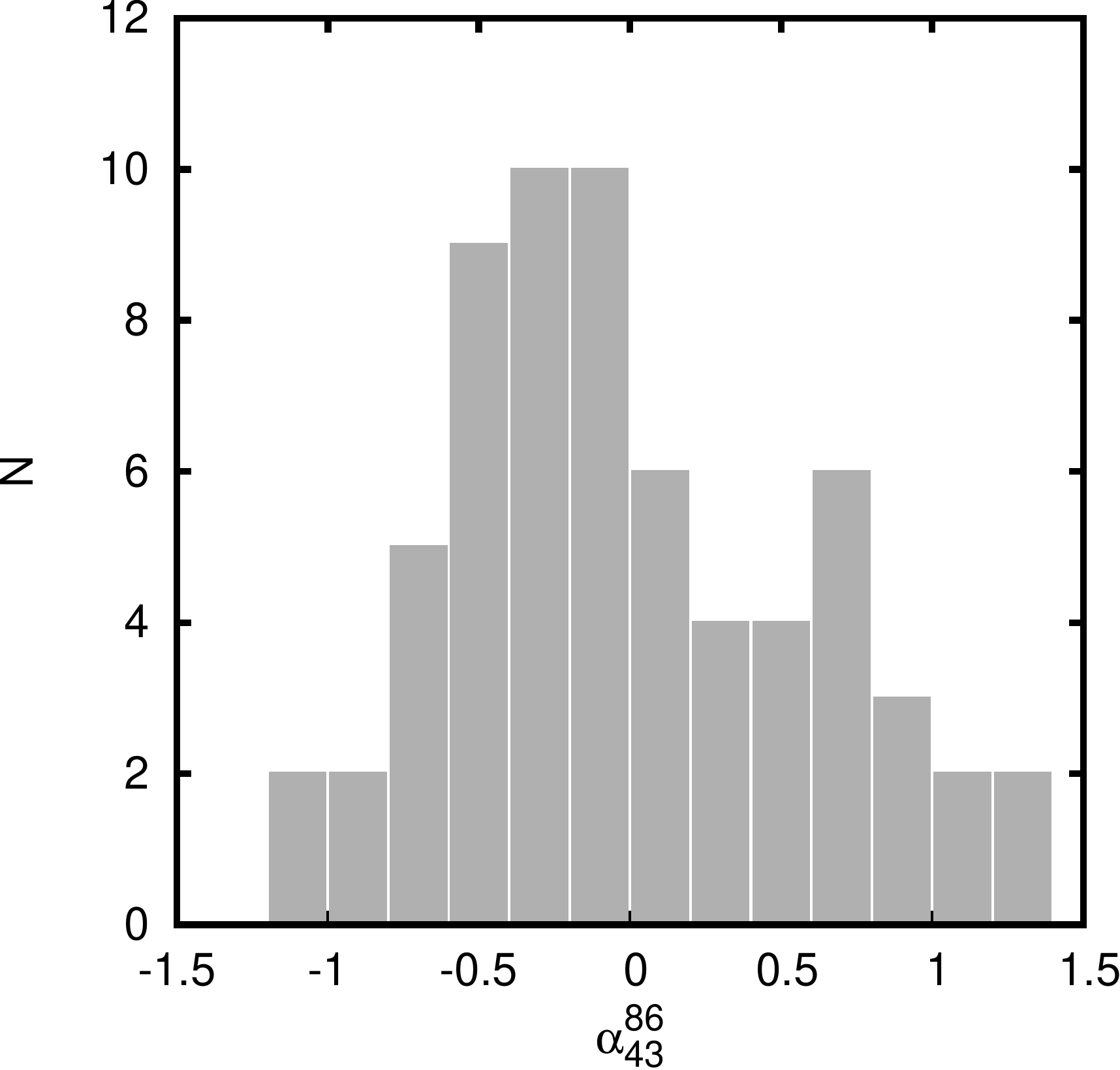}
\caption{On the left the two-colour diagram for total (circles) and polarized (triangles) intensity for 22-43-86~GHz. For total intensity this is very different to that presented in Fig.~\ref{fig:twocol} for 8.4-22-43~GHz. There is no obvious correlation between the two spectral indices. The two-colour diagram for polarized intensity is very similar to that for 8.4-22-43~GHz. On the right is a histogram of the total intensity spectral index between 43 and 86~GHz. The range of spectral indices presented here is substantially wider than seen between 22 and 43~GHz.}
\label{fig:wband}
\end{figure}

We have made a preliminary investigation of the correspondence between
our measurements at 22 and 43~GHz with those made at 86~GHz presented
in Agudo et al. (2009). In Fig.~\ref{fig:qvw} we present correlations
between the properties of the sources at 43 and 86~GHz. We see that
there are strong correlations between the total intensity, polarized
intensity and fractional polarization at the two frequencies indicating,
reassuringly, that the observations made with different telescopes, frequencies
and resolutions and, probably most importantly,
different times are related. Typically the intensity at 86~GHz is
lower than that at 43~GHz as would be expected for a spectrum which is
turning over, although there are a few sources for which the 86~GHz
flux density is larger than that at 43~GHz. The polarized intensities
at the two frequencies are comparable, albeit with a large
dispersion, and therefore the typical fractional polarization is
higher at 86~GHz. In addition, we see that for more than half of the
sources the polarization position angles are within $20^{\circ}$ indicating that
those measured at 43~GHz (and 86~GHz) are close to the intrinsic
position angle. We are, however, surprised by the tail of very large
position angle differences. This tail is more evident than in the
similar histogram for the difference in 22~GHz and 43~GHz polarization
position angles shown in Figure 6 of Jackson et al. (2010) thus
arguing against Faraday rotation as being the explanation. We will
return to this question below.

In Fig.~\ref{fig:wband} we present two-colour diagrams for the
spectral indices between 22-43-86~GHz for total and polarized
intensity, and a histogram for the intensity spectral index between 43
and 86~GHz. The two-colour diagram for polarized intensity is very
similar to that between 8.4-22-43~GHz, whereas that for total
intensity is very different. For 8.4-22-43~GHz there is a strong
correlation between $\alpha_{8.4}^{22}$ and $\alpha_{22}^{43}$
whereas no such correlation exists between $\alpha_{22}^{43}$ and
$\alpha_{43}^{86}$. We see that the range of values for
$\alpha_{43}^{86}$ in the histogram is much larger than that for
$\alpha_{22}^{43}$ with a reasonable fraction of sources having
$\alpha_{43}^{86}>0$ indicating that the spectrum is rising when it
was falling off between 22 and 43~GHz . This behaviour is unexpected
and is difficult to explain without appealing to the effects of
variability created by the non-contemporaneous observations.

\section{Discussion}

The results presented in the earlier sections suggest a number of
quantitative conclusions. The fractional polarization distribution is
relatively independent of the frequency at which it is measured. We have
also seen no evidence for any dependence on total intensity flux
density over the narrow range we have probed. In the subsequent
analyses we will assume that the median fractional polarization is
completely independent of frequency and adopt the value $100\Pi_{\rm
med}\approx 2.25$.  There is some dependence of the r.m.s. fractional
polarization for detected objects on the frequency of observation. If
we assume that this dependence is linear then
$100\langle\Pi^2\rangle^{1/2} =3.310+0.023(\nu/{\rm ~GHz})$ which fits
the data in the range $\nu=8.4$~GHz to 43~GHz. We note that while the
quoted values for the median take into account the undetected sources,
the computation of the r.m.s. uses only those which are detected, and
hence it could be biased to larger values.  We believe that this selection effect is
mainly responsible for the observed frequency dependence of the r.m.s. 
fractional polarization since the general trend is for 
polarized intensity to be weaker at higher frequencies and
hence less sources will actually be detected.

There is evidence for interesting behaviour revealed in the two-colour
diagrams for total intensity, polarized intensity and fractional
polarization. As expected there is a strong correlation between the
total intensity spectral indices; they satisfy the relation
$-0.81\alpha_{8.4}^{22}+0.58\alpha_{22}^{43}=-0.12\pm 0.17$. The
polarized intensity two-colour diagram is very different with
apparently little correlation between $\beta_{8.4}^{22}$ and
$\beta_{22}^{43}$. The combination of the strong correlation between
total intensity spectral indices and no correlation between the
polarization spectral indices leads to an anti-correlation between the
fractional polarization spectral indices. One symptom of this is
that there is fraction ($\sim 25\%$) of sources whose polarized
intensity spectra are upturning between 8.4 and 43~GHz, in marked
contrast to the case of total intensity for which there are no upturn
sources.

\begin{figure}
\centering
\includegraphics[scale=0.5]{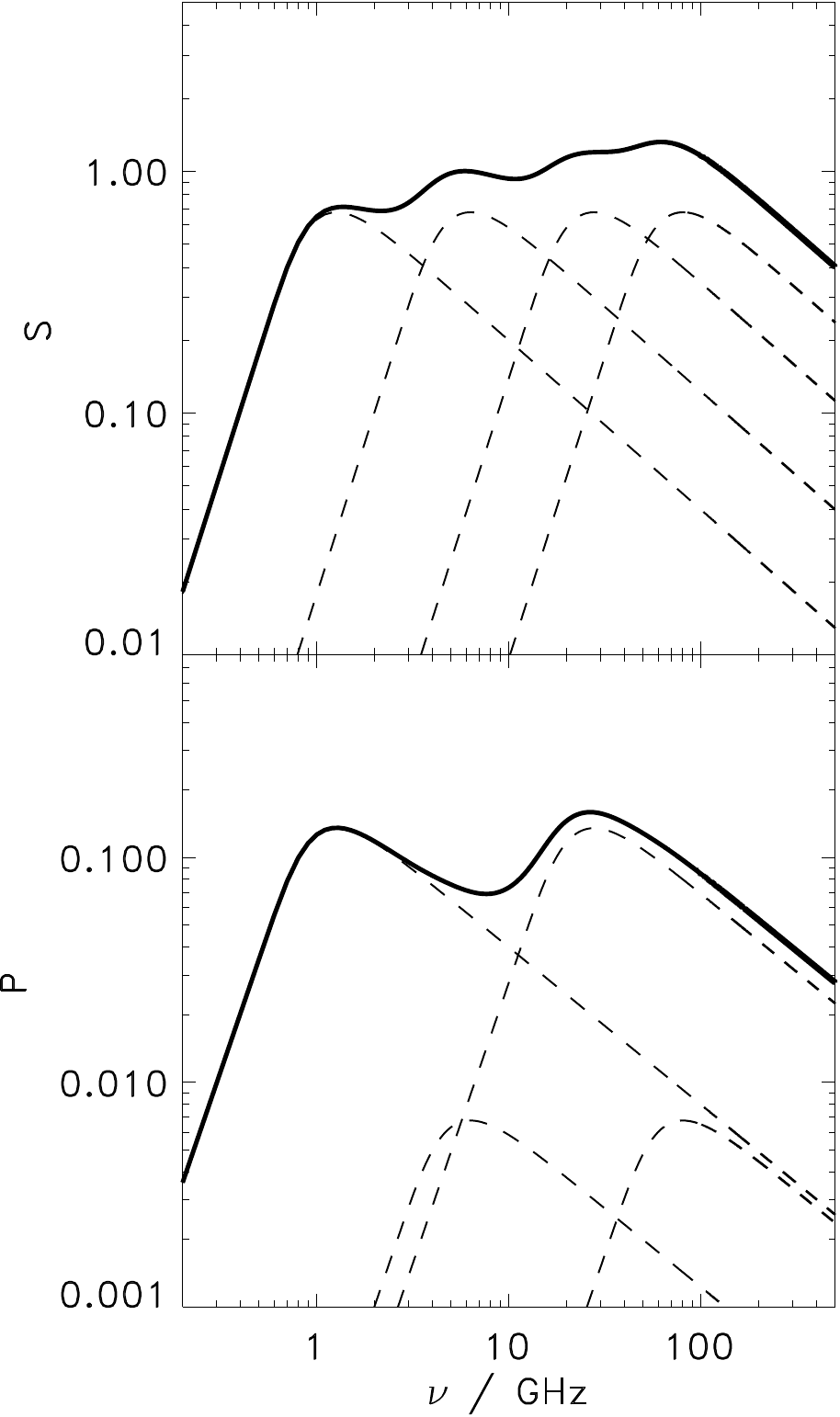}
\caption{A schematic of the total (top) and polarized (bottom)
intensity spectra of a flat-spectrum source in arbitrary units which
is made up of 4 components which have different fractional
polarizations - two with  $20$ per cent polarization and the other
two with $2$ per cent. In this example, the polarized intensity
spectrum has two well-defined peaks one around 1~GHz and the other
between 20 and 30~GHz. From the point of view of our observations,
this source would be a have a flat spectrum in total intensity and
would have a peaked spectrum in polarized intensity. One could have
easily created an example with an upturning spectrum in the range 8.4
to 43~GHz by exchanging the polarization fractions.}
\label{fig:bumpy}
\end{figure}

The lack of correlation between the polarization spectral indices
indicates that the polarized intensity spectra are less smooth than
the total intensity spectra, that is, there are significant
undulations in the polarized intensity spectra.  Most of the sources
in our sample have relatively flat total intensity spectra between 8.4
and 43~GHz due to their selection at 22~GHz. The standard explanation
for this (Cotton et al., 1981) is that the sources'
intensity spectra are made up of a number of synchrotron components
whose net effect adds up to something which is close to a flat
spectrum. This is illustrated in the left-hand panel of
Fig.~\ref{fig:bumpy}. The fractional polarizations and polarization
position angles of these components can vary with frequency for a
variety of physical reasons, e.g. changes in the magnetic field
direction, the degree of ordering of the field, spatially varying
Faraday rotation and whether the components are optically thin or
thick at the frequency of observation. All these effects can
contribute to polarization spectra being more ``bumpy'' than total
intensity spectra.  We have illustrated this in the right-hand
panel of Fig.~\ref{fig:bumpy} where we have assumed, for example, that
two of the components are $20$ per cent polarized and the others $2$ per cent
polarized. This extra degree of freedom will make the polarization
spectra less flat than their total intensity counterparts. The
specific make up of the components and their distribution of
fractional polarization could lead to the range of spectral type which
we observe. In addition, the low frequency spectrum of each of the
component could be cut-off sharply by synchrotron self-absorption. Optically thick
synchrotron emission is expected to be less polarized that optically thin
emission (Ginzburg and Syrovatskii, 1969).

It would be useful for the purpose of modelling the statistical properties
of the polarized source population if there was a dependence of
polarized flux upon total intensity spectral index. Such a dependence
might be expected since optically thin, steep spectrum, sources
generally have higher degrees of polarization at low frequencies and,
therefore, the absence of a detectable correlation between the total
intensity spectral index and the fractional polarization in our data
might be thought surprising. However, we suspect that there are just
too few steep spectrum sources in our sample of the strongest sources
in the sky to see such an effect. It may become more evident when
weaker samples in which the proportion of steep spectrum sources is
higher (e.g. \changed{Gawro\'nski et al. 2010}) are studied.


\section{Noise background for CMB polarization observations}

It is well understood that an ensemble of Poisson distributed point sources gives rise to a flat power spectrum in intensity (for example, Cleary et al., 2005 and references therein). For sources below some cut-off $S_{\rm cut}$, the amplitude of this white noise spectrum is given by 
\begin{equation}
C_{\ell}^{TT}=\left({dB\over dT}\right)^{-2}\langle S^2\rangle =\left({dB\over dT}\right)^{-2}\int_0^{S_{\rm cut}}dS\,S^2{dN\over dS}\,,
\end{equation}
where ${dN/dS}$ is the differential source count, $dB/dT= 2.5\times 10^{7}{\rm Jy}\,{\rm sr}^{-1}\,{\rm K}^{-1} \,x^4e^{x}/(e^{x}-1)^2$ and $x=\nu/56.8$~GHz.

One could define an equivalent expression for the polarization spectrum 
\begin{equation}
C_{\ell}^{P}=C_{\ell}^{EE}+C_{\ell}^{BB}=\left({dB\over dT}\right)^{-2}(\langle Q^2\rangle +\langle U^2\rangle)=\left({dB\over dT}\right)^{-2}\int_0^{P_{\rm cut}}dP\,P^2{dN\over dP}\,,
\end{equation}
in terms of the polarized source count $dN/dP$ and a cut off in polarized intensity $P_{\rm cut}$. In what follows we will assume that the emission will, on average, contribute equally to the EE and BB power spectra, that is $C_{\ell}^{EE}=C_{\ell}^{BB}={1\over 2}C_{\ell}^{\rm P}$. However, the polarized source counts are difficult to measure since the polarized signal is weak and most samples are defined by their completeness in terms of total intensity. It has been suggested (de Zotti et al. 1999, Mesa et al. 2002, Tucci et al. 2004) that one should use the intensity source count supplemented by information on the statistical properties of $\Pi$ defined by some probability density function ${\cal P}(\Pi)$ which for the purposes of our discussion we will assume is independent of the total intensity of the source. This is supported by our observations for a narrow range of flux densities $\sim 1{\rm Jy}$, but may not be true for lower values.

One approach (method A) is to define the noise spectrum in terms of a cut off in total intensity, $S_{\rm cut}$. This is likely to be most useful in practice since one could imagine having a high resolution survey complete to some level in total intensity, as opposed to the equivalent in polarized intensity. In this case one can define the power spectrum due to sources with a given fractional polarization, $\Pi$, as 
\begin{equation}
C_{\ell}^{P}(\Pi)=\left({dB\over dT}\right)^{-2}\int_0^{\Pi S_{\rm cut}}dP\, P^2{dN\over dP}=\left({dB\over dT}\right)^{-2}\Pi^2\int_0^{S_{\rm cut}}dS\, S^2{dN\over dS}\,.
\end{equation}
The spectrum due to all the sources is then given by 
\begin{equation}
C_{\ell}^{P}=\int_0^1 d\Pi\,{\cal P}(\Pi)C_{\ell}^{P}(\Pi)=\langle\Pi^2\rangle C_{\ell}^{T}\,,
\end{equation}
where $\langle\Pi^2\rangle=\int_0^{1}d\Pi\,\Pi^2{\cal P}(\Pi)$.

Another approach (method B) is to attempt to compute the noise spectrum due to sources in terms of  a cut-off in polarized intensity, $P_{\rm cut}$. If  we define the spectrum due to sources with fractional polarization $\Pi$ to be 
\begin{equation}
C_{\ell}^{P}(\Pi)=\left({dB\over dT}\right)^{-2}\int_0^{P_{\rm cut}}dP\,P^2{dN\over dP}=\left({dB\over dT}\right)^{-2}\Pi^2\int_0^{P_{\rm cut}/\Pi} dS\,S^2{dN\over dS}\,.
\end{equation}
The spectrum due to all sources is then given by 
\begin{equation}
C_{\ell}^P=\int_0^1d\Pi{\cal P}(\Pi)C_{\ell}^P(\Pi)=\left({dB\over dT}\right)^{-2}\int_0^{1}d\Pi\Pi^2{\cal P}(\Pi)\int_0^{P_{\rm cut}/\Pi} dS\,S^2{dN\over dS}\,,
\end{equation}
which can be rewritten as 
\begin{equation}
C_{\ell}^{P}=\left({dB\over dT}\right)^{-2}\bigg[\langle\Pi^2\rangle\int_0^{P_{\rm cut}} dS S^2{dN\over dS}+\int_{P_{\rm cut}}^{\infty}dS S^2{dN\over dS} F\left({P_{\rm cut}\over S}\right)\bigg]\,,
\end{equation}
where the weighting function 
\begin{equation}
F(x)=\int_0^{x} \Pi^2{\cal P}(\Pi) d\Pi\,,
\end{equation}
and we have that $F(1)=\langle\Pi^2\rangle$.

In order to estimate the error involved in using method A compared to method B, one can compute the ratio of the two noise power spectra (method B over method A) for $P_{\rm cut}=\langle\Pi\rangle S_{\rm cut}$. This is given by 
\begin{equation}
R={\int_0^{\infty}dS\, S^2{dN\over dS} G\left({\langle\Pi\rangle S_{\rm cut}\over S}\right)\over \int_0^{S_{\rm cut}} dS\, S^2{dN\over dS}}\,,
\end{equation}
where $G(x)=F(x)/F(1)$ for $x<1$ and $G(x)=1$ for $x>1$.

In view of the observed distributions of the fractional polarization (see Fig.~\ref{fig:frachist}), it should be possible to model ${\cal P}(\Pi)$ using a Gaussian in $\log\Pi$, that is 
\begin{equation}
{\cal P}(\Pi)={A\over\Pi}\exp\left[-{\left[\log\left({\Pi\over\Pi_0}\right)\right]^2\over 2\sigma^2}\right]\,,
\label{lognormal}
\end{equation}
in the range $0\le\Pi\le 1$. For a standard log-normal distribution, for which $0\le\Pi<\infty$, the normalization coefficient, $A$, can be computed to be $(2\pi\sigma^2)^{-1/2}$, $\Pi_0$ is the median and $\sigma$ is the standard deviation in log. Due to the fact that the upper limit of the distribution is $\Pi=1$, these need to be adjusted. For the particular choices of $\Pi_0$ and $\sigma$ which  we will choose, $\Pi=1$ will be sufficiently large so as to be effectively infinite and these adjustments will be sub-dominant, but we have included them below for completeness.

For this particular choice of probability density function, one finds that 
\begin{eqnarray}
A&=&\sqrt{1\over 2\pi\sigma^2}\left[{2\over 1+{\rm erf}\left({\log(1/\Pi_0)\over\sqrt{2}\sigma}\right)}\right]\,,\nonumber\\
\langle\Pi\rangle&=&\Pi_0e^{{1\over 2}\sigma^2}
\left[{ 
1+{\rm erf}\left({{\log(1/\Pi_0)\over\sqrt{2}\sigma}+{1\over\sqrt{2}}\sigma}\right) \over
1+{\rm erf}\left({\log(1/\Pi_0)\over\sqrt{2}\sigma}\right) 
}\right]\,,\nonumber\\
\langle\Pi^2\rangle &=& \Pi_0^2e^{2\sigma^2}
\left[{
1+{\rm erf}\left({{\log(1/\Pi_0)\over\sqrt{2}\sigma}+\sqrt{2}\sigma}\right) \over
1+{\rm erf}\left({\log(1/\Pi_0)\over\sqrt{2}\sigma}\right) 
}\right]\,,\\
\Pi_{\rm med}&=&\Pi_0\exp\left[\sqrt{2}\sigma{\rm erf}^{-1}\left[{1\over 2}\left({\rm erf}\left({\log(1/\Pi_0)\over\sqrt{2}\sigma}\right)-1\right)\right]\right]\,,
\end{eqnarray}
where the terms inside the square brackets in the first three expressions are the the corrections which would $\rightarrow 1$ if $\log(1/\Pi_0)\rightarrow\infty$ and the argument for the exponential will $\rightarrow 0$ in the same limit. One can also compute the weighting function
\begin{equation}
F(x)=\Pi_0^2e^{2\sigma^2}
\left[{
1+{\rm erf}\left({{\log(x/\Pi_0)\over\sqrt{2}\sigma}+\sqrt{2}\sigma}\right) \over
1+{\rm erf}\left({\log(1/\Pi_0)\over\sqrt{2}\sigma}\right) 
}\right]\,.
\end{equation}

\section{Predictions for the noise background}
\label{sec:prediction}

\begin{figure}
\centering
\includegraphics[width=5cm,height=5cm]{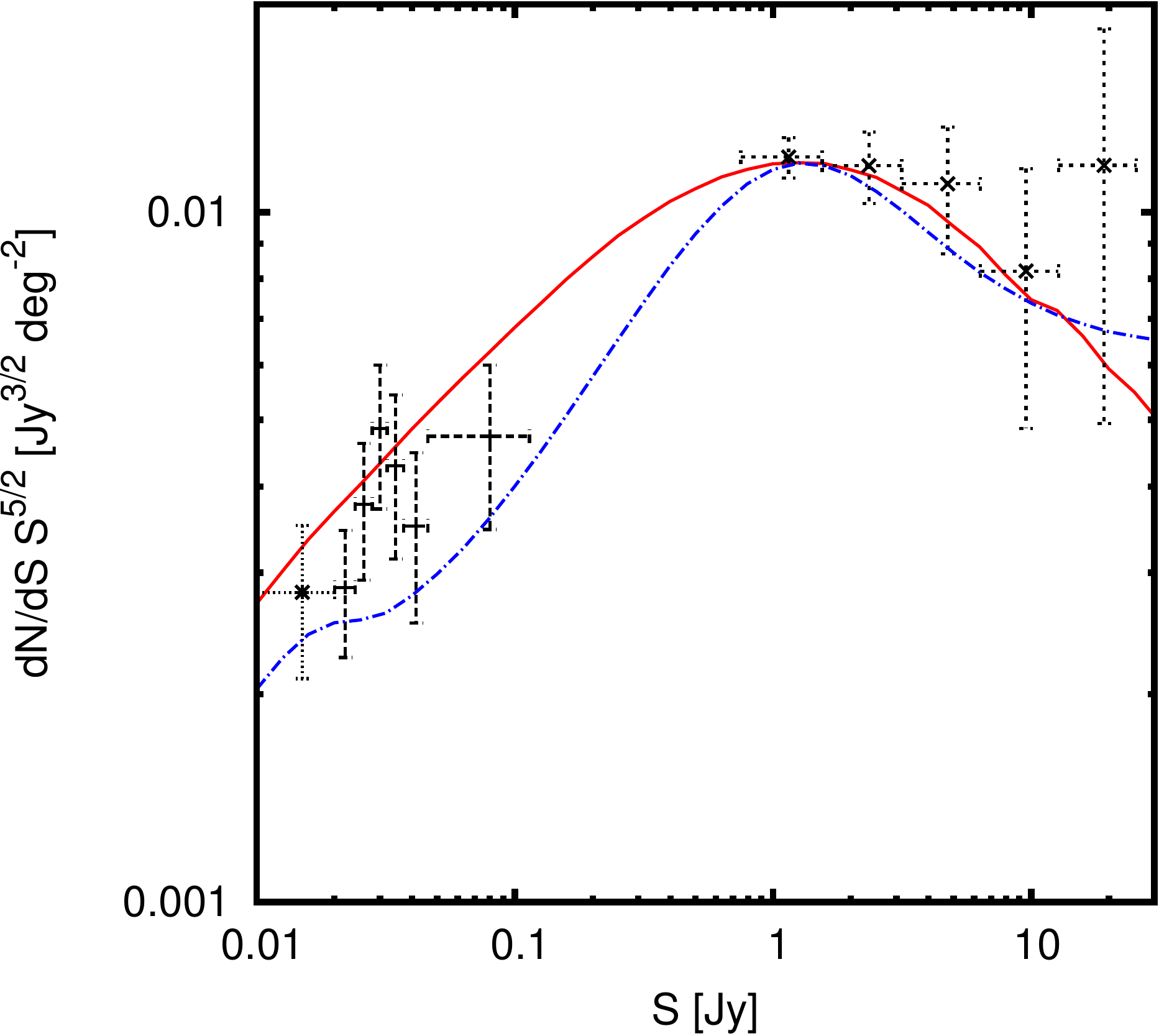}
\includegraphics[width=5cm,height=5cm]{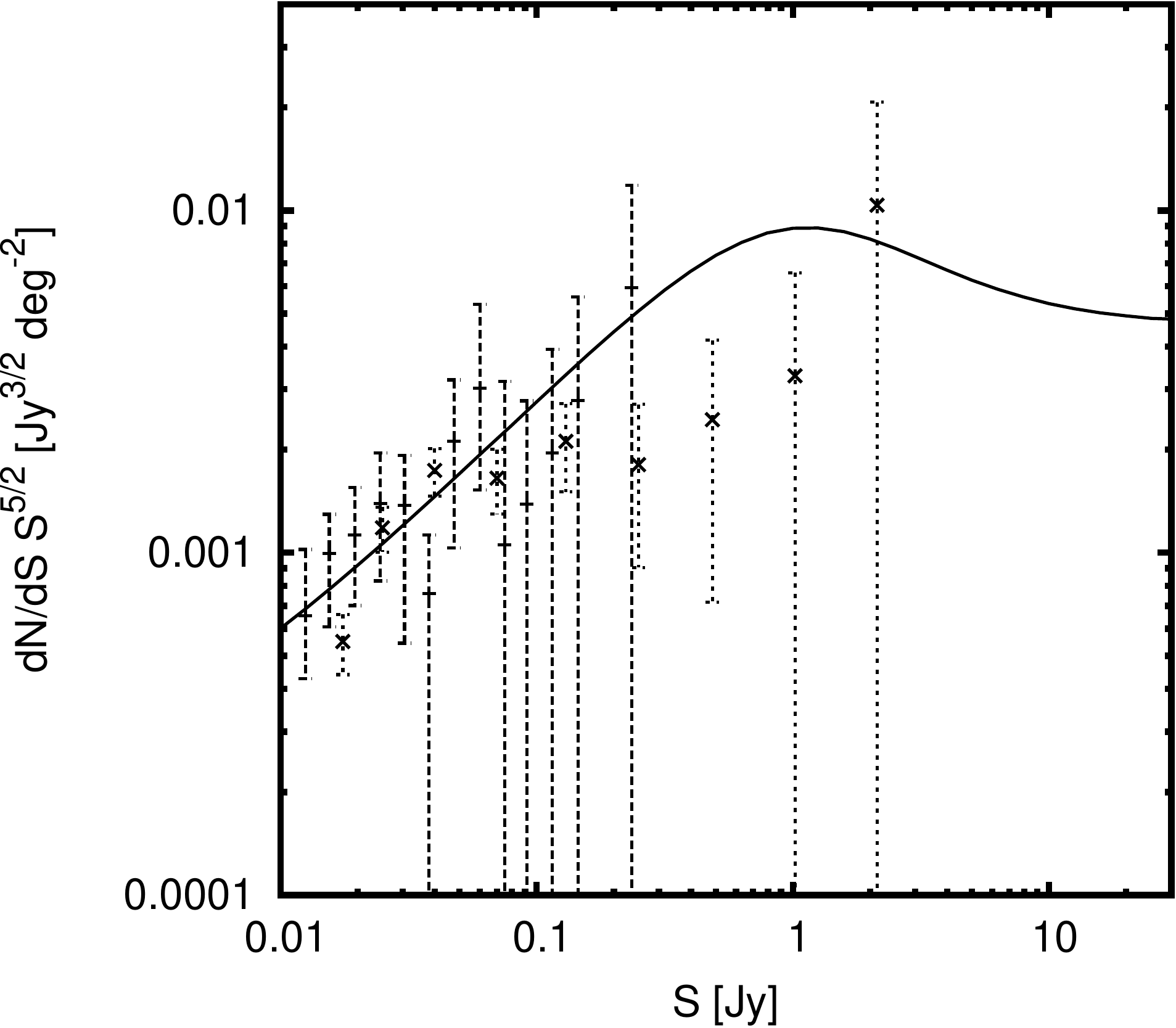}
\includegraphics[width=5cm,height=5cm]{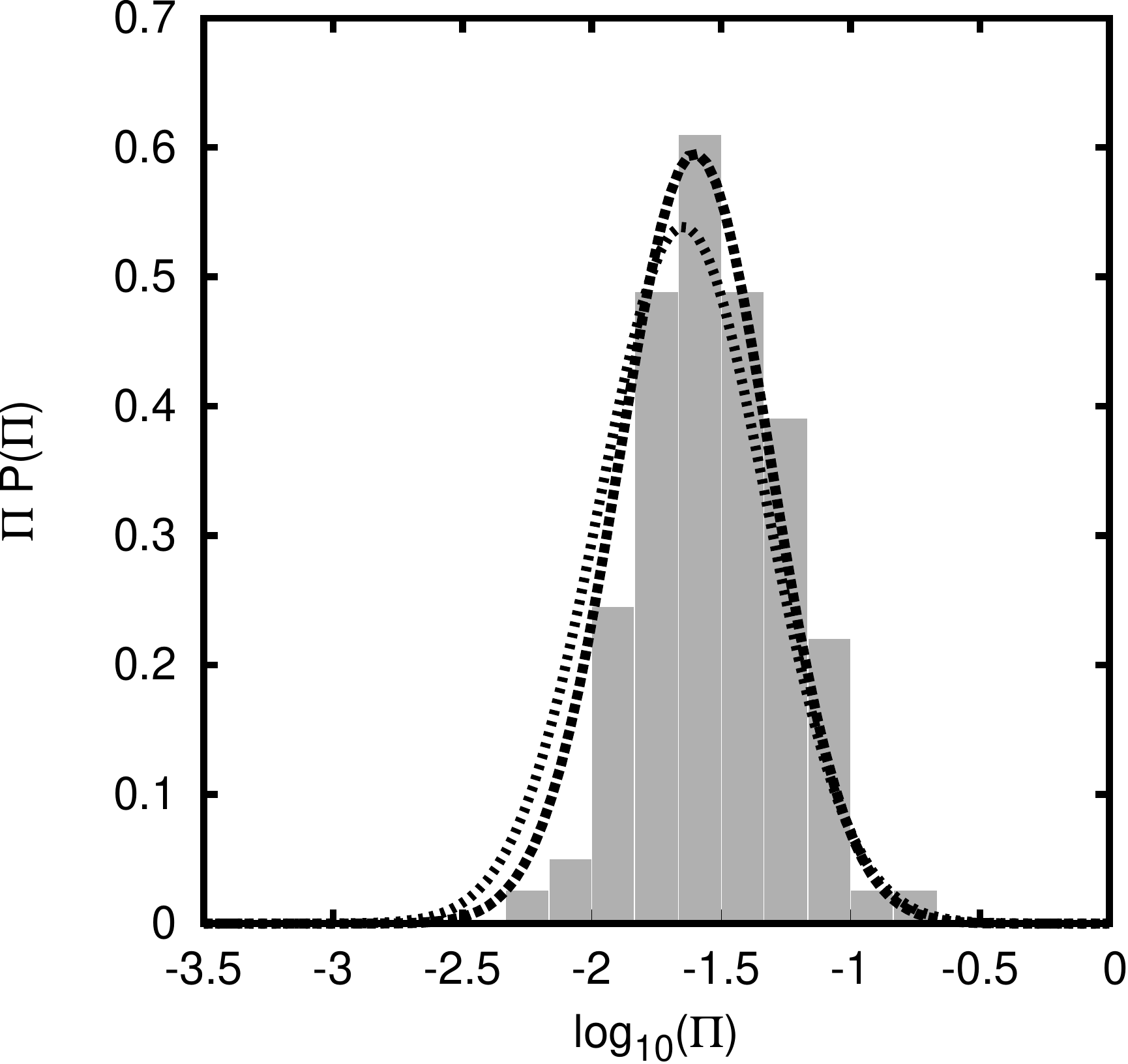}
\caption{Left: the source surface density at 30~GHz against flux density \changed{from the dZ05 model (blue) and the T98 model (red). Note that they agree in the region of $1{\rm Jy}$ but differ significantly at lower flux densities.} Included also are measured source counts from Wright et al. (2009), Cleary et al. (2005) and \changed{Gawro\'nski et al. (2010). Middle: the source surface density at 150GHz for the dZ05 model. Included also are the measured source counts from Viera et al (2010) and Marriage et al (2010)}. Right: the probability distribution (for logarithmic bins) for the fractional polarization with $100\Pi_0=2.25$, $\sigma=0.74$ (dotted line) and $100\Pi_0=2.5$, $\sigma=0.67$ (dashed line) along with the histogram for detected sources at 22~GHz.}
\label{fig:dnds_prop}
\end{figure}

\begin{figure}
\centering
\includegraphics[width=6cm,height=6cm]{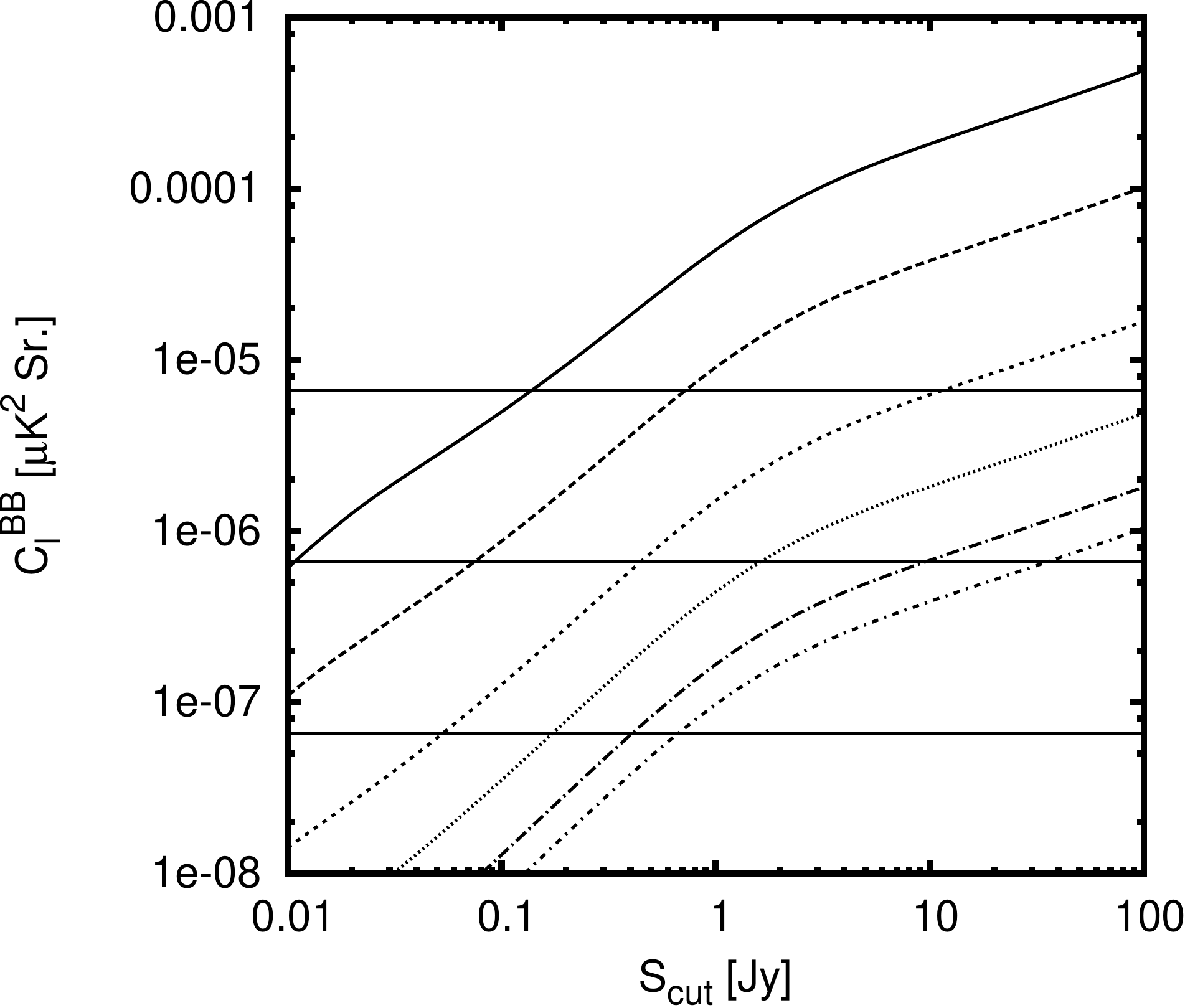}
\includegraphics[width=6cm,height=6cm]{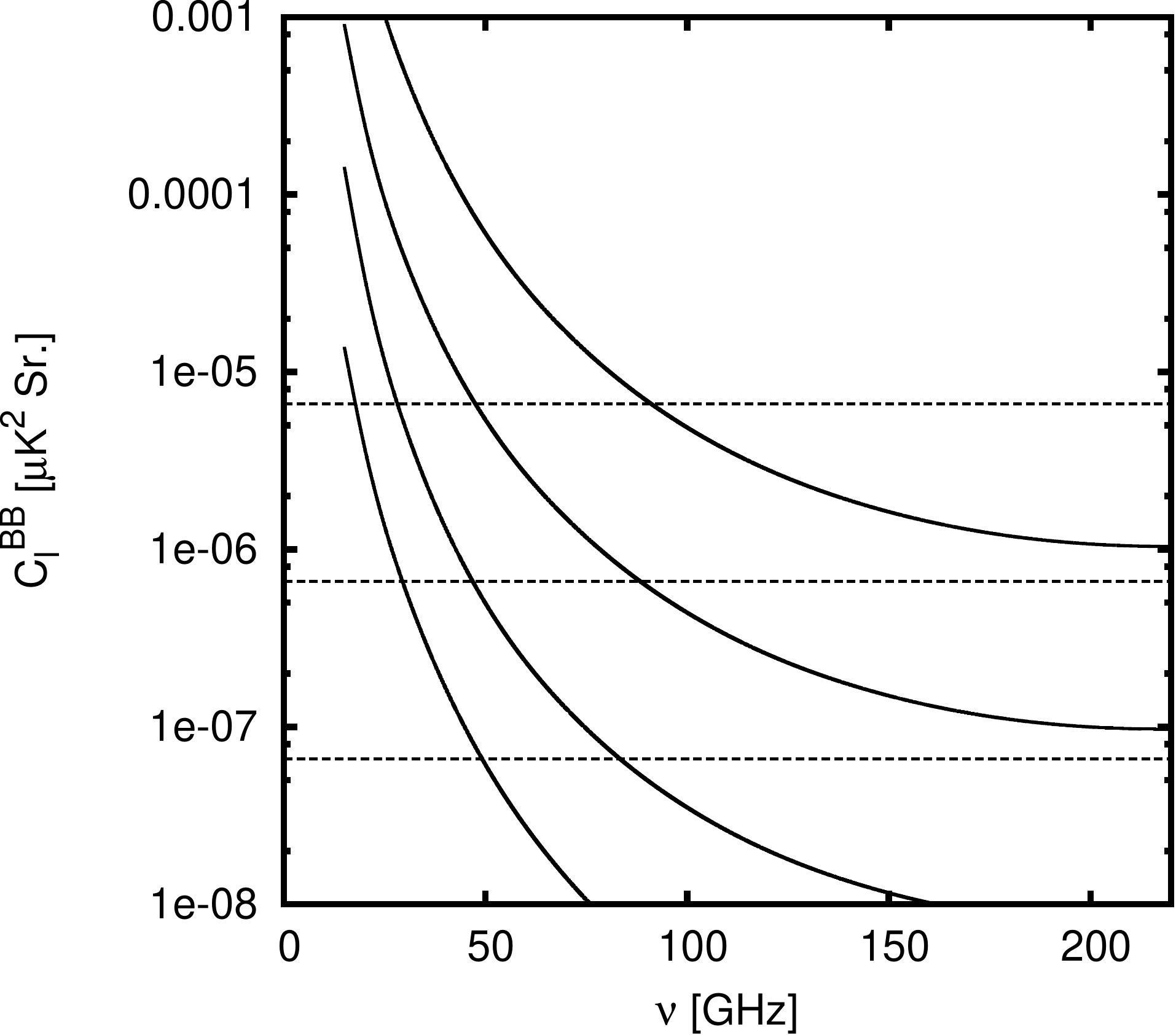}
\caption{Left: the projected point source noise spectrum, $C_{\ell}^{BB}$, as function of $S_{\rm cut}$ for a range of frequencies (30~GHz - solid line, 44~GHz - dashed line, 70~GHz - coarse dotted line, 100~GHz - fine dotted line, 150~GHz - long-dash dotted line, 220~GHz - short-dash dotted line. The 3 horizontal lines correspond to the power spectrum amplitudes at the maximum, $\ell=80$) of the B-mode spectrum for $r=0.1$, $0.01$ and $0.001$ from top to bottom. Right: the point source noise spectrum for $S_{\rm cut}=100, 1, 0.1, 0.01{\rm Jy}$ (top to bottom) as a function of frequency. $S_{\rm cut}=100{\rm Jy}$ corresponds to the level of noise we predict for no source subtraction.}
\label{fig:ns}
\end{figure}

\changed{In order to make predictions of the contamination of CMB polarization observations from point sources we need to have a model for the differential source count $n(S)=dN/dS$ as function of frequency. We have used the dZ05 model (de Zotti et al, 2005) which is available at a range of suitable frequencies.\footnote{See \url{http://web.oapd.inaf.it/rstools/srccnt/srccnt_tables.html}} . We have adopted the log-normal probability distribution (\ref{lognormal}) with $100\Pi_0=2.25$ and $\sigma=0.74$, independent of frequency, which gives $100\langle\Pi\rangle=2.9$ and $100\langle\Pi^2\rangle^{1/2}=3.9$ compatible with the observed values at 22~GHz. The differential source counts at 30~GHz and 150~GHz, along with measured source counts from Wright et al. (2009), Cleary et al. (2005) and Gawro\'nski et al. (2010) at 30GHz and Viera et al. (2010) and Marriage et al. (2010) at $\sim 150$GHz, and the probability distributions are presented in Fig.~\ref{fig:dnds_prop}. We also included predicted source counts at 30GHz from Toffolatti et al. (1998, T98) for comparison. The values of $\Pi_0$ and $\sigma$ were chosen to give the measured values for the median and the r.m.s. $\langle\Pi^2\rangle^{1/2}$. We also include a curve using $100\Pi_0=2.5$ and $\sigma=0.67$, values which are chosen to give the measured values of $\langle\Pi\rangle$ and $\langle\Pi^2\rangle^{1/2}$ for the contemporaneously detected sample at 22~GHz. Note that this \changed{second} set of values agree well with the histogram of {\it detected} sources, but the values we have chosen for the subsequent analysis are more realistic since they taken into account undetected sources.}

\begin{figure}
\centering
\includegraphics[scale=0.4]{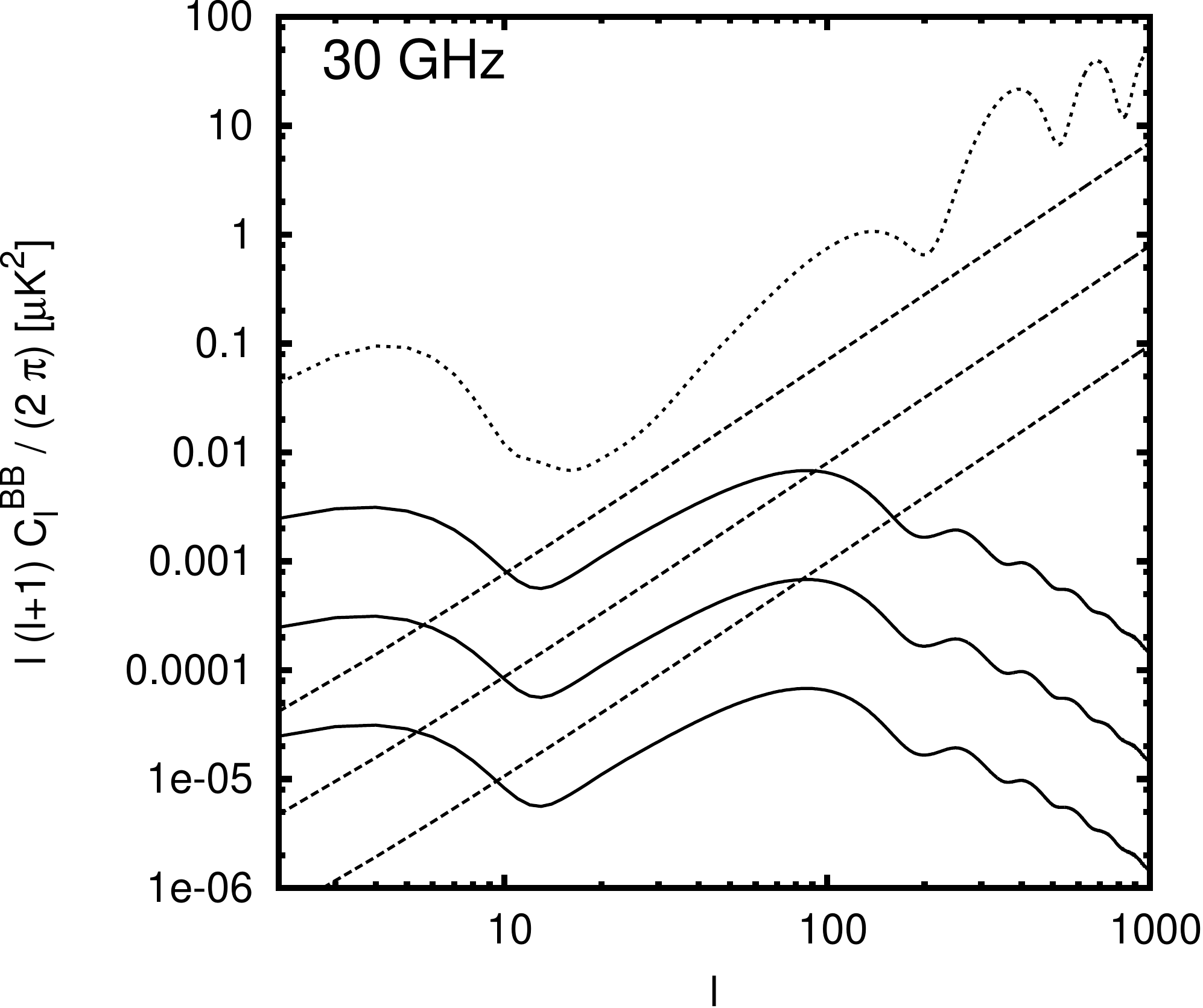}
\includegraphics[scale=0.4]{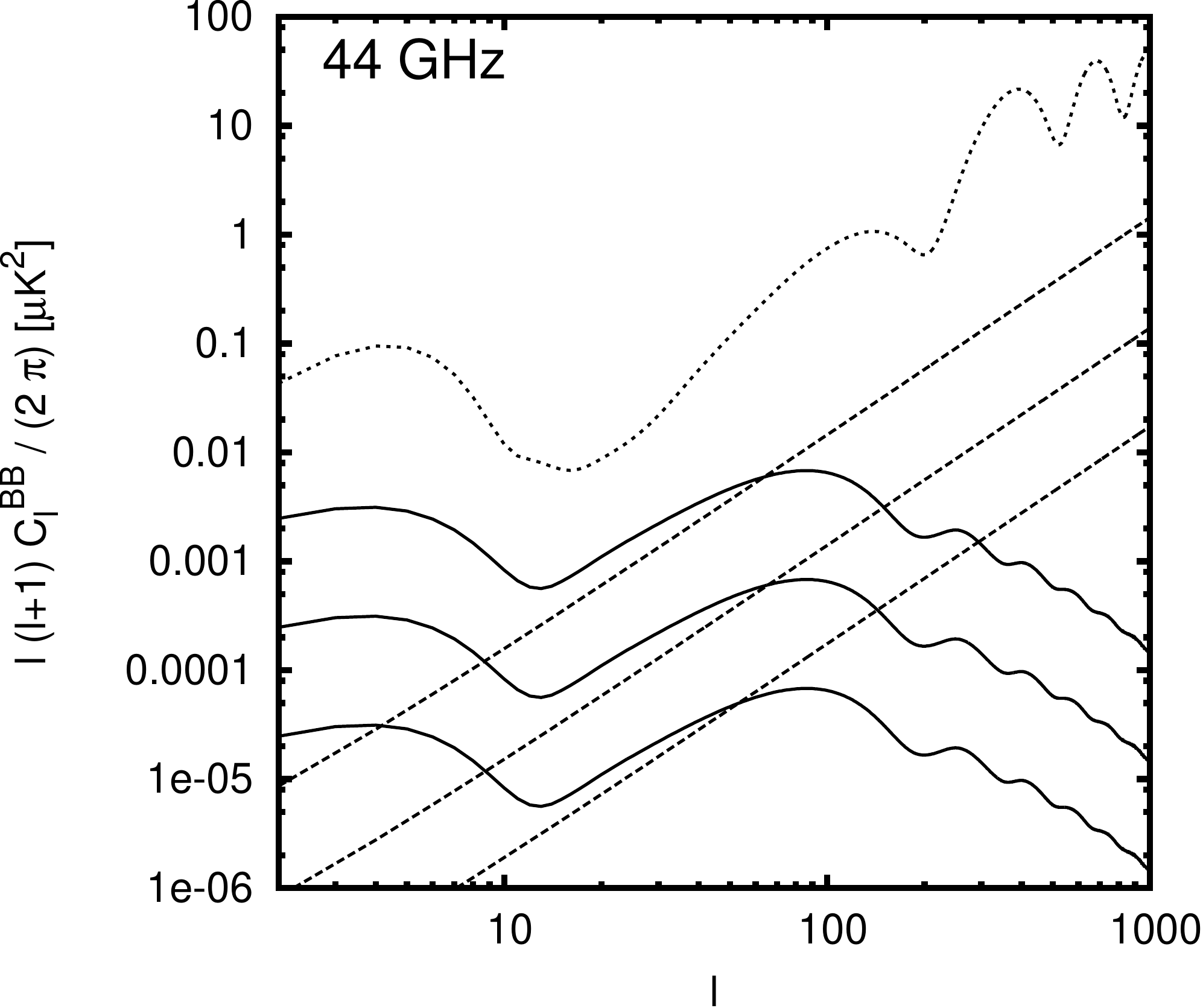}\\
\includegraphics[scale=0.4]{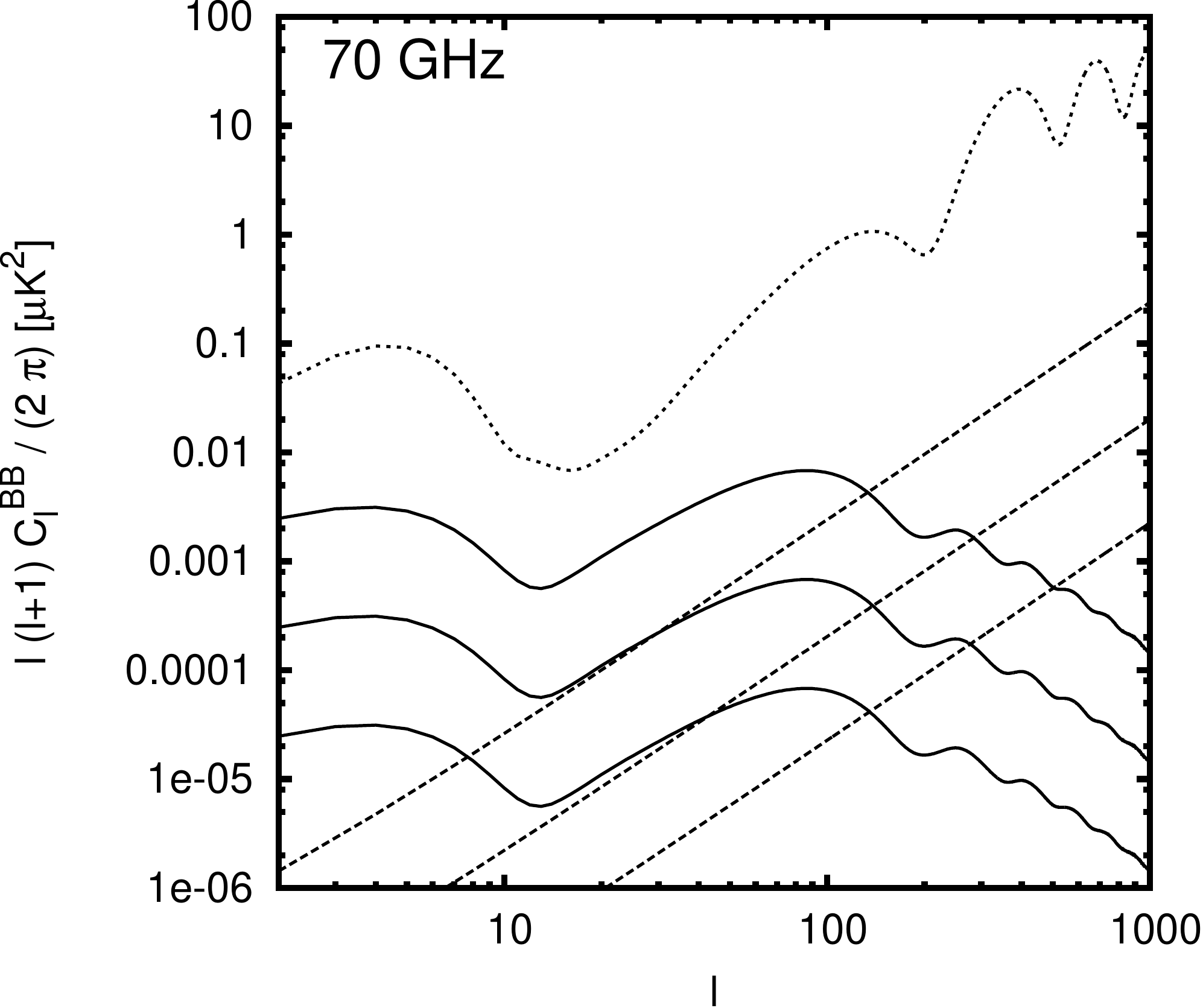}
\includegraphics[scale=0.4]{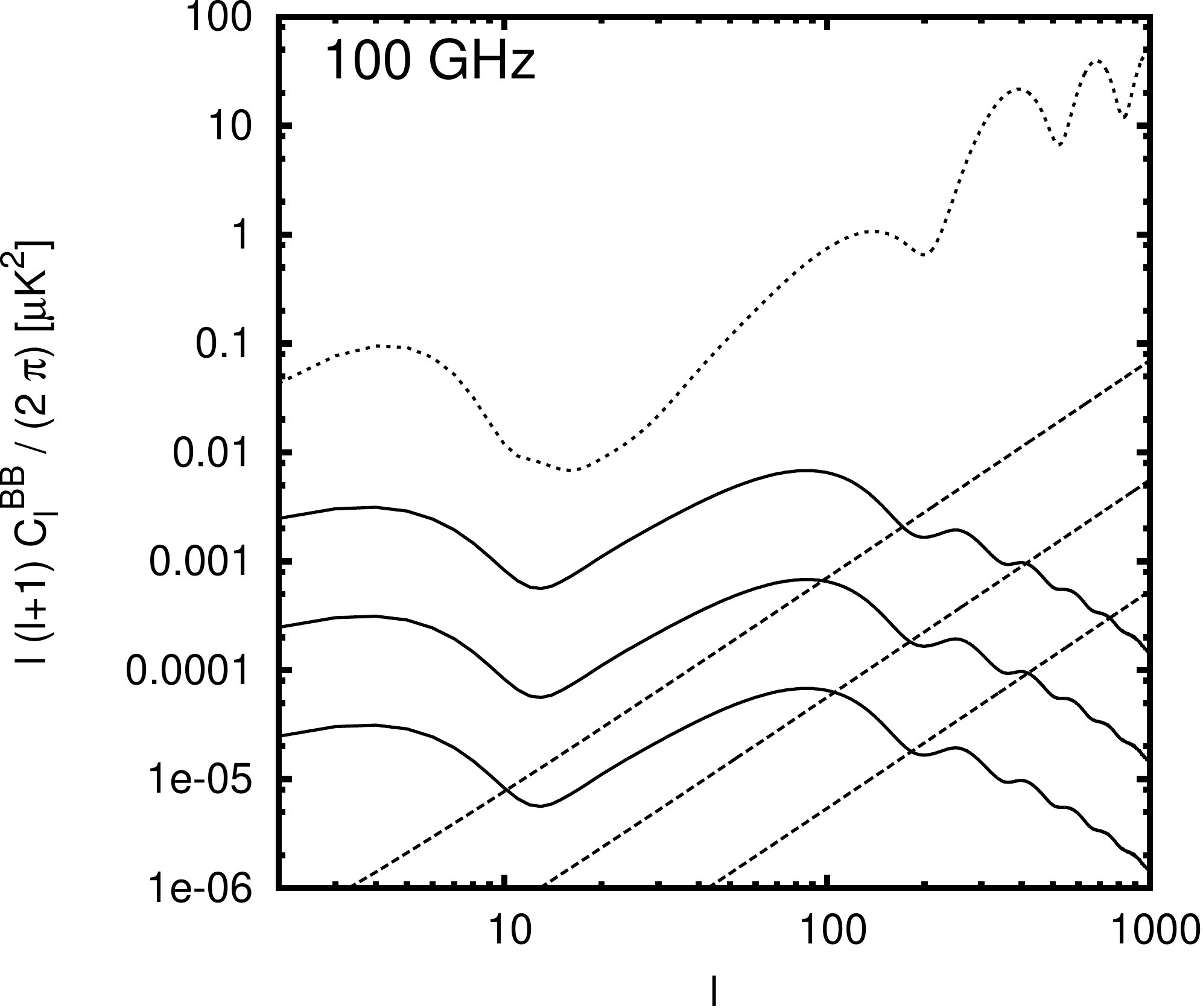}\\
\includegraphics[scale=0.4]{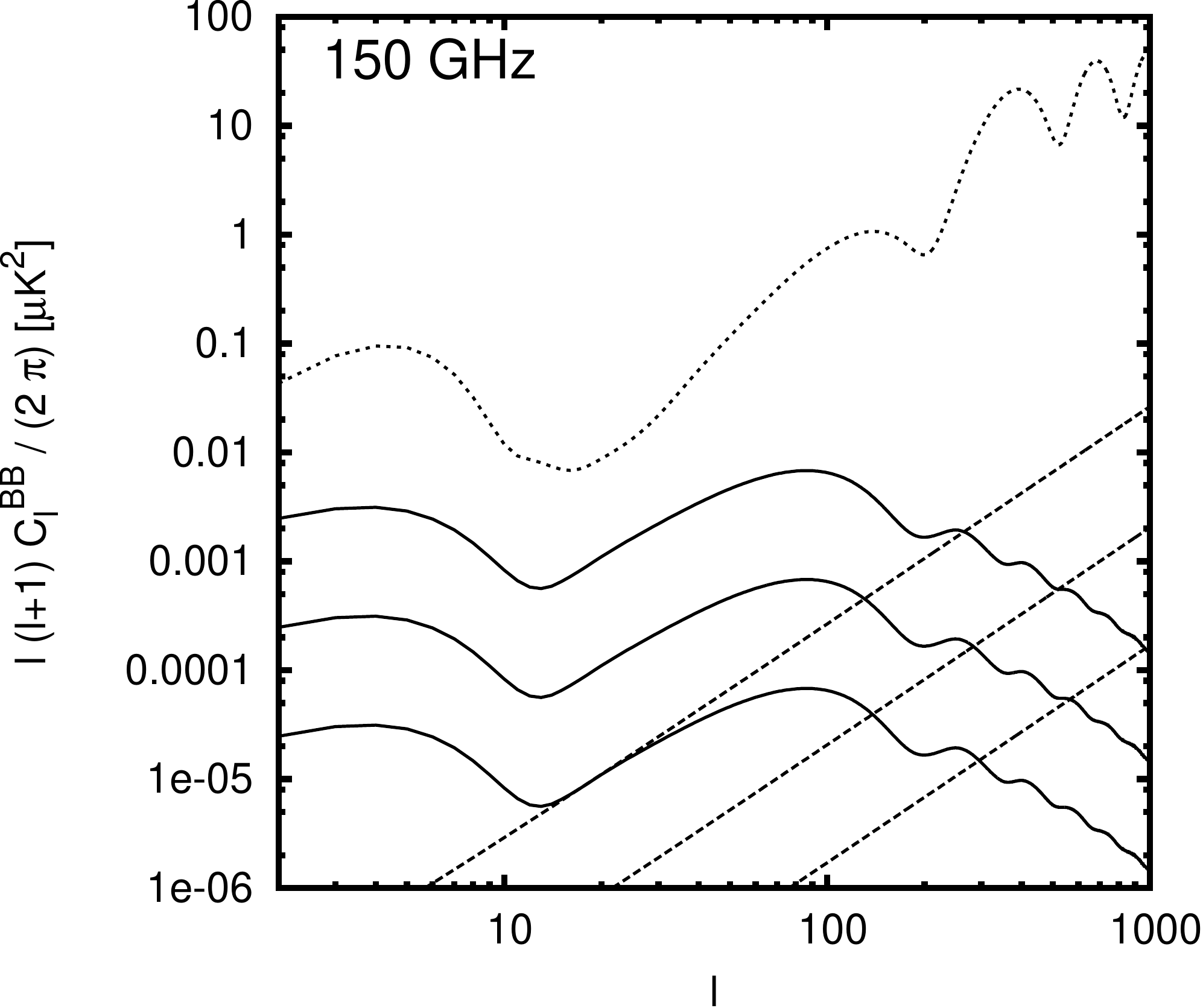}
\includegraphics[scale=0.4]{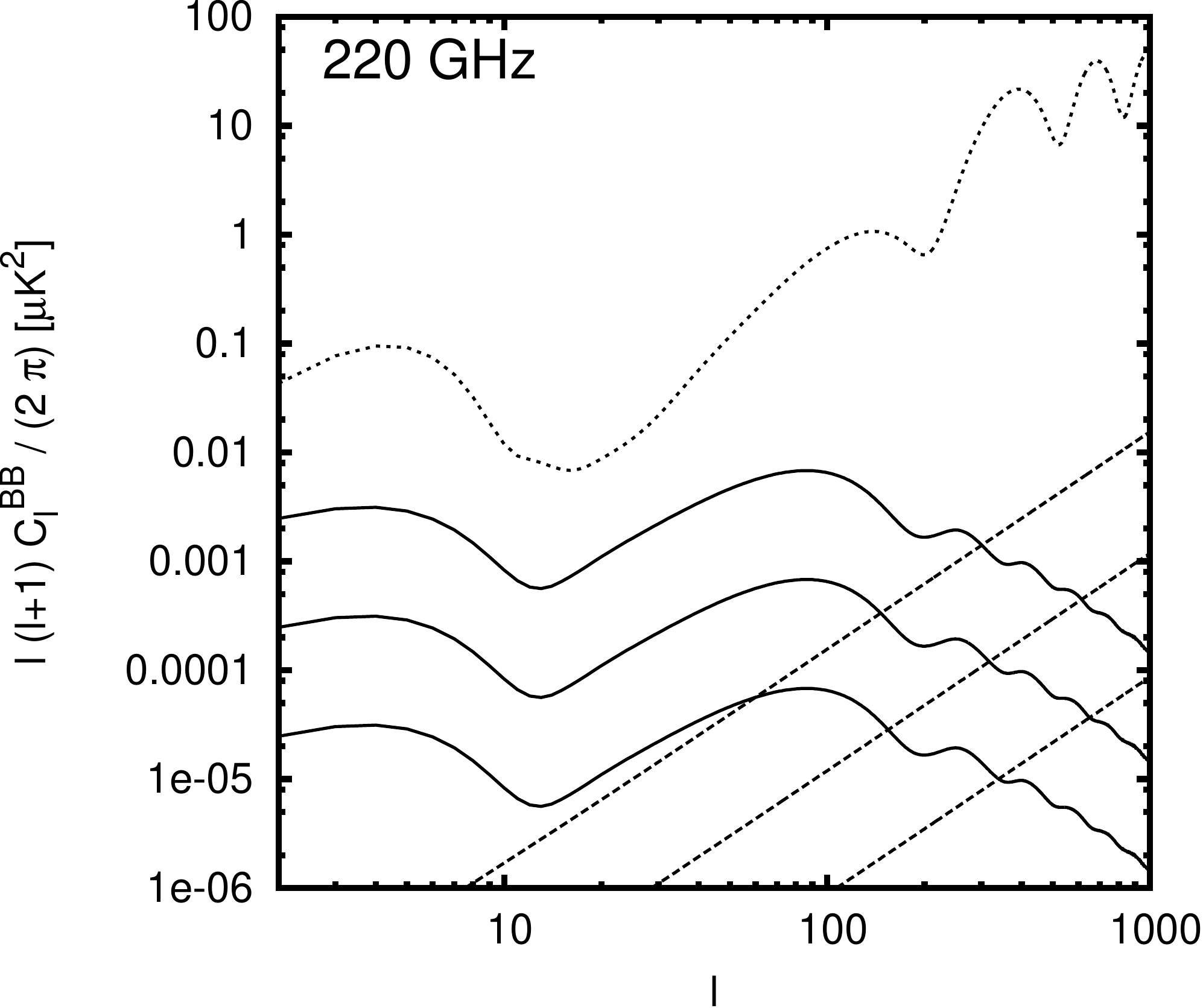}
\caption{The projected noise spectra for a range of frequencies and $S_{\rm cut}$. On each plot there are spectra for  three values of $S_{\rm cut}=1{\rm Jy}$, $100{\rm mJy}$ and $10{\rm mJy}$. Included also are a typical E-mode spectrum and  B-mode spectra for $r=0.1$ $0.01$ and $0.001$. The plots are for 30~GHz (top left), 44~GHz (top right), 70~GHz (middle left), 100~GHz (middle right), 150~GHz (bottom left) and 220~GHz (bottom right).}
\label{fig:bbmontage}
\end{figure}

We have presented the results using method A in the left hand panel of Fig.~\ref{fig:ns} as function of $S_{\rm cut}$  for a range of frequencies popular for CMB observations. We plot $C_{\ell}^{BB}=C_{\ell}^{P}/2$ and have included lines which represent the expected power spectrum amplitude for $r=0.1$, 0.01 and 0.001 at the maximum of the B-mode power spectrum ($\ell\approx 80$).  This is indicative of the level of foreground source subtraction required so as not to inhibit the detection of a gravitational wave induced B-mode polarization signal. It has been chosen as reference point, not as the precisely calculated requirement, since this would depend on the resolution of the instrument. As expected the point source noise level increases with $S_{\rm cut}$, asymptoting at higher  values and is larger for lower frequencies.  

We have computed the value of $C_{\ell}^{BB}$ as a function of frequency for $S_{\rm cut}=100, 1, 0.1, 0.01{\rm Jy}$. The cut-off of $100{\rm Jy}$ is effectively infinite as there are no sources with flux density greater than this. This is presented in the right hand panel of Fig.~\ref{fig:ns}  and illustrates the level of point source noise one would expect if no source subtraction is performed. It  appears to show that it would be possible to detect $r=0.1$ at frequencies greater than around 100~GHz without any subtraction. Some of the same information is presented in Fig.~\ref{fig:bbmontage} which shows the power spectra from the point sources and those expected from the primordial CMB signal. This gives a more detailed picture of what level of source subtraction is required to achieve a detection at a given value of $r$ as a function of resolution.

\begin{figure}
\centering
\includegraphics[width=6cm]{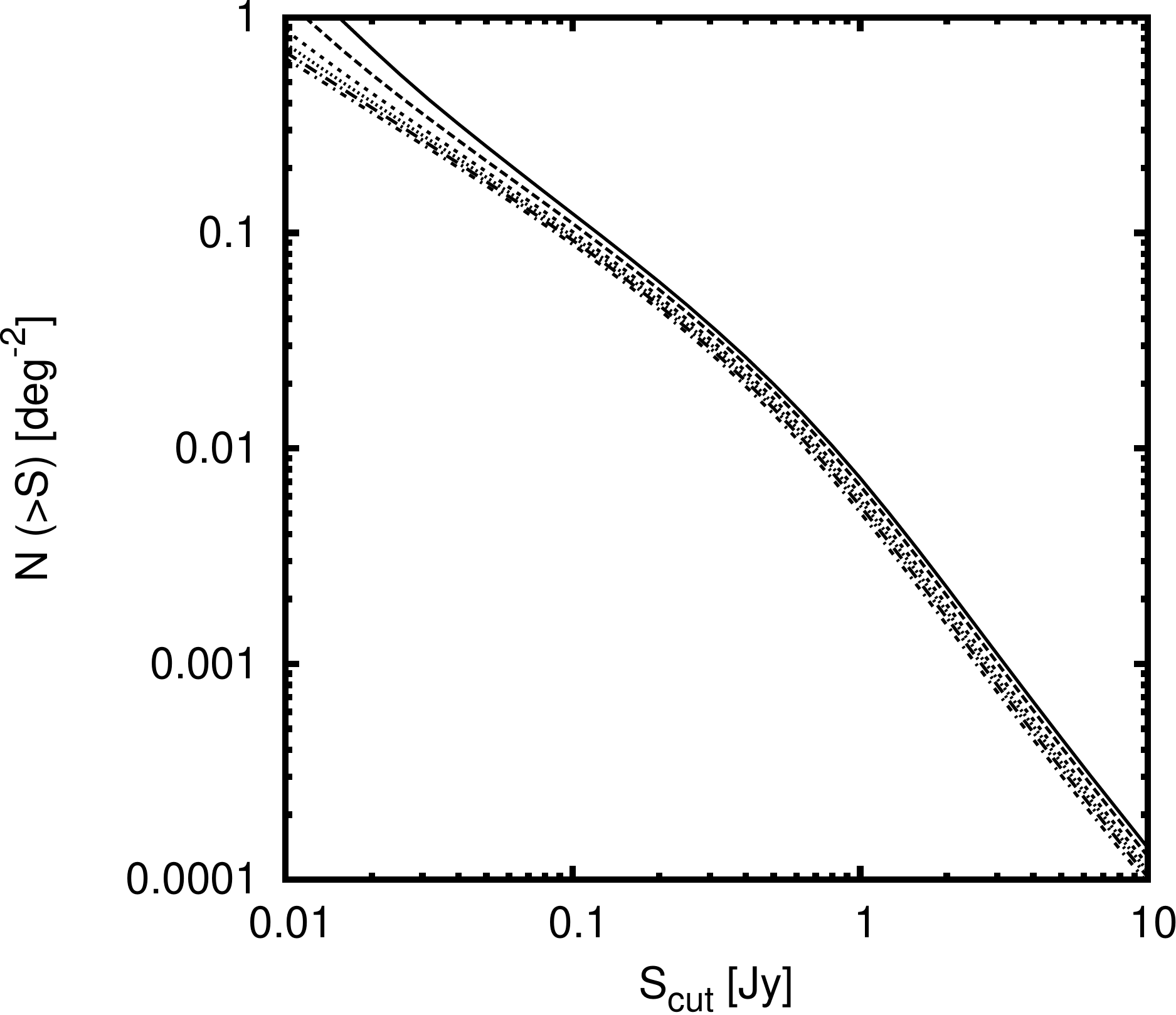}
\caption{The number of sources per square degree which need to be removed from a map for a given value of $S_{\rm cut}$. Curves are included for each of the frequencies 30, 40, 70, 100, 150, 220~GHz used in the earlier plots using the same line convention as in Fig.~\ref{fig:bbmontage}. We note that they are all very similar. }
\label{fig:ngts}
\end{figure}

In order to assess the feasibility of source subtraction it is important to know how many sources one might have to identify and extract from a map. In Fig.~\ref{fig:ngts} we have computed the expected number of sources above $S_{\rm cut}$ for $\nu=30-220$~GHz. The first thing to note is that this is relatively independent of $\nu$; there is less than a factor of 2 between the prediction at 30~GHz and that at 220~GHz. This may seem a little paradoxical since we have seen that the effect of the sources on the power spectrum is strongly dependent on frequency. The reason for this is frequency dependence of the conversion factor between brightness temperature and flux density, $dB/dT$ : sources $\sim 1{\rm Jy}$ correspond to substantially higher thermodynamic brightness temperatures at low frequencies.   For subtraction at the level of $S_{\rm cut}=1{\rm Jy}$, where most of the sources are likely to have been already identified, there is $\approx 1$ source per $200\,{\rm deg}^2$  which would have to be removed.  For $S_{\rm cut}=100{\rm mJy}$ for which a moderate sized program would have to be mounted using, for example, a dedicated \changed{instrument} or the VLA, there would be $\approx 1$ source per $8\,{\rm deg}^2$.

We note that the estimate above only includes the effects of point sources which are dominated by synchrotron emission and hence can be identified at low frequencies. It is well known that CMB observations at higher frequencies \changed{(greater than $\sim$100~GHz)} are contaminated by dust emission from star-forming galaxies at high redshift (Scott \& White, 1999). There is very little information available about the polarization of such objects but one would expect it to be relatively low since the alignment between the magnetic fields and dust grains in these objects is likely to be disordered, leading to low fractional polarization when integrated over the whole galaxy. The only information available at present for this population is for the local \changed{starburst galaxies} Arp220 for which there is an upper limit of $100\Pi<1.5$ at 353~GHz (Sieffert et al., 2007) \changed{and M82 which has been estimated to have $100\Pi\approx 0.4$ at the same frequency (Greaves and Holland, 2002).}

In order to estimate the confusion noise from this \changed{component we have to not only included the contribution from Poisson fluctuations, but also that from the clustering of the sources. We have been able to make an estimate for the Poission contribution by making the} assumption that $100\langle\Pi^2\rangle^{1/2}=1$  and have used the differential source count at 350~GHz suggested in Borys et al. (2003)
\begin{equation}
{dN\over dS}={N_0/S_0}\left[\left({S\over S_0}\right)^{\alpha}+\left({S\over S_0}\right)^{\beta}\right]^{-1}\,,
\end{equation}
where $S_0=1.8{\rm mJy}$, $\alpha=1.0$, $\beta=3.3$ and $N_0=1.5\times 10^4{\rm deg}^{-2}$. The extrapolation to lower frequencies is made using $S\propto\nu^{\gamma}$ where $\gamma\approx 3.5$ compatible with the measurements of Hall et al (2010). We find that $C_{\ell}^{P}=9\times 10^{-9}\mu{\rm K}^2{\rm sr}$ at 150~GHz, $2\times 10^{-8}\mu{\rm K}^2{\rm sr}$ at 220~GHz and $3\times 10^{-7}\mu{\rm K}^2{\rm sr}$ at 350~GHz compatible with the values presented in Seiffert et al. (2007). It appears that \changed{Poissonian contribution of these dusty galaxies is} sub-dominant to our estimate of the contribution from synchrotron dominated sources in this frequency range, though one should note our earlier remark that the radio source contribution may be over-estimated. Irrespective of this it seems unlikely that \changed{the Poissonian contribution} will have a significant impact on our ability to detect B-mode polarization for $r>0.01$ below 220~GHz unless there is a significant population of sources with $100\Pi\gg 1$. \changed{However, the galaxies are expected to be significantly clustered.  At present the scale of this effect is somewhat uncertain, but most studies (for example, Negrello et al 2007) appear to find a significant contribution to the confusion noise in total intensity, and observations are starting to constrain its amplitude (see Hall et al. 2010). The exact impact of this effect will be better understood when the results from the Planck satellite become available, which will significantly increase our knowledge of this population of galaxies over a range of frequencies.}

\begin{figure}
\centering
\includegraphics[width=6cm]{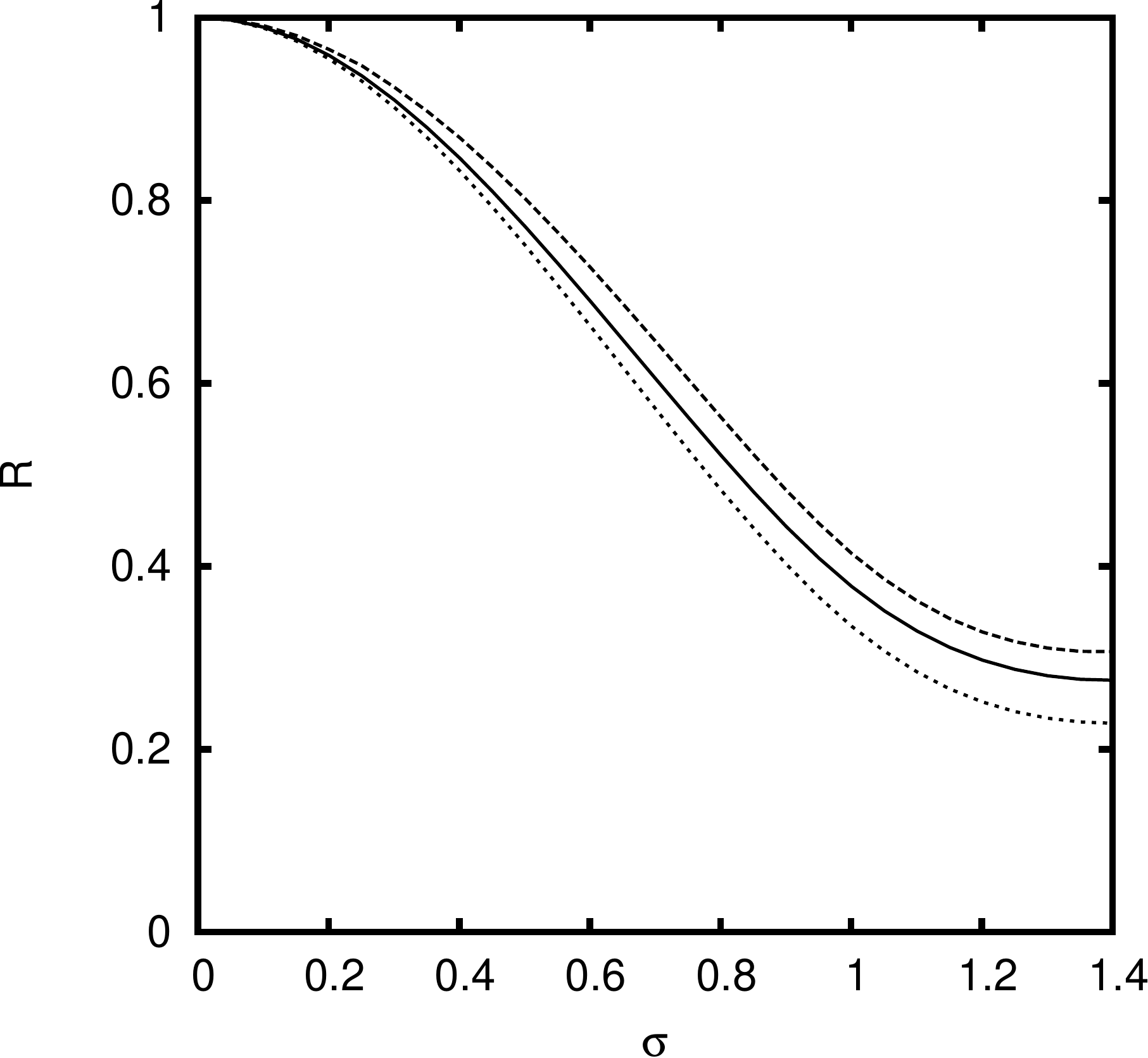}
\caption{The ratio R of the noise spectrum from method A and method B as a function of the scatter $\sigma$ for $\nu=30$~GHz. The three different curves are for $S_{\rm cut}=1{\rm Jy}$ (solid line), 100mJy (dashed line) and 10mJy (dotted line).}
\label{fig:ratio}
\end{figure}

Finally we have investigated the possible \changed{uncertainties} in using method A by setting $P_{\rm cut}=\langle\Pi\rangle S_{\rm cut}$ in method B. We have  computed the ratio $R$ (the \changed{confusion} noise from method B divided by that from method A) as a function of the scatter $\sigma$. For $\sigma=0$ the two methods give the same answer, that is, $R=1$ as expected, whereas for large values of $\sigma$ we find that $R<1$, that is, one finds that method A overpredicts the level of point source contamination when compared to model B. This might seem counterintuitive since one might expect a  number of sources with unusually large fractional polarizations to scatter upwards. One can confirm this behaviour by computing $R$ for a power-law differential source count $dN/dS\propto S^{-\alpha}$ for $\alpha<3$. We find that $R\propto\exp[\sigma^2\alpha(\alpha-3)/2]$ which decreases as $\sigma$ increases.

\section{Comparison with Planck Sky Model}
\label{sec:psm}

In order to make a comparision with previous work we have investigated the power spectrum of  polarized extragalactic sources in  the {\it Planck} Sky Model (PSM) version 1.6.6\footnote{http://www.apc.univ-paris7.fr/APC\_CS/Recherche/Adamis/PSM/psky-en.php}. The radio source model is based on sources from low frequency surveys extrapolated to higher frequencies with simple power-laws for each source. The total intensity spectra are adjusted above 20~GHz to agree with WMAP measurements as well as remaining consistent with the source count model of de Zotti et al. (2005). Polarization properties are based on the polarization fraction distribution for a sample of sources observed with ATCA at 18.5~GHz (Ricci et al. 2004) and an assumption of random position angles. A summary of the complete source model is given in Massardi et al. (2006).  

\begin{figure}
\centering
\includegraphics[height=6cm]{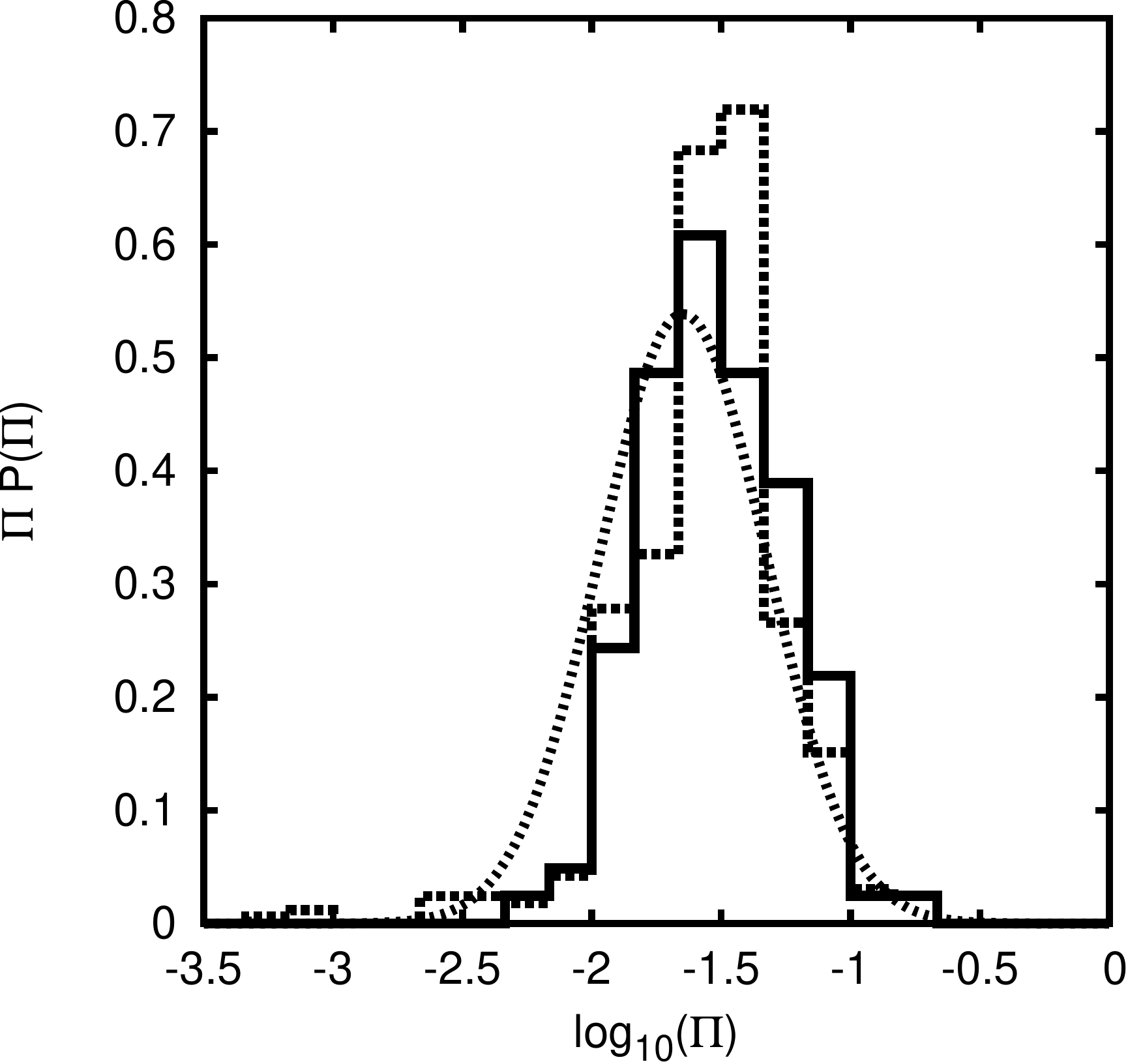}
\includegraphics[height=6cm]{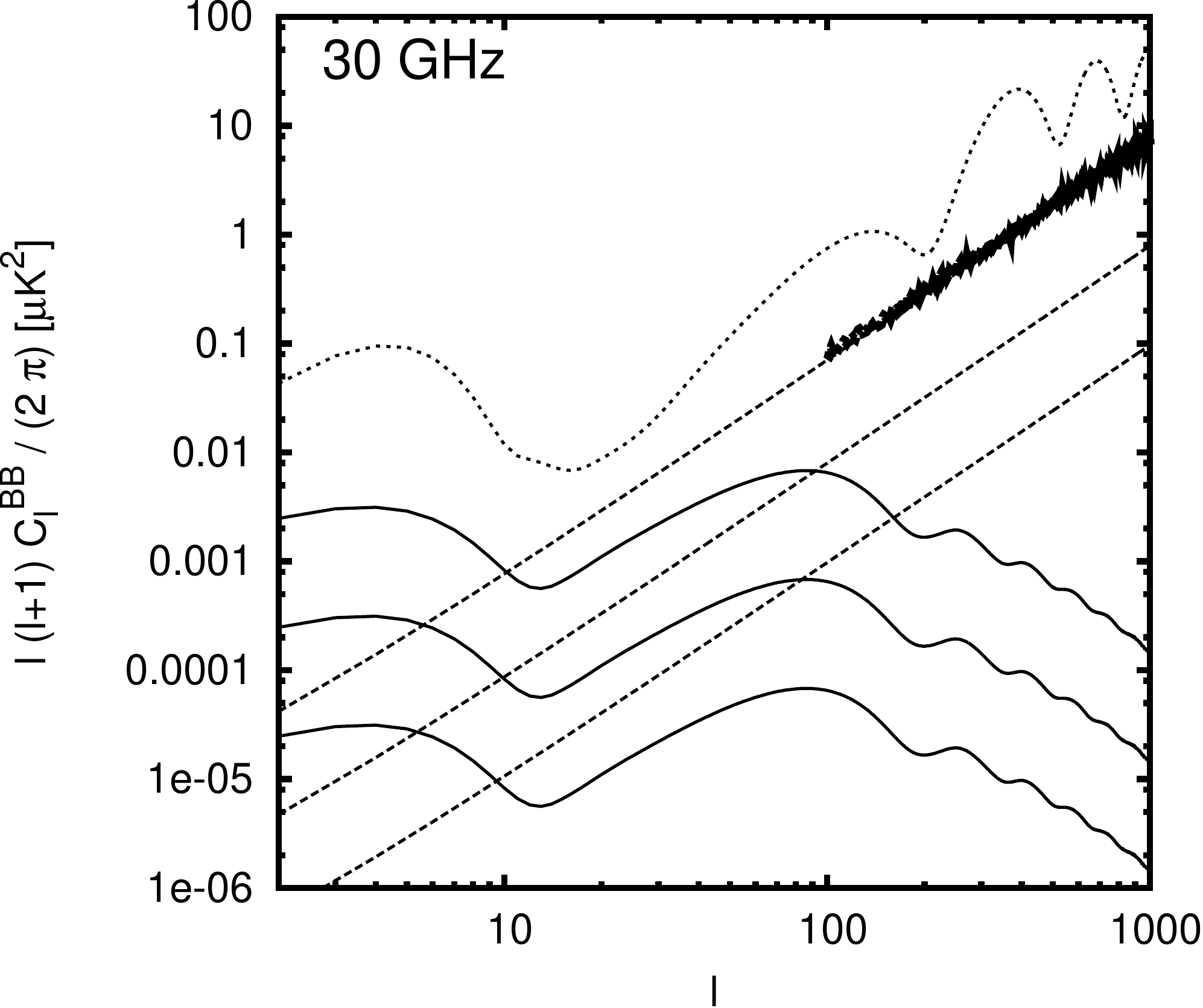}
\caption{On the left the fractional polarization distribution for source with flux density $>1{\rm Jy}$ from the PSM compared to that from our sample selected at 22~GHz and the distribution we used in section~\ref{sec:prediction}. On the right is the top left hand panel of Fig.~\ref{fig:bbmontage} with the reconstructed power spectrum from the PSM at 30~GHz for sources less than $1{\rm Jy}$ included.}
\label{fig:psm}
\end{figure}

In the left-hand panel of Fig.~\ref{fig:psm} we present the fractional polarization distribution for simulated sources extracted from the PSM with flux density greater than $1{\rm Jy}$ along with the model for $P(\Pi)$ we have used in section~\ref{sec:prediction}. There appears to be good agreement with that found in our work implying that there are similarities between fractional polarization distribution of our sample and that of Ricci et al (2004).

A full-sky realization of extragalactic sources was made at 30~GHz with a limiting flux cut-off of 1~Jy. The power spectrum was calculated using the {\it PolSpice} software v02-06-09\footnote{http://www.planck.fr/article141.html} (Chon et al. 2004) for Galactic latitudes $>5^{\circ}$. Fig.~\ref{fig:psm} (right-hand panel) shows the reconstructed B-mode spectrum from the PSM simulation in the range $\ell=100-1000$ for a flux cut-off of 1~Jy along with the predicted noise spectra from our model. The agreement with that presented in Fig.~\ref{fig:bbmontage} is extremely good by virtue of the polarization fraction distribution being close to ours and the source count in the PSM, measured to be $S^{5/2}{dN/dS}\approx 0.012 {\rm Jy}^{3/2}{\rm deg}^{-2}$, being very similar to the dZ05 model around $1{\rm Jy}$. In the range of flux densities where our observations are relevant the two models appear to agree. We do not compare lower flux density cuts since the PSM is incomplete below $\sim 100{\rm mJy}$.

\section{Summary and conclusions}

We have presented some statistical properties of polarized radio sources from our observations of the WMAP point source catalogue at 8.4, 22 and 43~GHz. The main results are:

\begin{itemize}

\item the median level of fractional polarization is around $2-2.5$ per cent and there is no evidence for any dependence on frequency in the range 8.4 to 43~GHz;

\item we have argued that our data are compatible with the  spectra in polarized intensity being less smooth than those in total intensity;

\item there is no evidence for a strong  correlation between a single intensity spectral index and the fractional polarization, although most of our sources are flat spectrum;

\item we were able to compute rotation measures for 45 out of the 105 sources which were detected contemporaneously at all three frequencies; 

\item there is statistical evidence for large rotation measures with a significant number of sources having rotation measures $>1000\,{\rm rad}\,{\rm m}^{-2}$.

\end{itemize}

We then used our results, in conjunction with the dZ05 model, to model the polarized source counts in order to compute the level of confusion noise expected in CMB polarization experiments designed to detect primordial gravitational waves from inflation. We find that 

\begin{itemize}

\item the source confusion due to jet-powered radio sources is likely to be the dominant contribution to the total up to relatively high frequencies - in our estimates they even dominate at 220~GHz but we have also pointed out that this could be an artefact of our assumption of a single spectral index for each source;

\item source subtraction will be important at frequencies below 100~GHz if one is to detect $r\sim 0.1$;

\item if one is to detect $r\sim 0.01$ some source mitigation strategy will be required at all frequencies.

\end{itemize}
In addition, 
there appears to be good agreement with our modelling and that in the PSM at 30~GHz for flux densities $\approx 1{\rm Jy}$. 

There is one final point we should make. Many of the sources, particularly those with high flux densities, are expected to be significantly variable. This will manifest itself not only in variable total and polarized flux densities, but also in the polarization position angle. This could make it difficult to subtract the effects of source confusion from the CMB polarization signal. If the observations of a particular CMB field are performed over a timescale which is shorter than the timescale of variability, this will require the high resolution observations performed to assist source subtraction to take place contemporaneously in order to make an accurate subtraction. Conversely, if the timescale for observations is longer than the variability, for example, if observations on a particular region are built up  a series of short integrations over many days, then the variability of the source polarization could easily average out to a significantly lower observed polarization when the individual integrations are stacked. This effect will make accurate source subtraction difficult since it is likely to be impractical to monitor the level of variability for a substantial number of sources.
 
\section*{Acknowledgements}

The authors acknowledge the use of the {\it Planck} Sky Model, developed by the Component Separation Working Group (WG2) of the {\it Planck} Collaboration. We thank G. de Zotti and M. Massardi for helpful clarifications. CD acknowledges an STFC advanced fellowship and ERC grant under FP7.

\section*{Appendix: Correlations between observed quantities}

\begin{figure}
\includegraphics[scale=0.23]{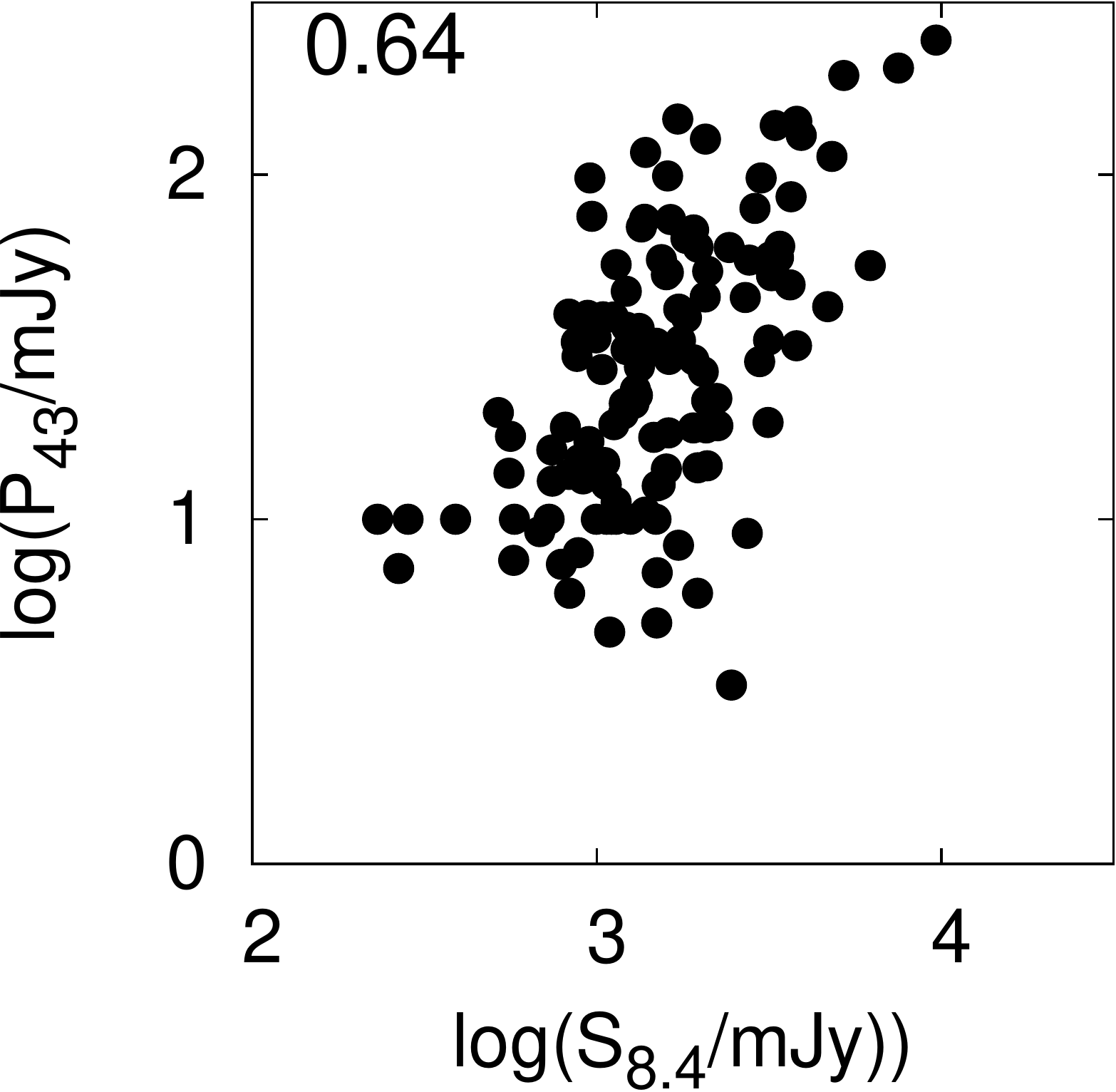}
\includegraphics[scale=0.23]{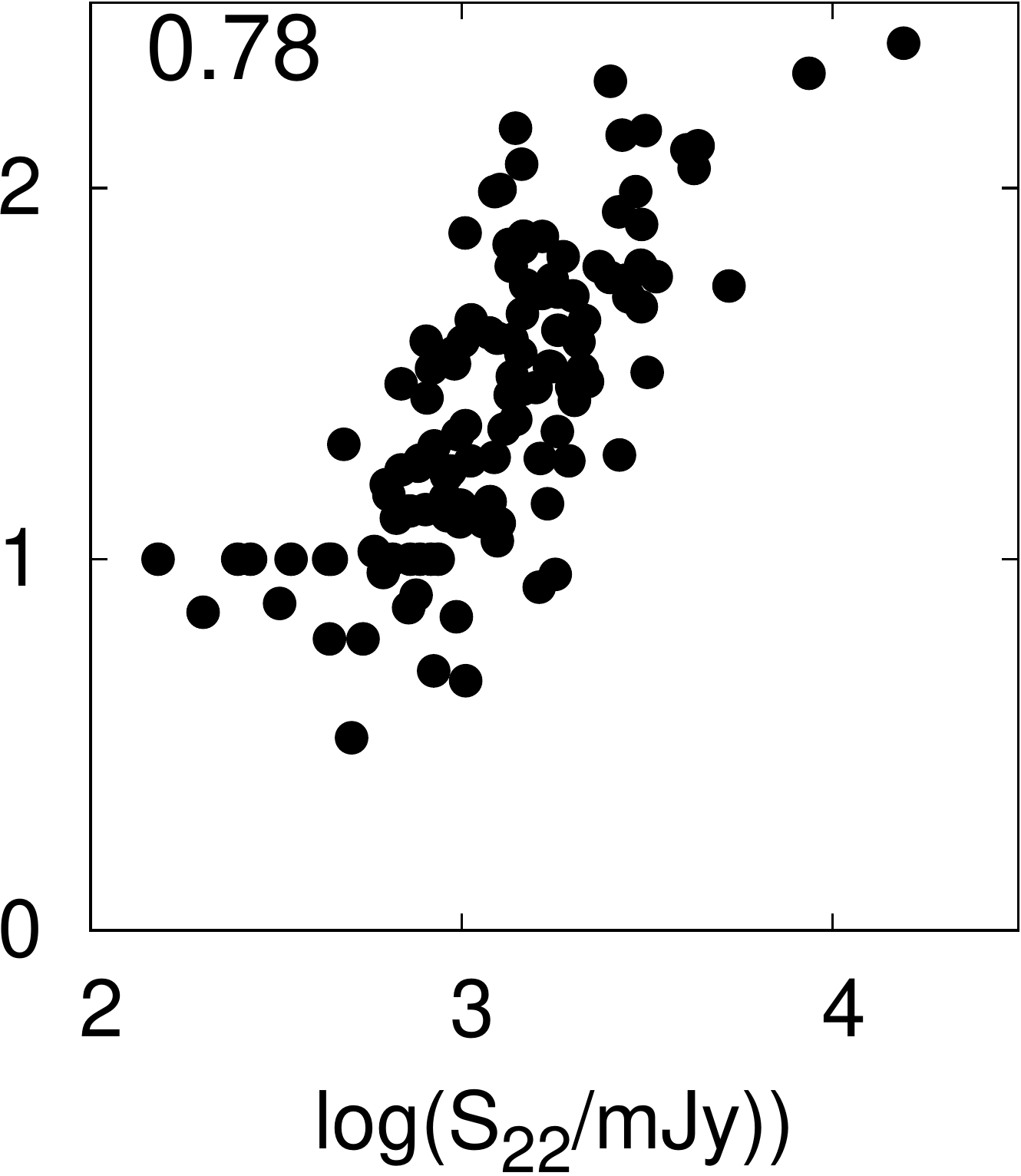}
\includegraphics[scale=0.23]{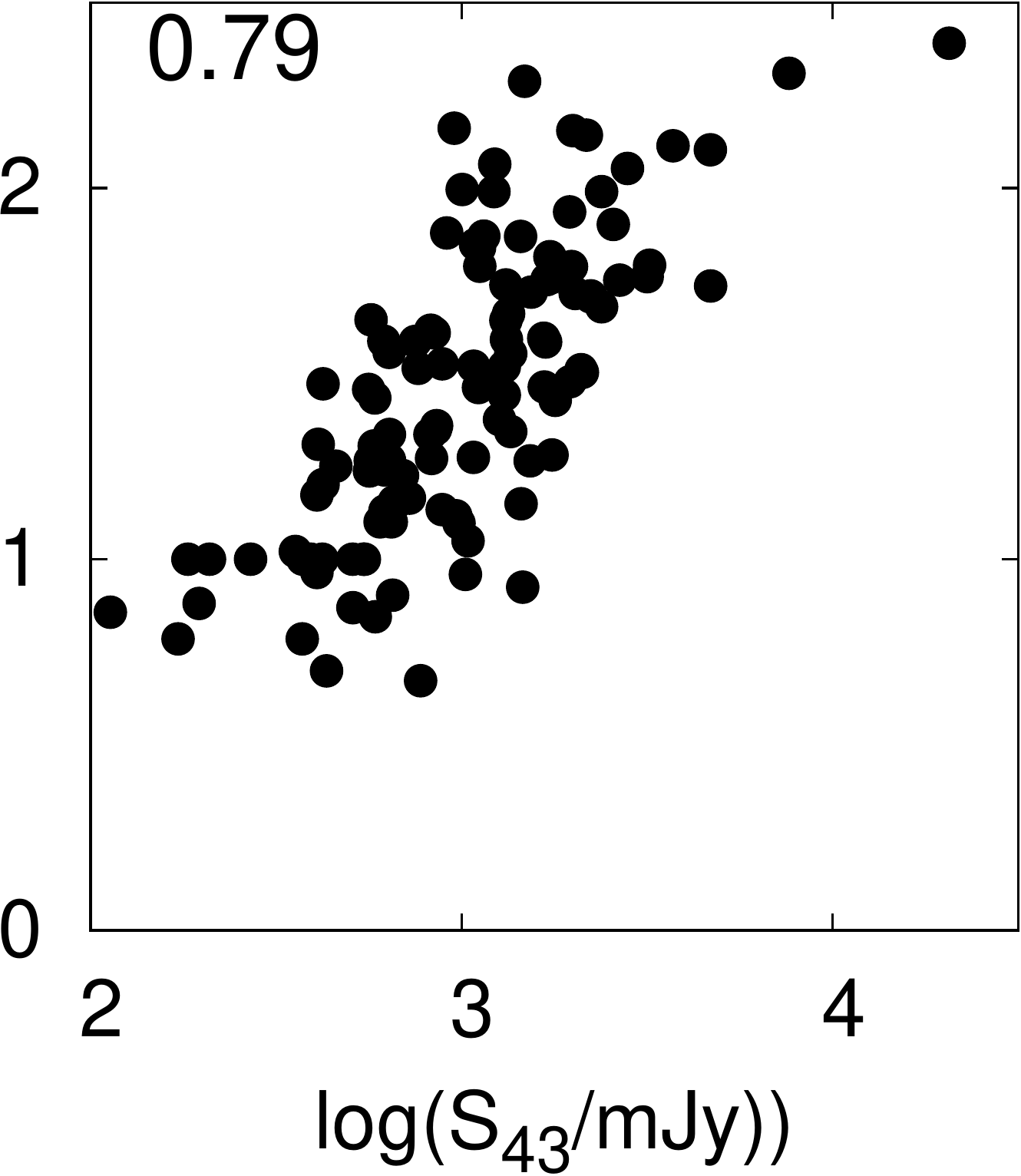}
\includegraphics[scale=0.23]{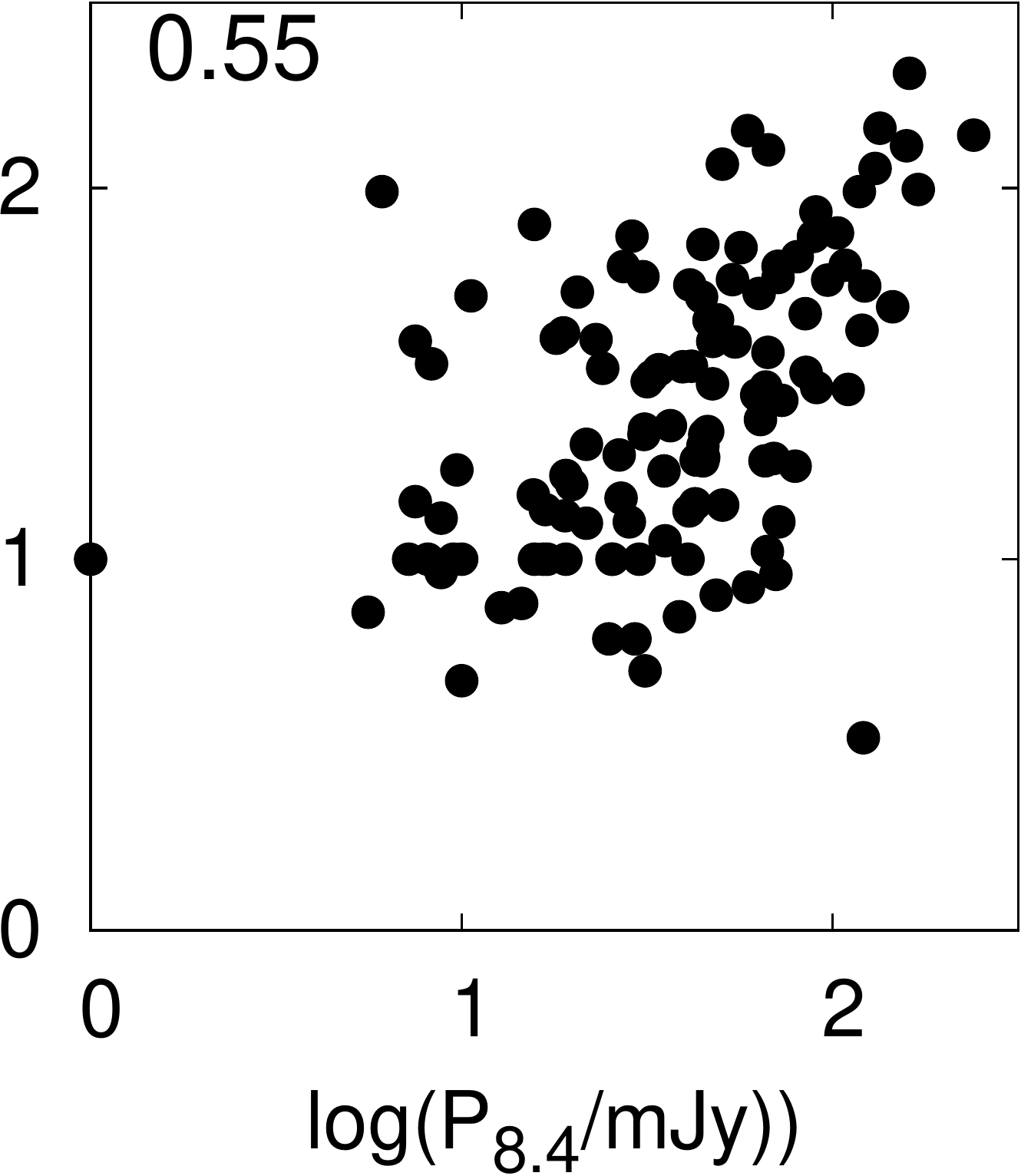}
\includegraphics[scale=0.23]{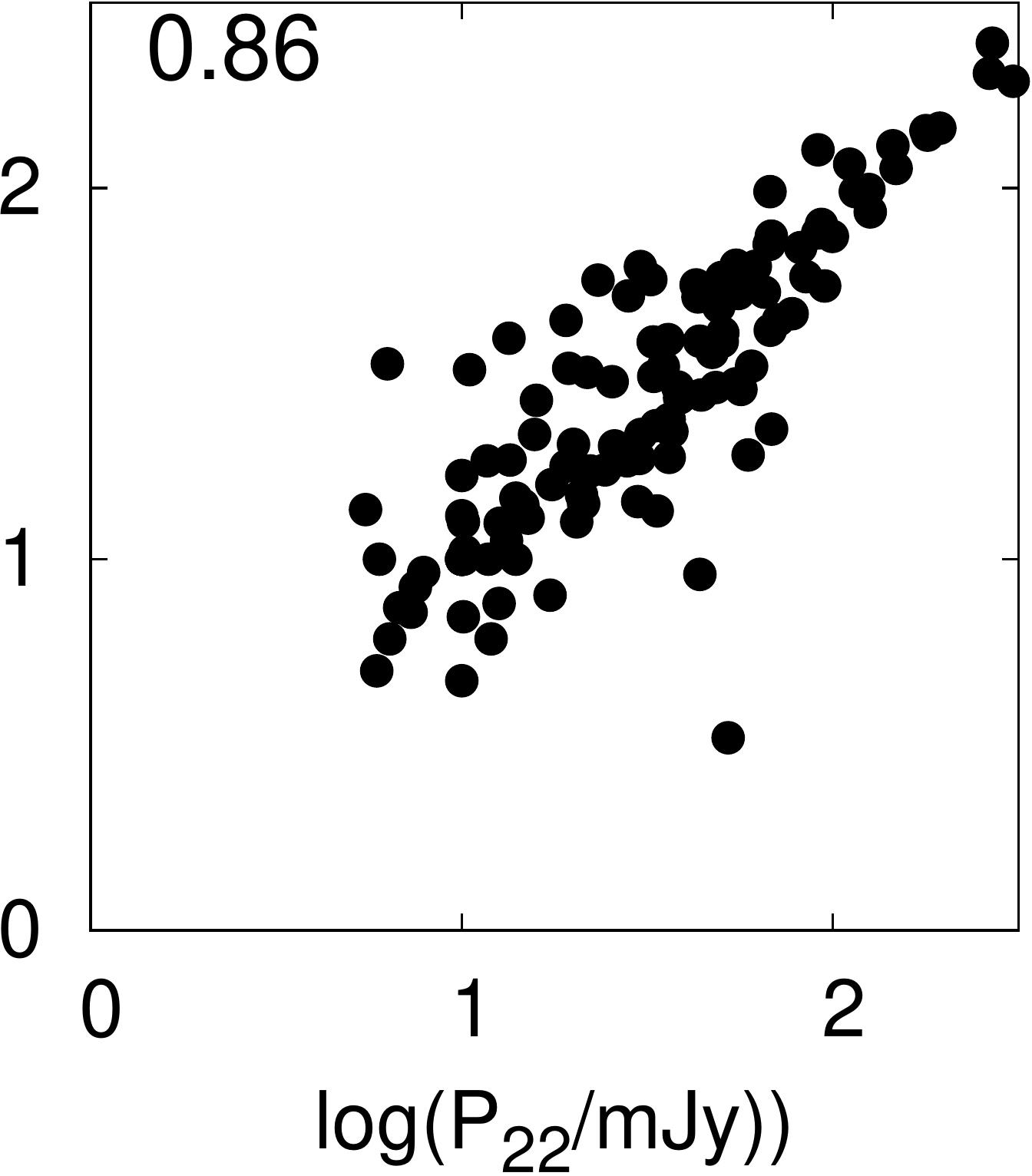}\\
\includegraphics[scale=0.23]{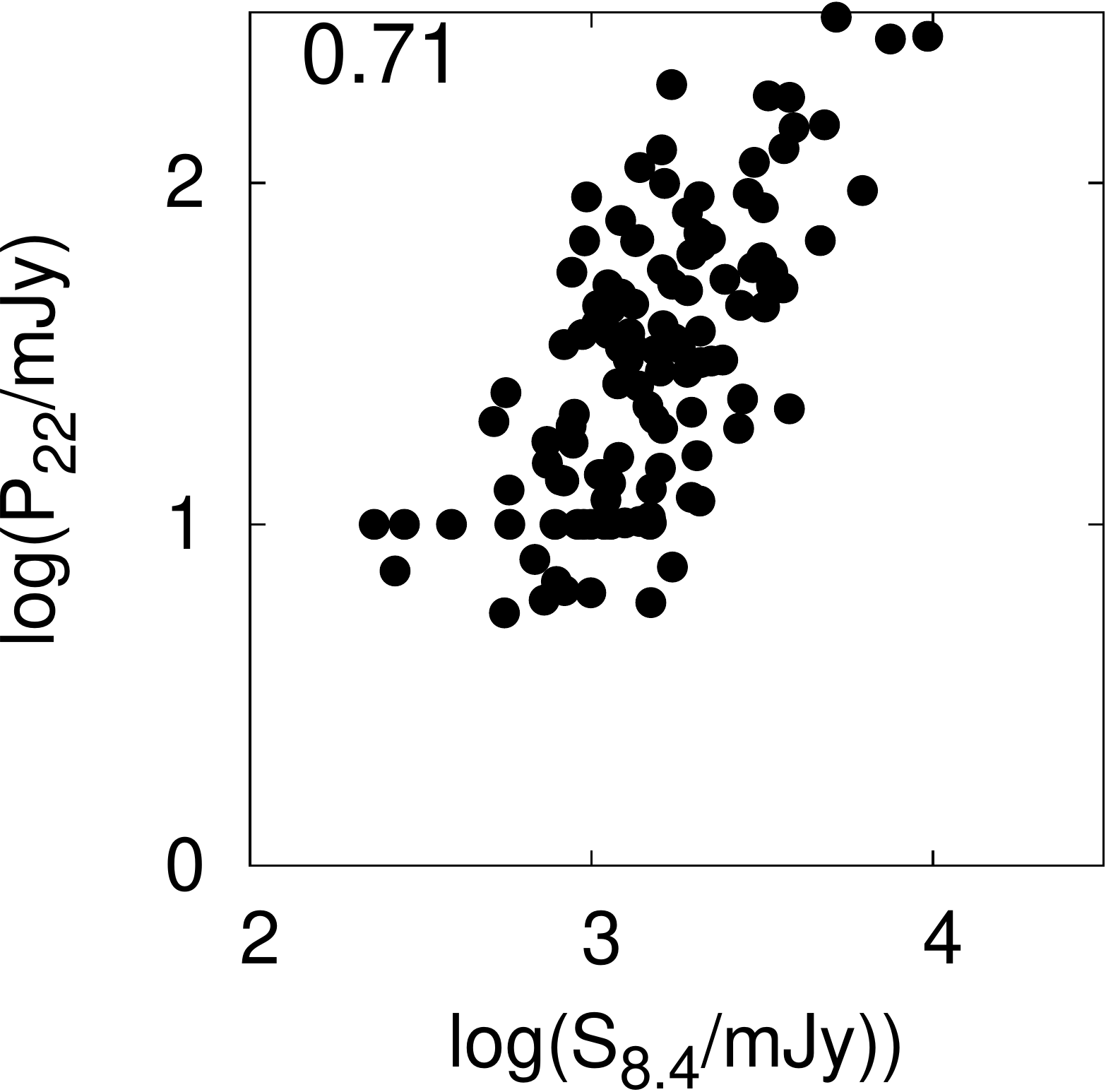}
\includegraphics[scale=0.23]{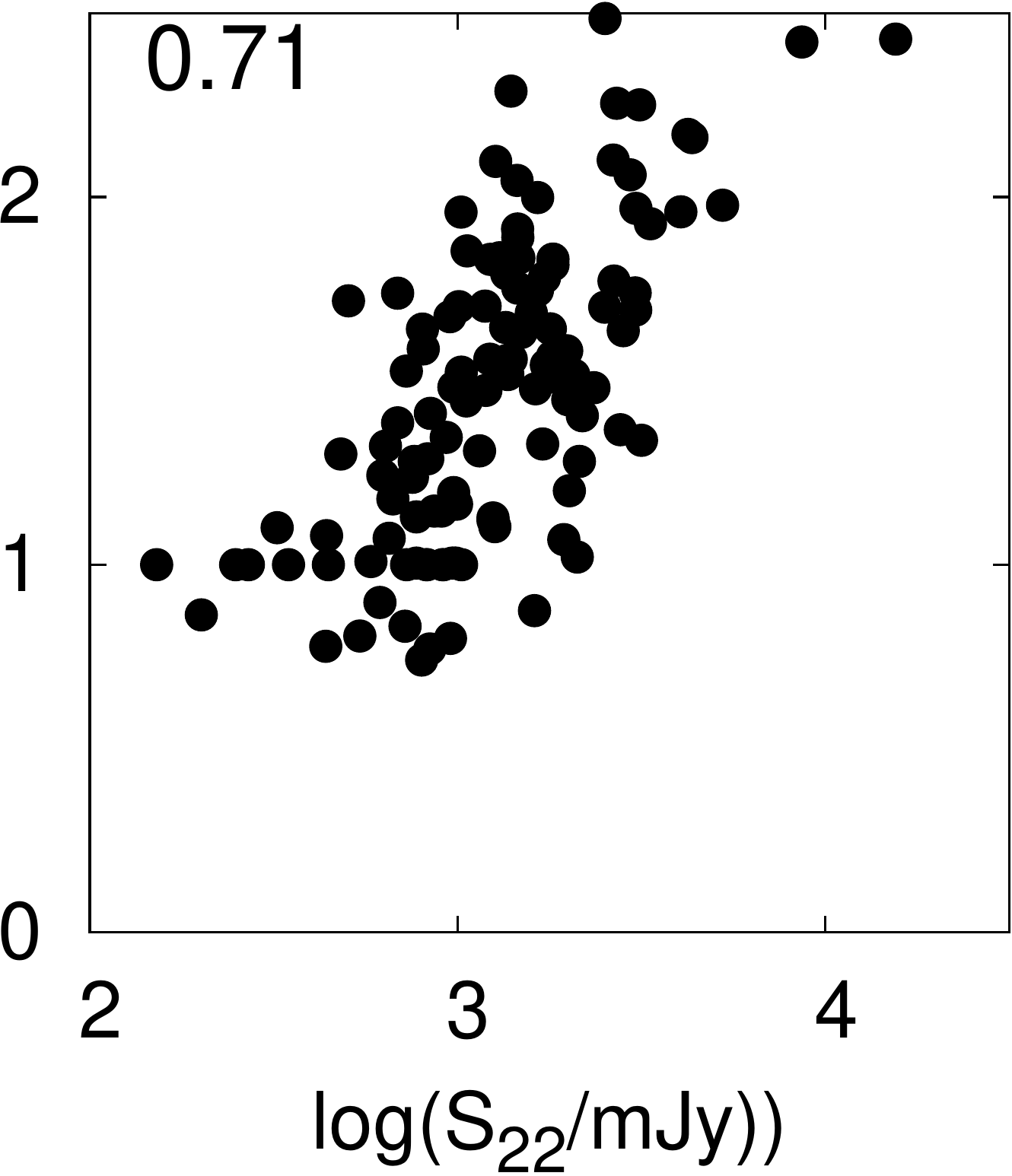}
\includegraphics[scale=0.23]{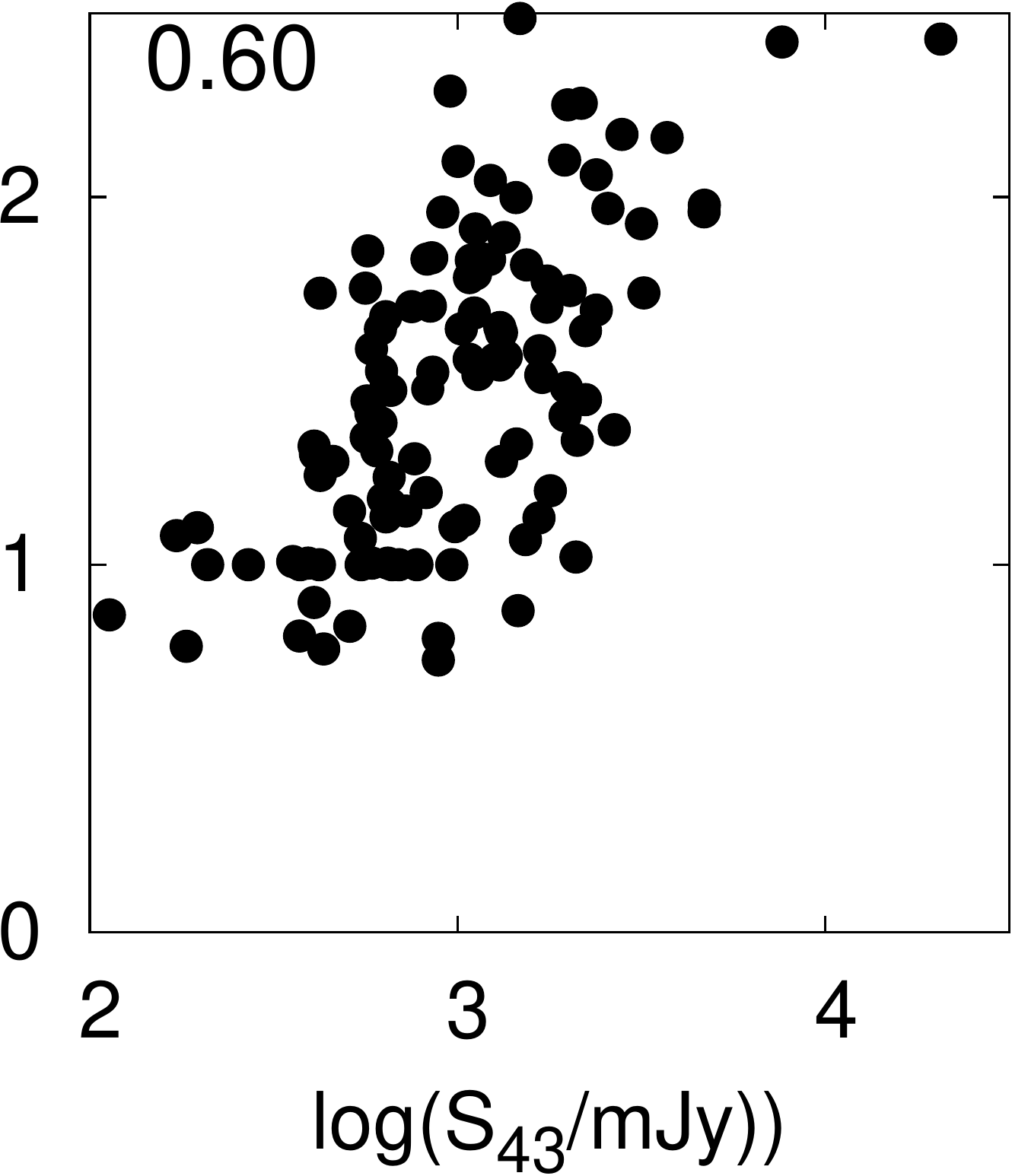}
\includegraphics[scale=0.23]{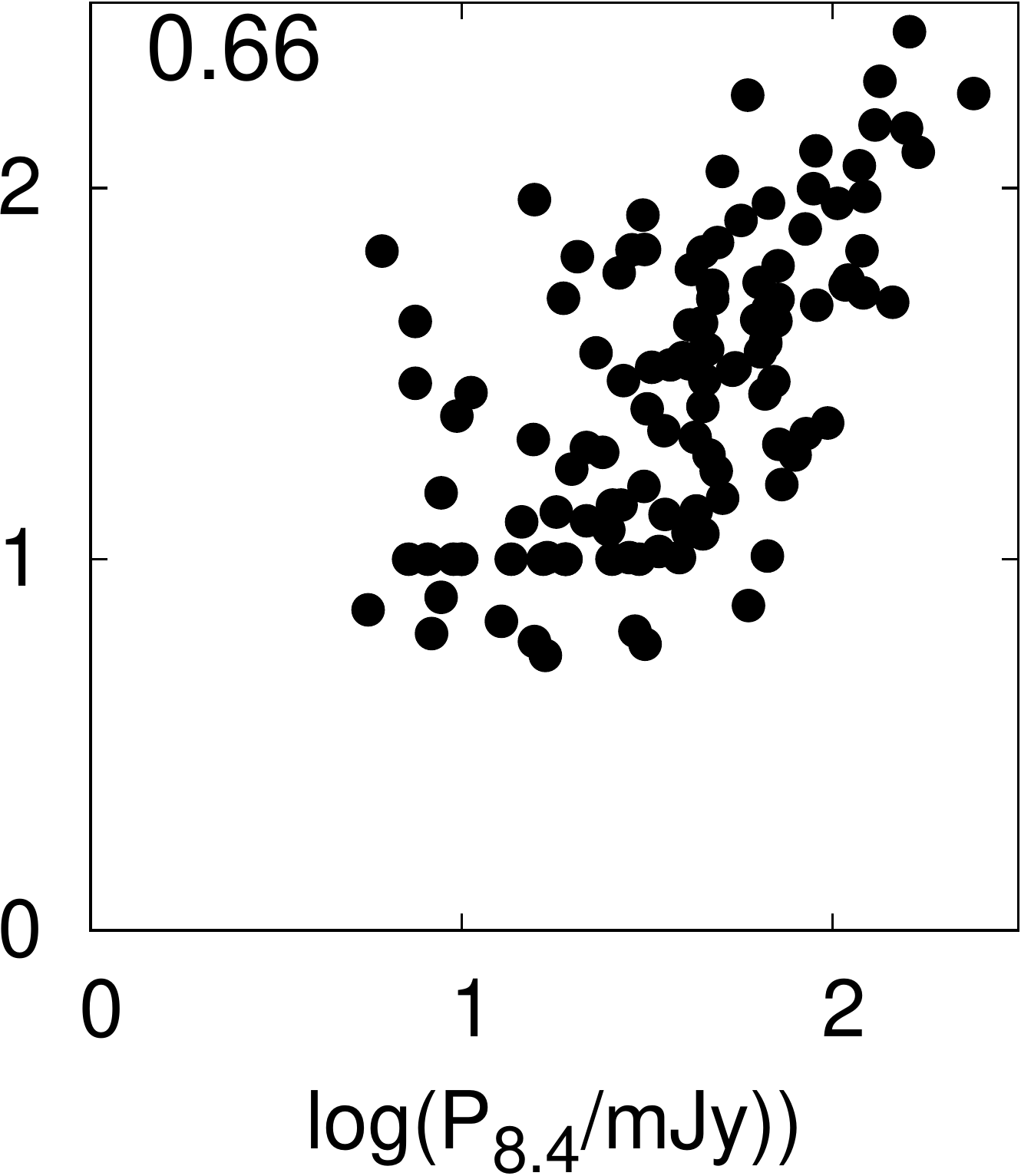}\\
\includegraphics[scale=0.23]{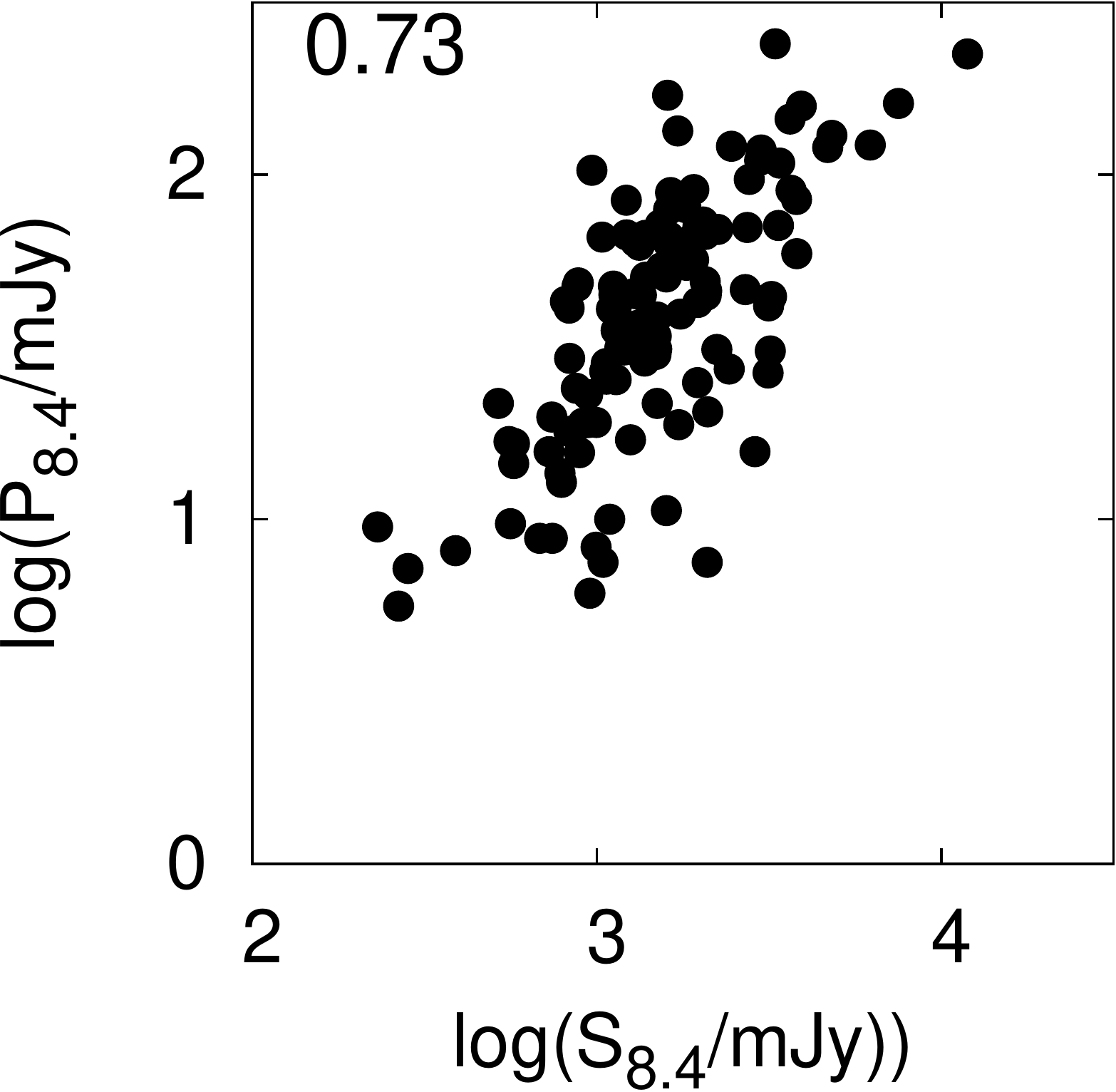}
\includegraphics[scale=0.23]{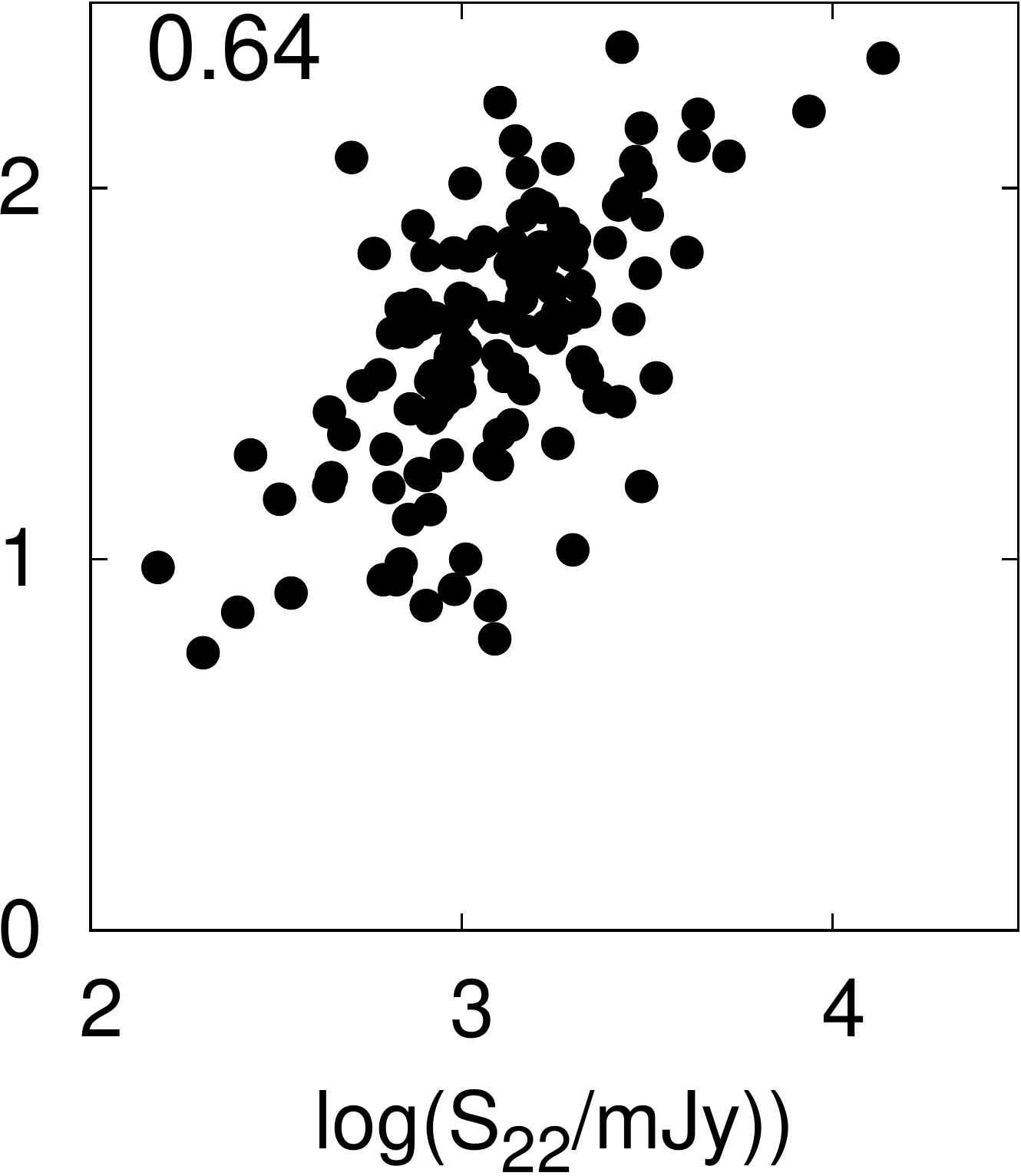}
\includegraphics[scale=0.23]{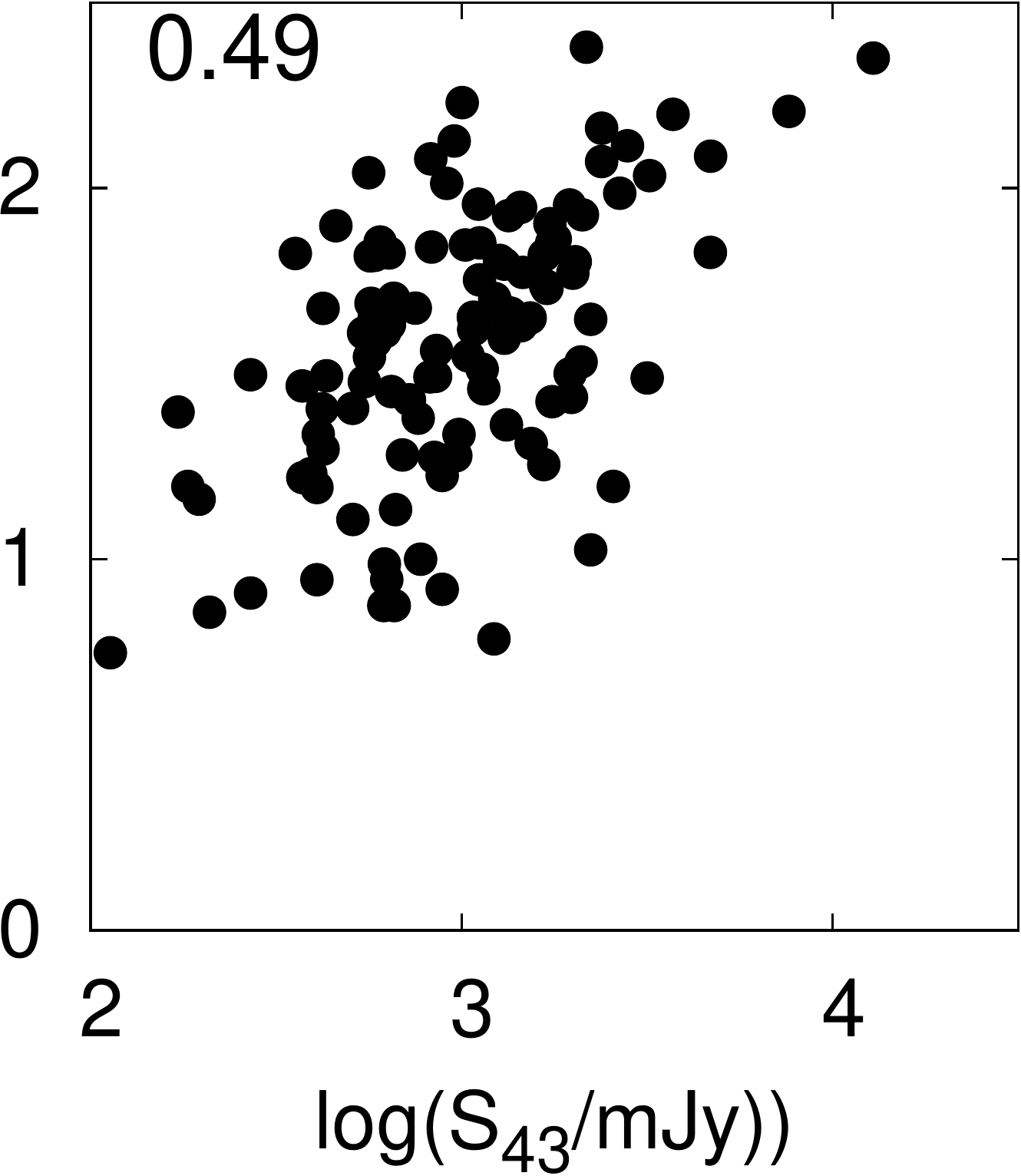}\\
\includegraphics[scale=0.23]{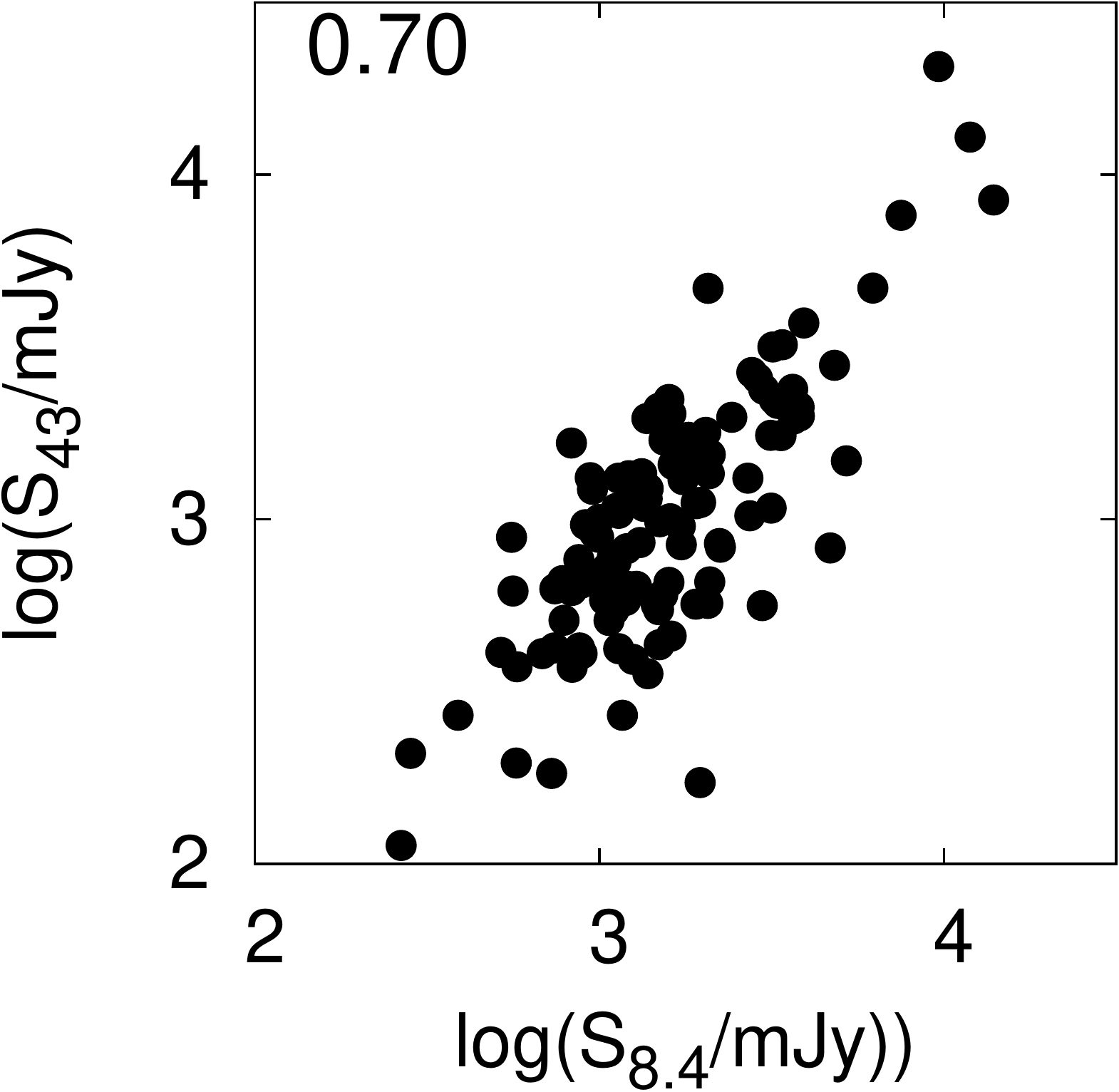}
\includegraphics[scale=0.23]{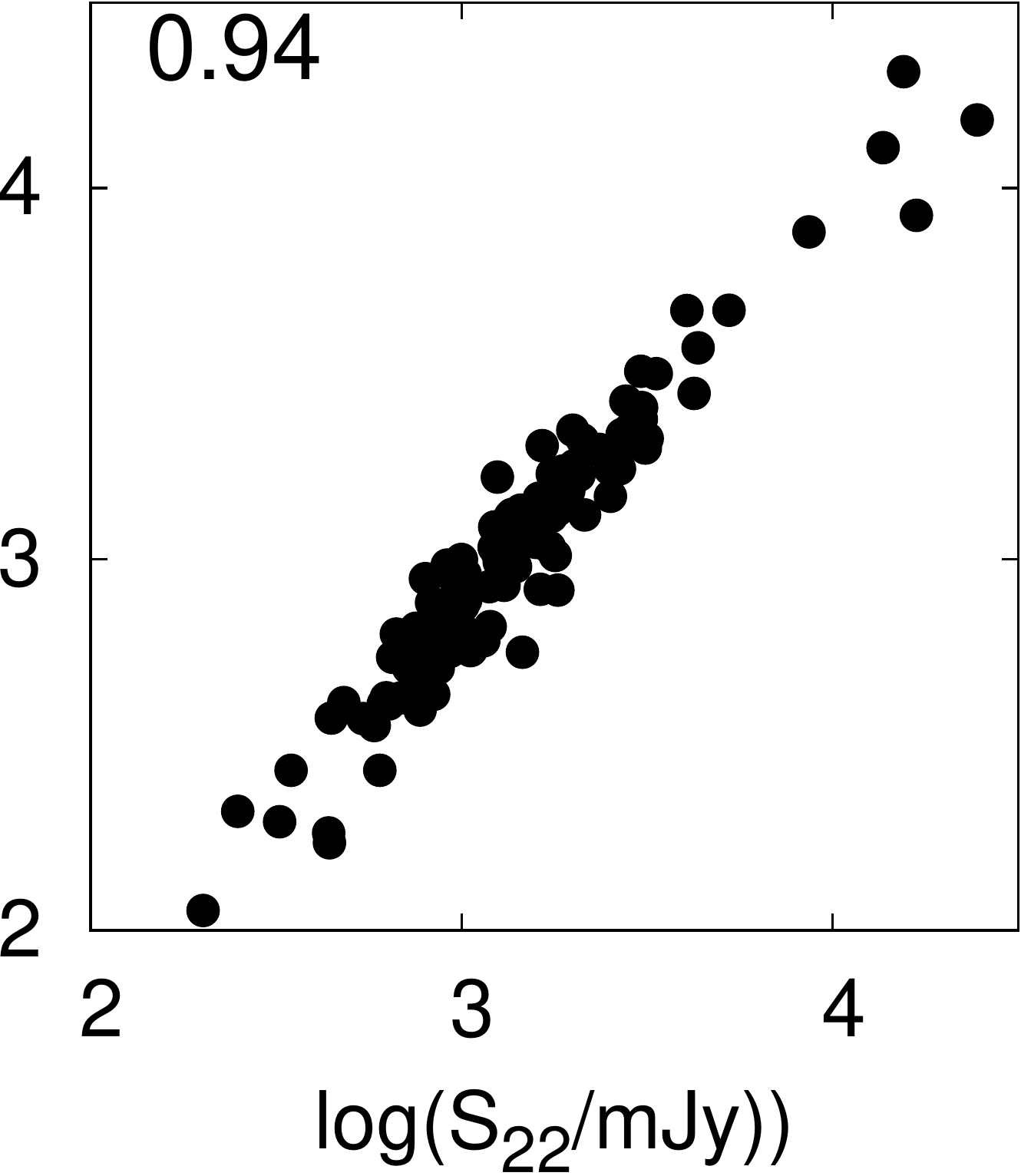}\\
\includegraphics[scale=0.23]{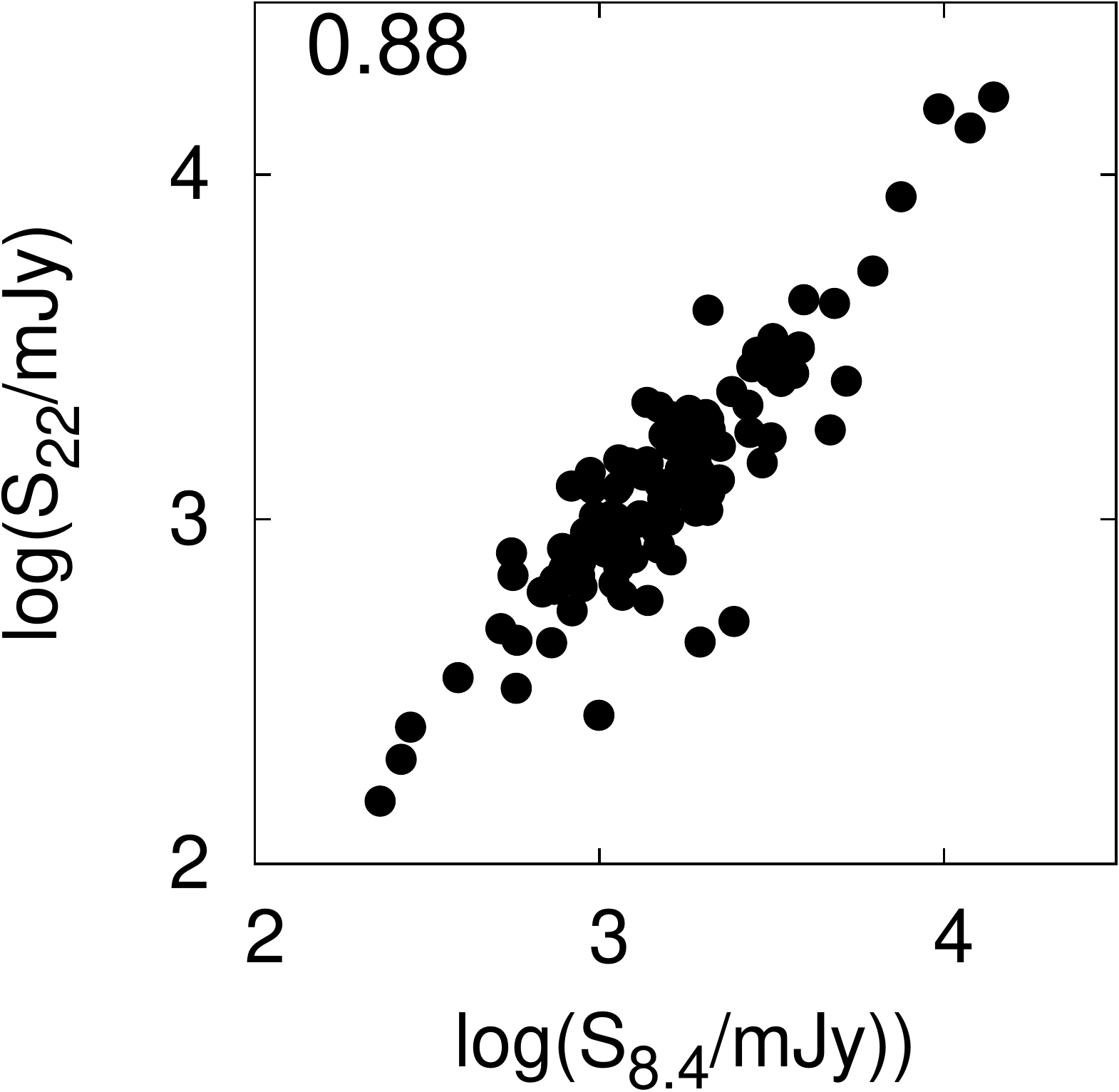}
\caption{Correlations between the 6 observed quantites, the three intensities and the three polarized intensities, for the contemporaneous sample, along with a table of the correlation coefficients. It is clear that there are strong correlations between most of the quantities.}
\label{fig:corrobs}
\end{figure}

In this appendix we present an analysis of the correlations between $S_{\nu_i}$ and $P_{\nu_i}$ for each of the 3 frequencies for the contemporaneous sample. In Fig.~\ref{fig:corrobs} we present a series of plots exhibiting the correlations between the 6 observed quantities. There are strong correlation (the correlation coefficient $>0.5$) between all 6 quantities with some, for example, those between the intensity at the three frequencies having very high correlation coefficients ($>0.7$). These correlations lend support to our assumption that the sample of data contains considerable information and the polarization measurements are not dominated by noise.

\end{document}

%% file: det.tex
0029+059 & $ 116.0 \pm   0.5$ & $ 129.8 \pm   3.9 $ & $ 146.9 \pm   2.5 $ & $ -3116 \pm    63$ & $ 138.2 \pm   2.5$ \\
0050-068 & $ 144.8 \pm   0.4$ & $ 145.2 \pm   5.1 $ & $ 127.9 \pm   1.1 $ & $  2887 \pm    82$ & $ 136.0 \pm   1.1$ \\ 
0121+118 & $  90.3 \pm   0.1$ & $  15.8 \pm   0.9 $ & $   8.0 \pm   0.3 $ & $  1199 \pm    15$ & $  11.3 \pm   0.3$ \\
0308+040 & $ 105.0 \pm   0.6$ & $  63.3 \pm   5.1 $ & $ 115.8 \pm   5.2 $ & $ -5146 \pm    83$ & $ 101.4 \pm   5.2$ \\ 
0329-239 & $ 157.0 \pm   0.5$ & $ 103.7 \pm   4.0 $ & $ 110.3 \pm   5.0 $ & $ -2037 \pm    65$ & $ 104.6 \pm   5.0$ \\ 
0340-213 & $ 132.3 \pm   0.4$ & $  38.8 \pm   3.8 $ & $  30.3 \pm   2.8 $ & $  1503 \pm    62$ & $  34.5 \pm   2.8$ \\ 
0405-130 & $ 157.9 \pm   0.5$ & $ 169.4 \pm   1.0 $ & $ 166.4 \pm   2.2 $ & $  -184 \pm    18$ & $ 165.9 \pm   2.2$ \\ 
0423+023 & $ 126.0 \pm   2.4$ & $ 123.8 \pm   3.9 $ & $ 115.9 \pm   5.2 $ & $    43 \pm    74$ & $ 116.0 \pm   5.2$ \\ 
0453-281 & $ 123.8 \pm   2.4$ & $  62.6 \pm   4.9 $ & $  11.5 \pm   5.8 $ & $  6791 \pm    88$ & $  30.5 \pm   5.8$ \\ 
0501-019 & $ 109.7 \pm   0.9$ & $  33.4 \pm   6.0 $ & $  34.1 \pm   0.8 $ & $  1210 \pm    98$ & $  37.5 \pm   0.8$ \\ 
0513-219 & $ 122.1 \pm   1.7$ & $ 135.9 \pm   4.1 $ & $ 134.1 \pm   1.2 $ & $  -215 \pm    72$ & $ 133.5 \pm   1.2$ \\ 
0608-223 & $  89.2 \pm   0.4$ & $  50.0 \pm   2.2 $ & $  45.3 \pm   0.8 $ & $   632 \pm    36$ & $  47.0 \pm   0.8$ \\ 
0721+713 & $ 180.0 \pm   0.1$ & $  28.5 \pm   5.6 $ & $  52.0 \pm   0.3 $ & $ -3357 \pm    90$ & $  42.6 \pm   0.4$ \\ 
0738+177 & $ 147.4 \pm   0.8$ & $ 160.7 \pm   2.3 $ & $ 154.7 \pm   4.0 $ & $  -211 \pm    40$ & $ 154.1 \pm   4.0$ \\ 
0750+125 & $  69.4 \pm   0.4$ & $  33.2 \pm   0.8 $ & $  27.1 \pm   0.5 $ & $   588 \pm    14$ & $  28.7 \pm   0.5$ \\ 
0757+099 & $  46.7 \pm   0.2$ & $  23.0 \pm   1.3 $ & $  17.8 \pm   0.8 $ & $   386 \pm    22$ & $  18.9 \pm   0.8$ \\ 
0813+482 & $  11.9 \pm   0.5$ & $ 134.4 \pm   2.2 $ & $ 119.2 \pm   3.9 $ & $   931 \pm    36$ & $ 121.8 \pm   3.9$ \\ 
0902-142 & $  84.7 \pm   0.2$ & $  56.8 \pm   4.1 $ & $ 102.1 \pm   4.0 $ & $ -5361 \pm    66$ & $  87.1 \pm   4.0$ \\ 
0914+028 & $ 104.6 \pm   0.4$ & $ 119.7 \pm   2.5 $ & $ 119.7 \pm   2.3 $ & $  -241 \pm    40$ & $ 119.0 \pm   2.3$ \\ 
0920+446 & $ 162.6 \pm   0.4$ & $ 126.3 \pm   2.0 $ & $  93.3 \pm   2.5 $ & $  3492 \pm    32$ & $ 103.1 \pm   2.5$ \\ 
0958+473 & $  19.5 \pm   0.5$ & $ 150.0 \pm   3.0 $ & $ 156.4 \pm   4.6 $ & $ -2099 \pm    49$ & $ 150.5 \pm   4.6$ \\ 
1041+061 & $  80.1 \pm   0.3$ & $  36.3 \pm   2.8 $ & $  34.1 \pm   2.6 $ & $   703 \pm    46$ & $  36.0 \pm   2.6$ \\ 
1130-148 & $ 131.6 \pm   0.1$ & $ 167.8 \pm   2.8 $ & $ 157.9 \pm   4.0 $ & $  2314 \pm    45$ & $ 164.4 \pm   4.0$ \\ 
1153+495 & $  90.6 \pm   0.2$ & $  14.3 \pm   3.3 $ & $  26.0 \pm   4.9 $ & $ -1671 \pm    54$ & $  21.3 \pm   4.9$ \\ 
1155+810 & $  74.0 \pm   0.4$ & $  86.0 \pm   2.3 $ & $  73.6 \pm   4.6 $ & $  2705 \pm    38$ & $  81.1 \pm   4.6$ \\ 
1331+305 & $  33.0 \pm   0.3$ & $  32.7 \pm   0.8 $ & $  32.6 \pm   0.4 $ & $     5 \pm    13$ & $  32.6 \pm   0.4$ \\ 
1354-106 & $ 152.7 \pm   0.4$ & $  35.9 \pm   3.7 $ & $  14.5 \pm   4.4 $ & $  1890 \pm    60$ & $  19.8 \pm   4.4$ \\ 
1408-078 & $ 166.2 \pm   0.4$ & $  66.2 \pm   3.2 $ & $  24.9 \pm   1.9 $ & $  4526 \pm    52$ & $  37.6 \pm   1.9$ \\  
1504+105 & $ 155.9 \pm   0.4$ & $ 105.0 \pm   1.8 $ & $  72.9 \pm   0.7 $ & $  3730 \pm    29$ & $  83.3 \pm   0.7$ \\ 
1512-090 & $ 117.4 \pm   0.7$ & $ 112.1 \pm   1.5 $ & $ 114.6 \pm   0.7 $ & $    80 \pm    26$ & $ 114.8 \pm   0.7$ \\ 
1513-100 & $ 157.5 \pm   1.5$ & $ 154.2 \pm   1.9 $ & $ 154.6 \pm   1.9 $ & $    52 \pm    38$ & $ 154.8 \pm   1.9$ \\ 
1540+147 & $ 154.8 \pm   0.4$ & $ 142.5 \pm   0.6 $ & $ 140.5 \pm   0.3 $ & $   200 \pm    11$ & $ 141.1 \pm   0.3$ \\ 
1550+054 & $ 144.9 \pm   0.3$ & $ 141.3 \pm   0.5 $ & $ 139.8 \pm   0.3 $ & $    60 \pm     9$ & $ 140.0 \pm   0.3$ \\ 
1608+104 & $ 161.7 \pm   0.6$ & $  74.9 \pm   2.6 $ & $  65.7 \pm   2.2 $ & $  1399 \pm    43$ & $  69.6 \pm   2.2$ \\ 
1657+479 & $  71.5 \pm   0.7$ & $ 148.5 \pm   1.3 $ & $ 135.4 \pm   8.4 $ & $  1661 \pm    23$ & $ 140.0 \pm   8.4$ \\ 
1800+784 & $ 111.7 \pm   0.4$ & $  99.9 \pm   0.8 $ & $  97.2 \pm   0.4 $ & $   192 \pm    14$ & $  97.8 \pm   0.4$ \\ 
1801+440 & $  49.9 \pm   0.2$ & $  44.2 \pm   0.5 $ & $  48.5 \pm   2.6 $ & $    91 \pm     9$ & $  48.8 \pm   2.6$ \\ 
1824+568 & $  33.2 \pm   0.3$ & $  19.4 \pm   1.1 $ & $  15.4 \pm   0.4 $ & $   227 \pm    18$ & $  16.1 \pm   0.4$ \\ 
1849+670 & $ 143.9 \pm   0.4$ & $ 116.5 \pm   3.6 $ & $ 155.4 \pm   0.9 $ & $ -5359 \pm    57$ & $ 140.4 \pm   0.9$ \\ 
1902+318 & $  44.2 \pm   0.2$ & $  27.9 \pm   2.4 $ & $  43.2 \pm   2.0 $ & $ -2633 \pm    39$ & $  35.9 \pm   2.0$ \\ 
1939-154 & $ 157.0 \pm   0.5$ & $  61.6 \pm   5.9 $ & $  29.8 \pm   3.1 $ & $  4437 \pm    94$ & $  42.2 \pm   3.1$ \\ 
2011-157 & $ 149.7 \pm   0.9$ & $ 124.6 \pm   3.1 $ & $  95.0 \pm   8.3 $ & $  3309 \pm    52$ & $ 104.3 \pm   8.3$ \\ 
2143+176 & $ 142.4 \pm   0.8$ & $  11.4 \pm   1.7 $ & $  13.7 \pm   2.0 $ & $  -786 \pm    30$ & $  11.5 \pm   2.0$ \\ 
2346+094 & $ 130.5 \pm   0.2$ & $ 150.3 \pm   1.7 $ & $ 154.0 \pm   1.5 $ & $  -322 \pm    27$ & $ 153.1 \pm   1.5$ \\ 
2358-102 & $ 122.8 \pm   0.3$ & $  32.0 \pm   5.5 $ & $  17.1 \pm   3.6 $ & $  1470 \pm    88$ & $  21.2 \pm   3.6$ \\ 